\documentclass[a4paper,11pt]{article}
\pdfoutput=1 
\date{}
\usepackage{jheppub}
\usepackage[utf8]{inputenc}
\usepackage[english]{babel}
\usepackage{amsmath,amssymb,amsfonts,amsxtra, mathrsfs,graphics,graphicx,amsthm,epsfig, youngtab,bm,longtable,float,tikz,empheq}
\usepackage[margin=0.5cm]{caption}
\usepackage{thmtools}
\declaretheorem[numbered=no]{result}
\declaretheorem[numbered=no]{definition}
\allowdisplaybreaks[1]
\usetikzlibrary{positioning}
\usetikzlibrary{automata}
\usetikzlibrary{arrows}
\usetikzlibrary{calc}
\usetikzlibrary{decorations.markings}
\usetikzlibrary{decorations.pathreplacing}
\usetikzlibrary{intersections}
\usetikzlibrary{positioning}
\usetikzlibrary{topaths}
\usetikzlibrary{shapes.geometric}
\usetikzlibrary{shapes.misc}
\tikzset{cf-group/.style = {
    shape = rounded rectangle, minimum size=1.0cm,
    rotate=90,
    rounded rectangle right arc = none,
    draw}}
\tikzset{cross/.style={path picture={ 
  \draw[black]
(path picture bounding box.south east) -- (path picture bounding box.north west) (path picture bounding box.south west) -- (path picture bounding box.north east);
}}}

\newcommand*\widefbox[1]{\fbox{\hspace{2em}#1\hspace{2em}}}

\newcommand{\be}{\begin{equation}}
\newcommand{\ee}{\end{equation}}
\newcommand{\ba}{\begin{array}}
\newcommand{\ea}{\end{array}} 
\newcommand{\bi}{\begin{itemize}}
\newcommand{\ei}{\end{itemize}}
\def\vec#1{\bm{#1}}
\def\bea#1\eea{\allowdisplaybreaMs \begin{align}#1\end{align}}
 \newcommand{\ben}{\begin{enumerate}}
\newcommand{\een}{\end{enumerate}}
\newcommand{\bean}{\begin{eqnarray*}}
\newcommand{\eean}{\end{eqnarray*}}
\newcommand{\eref}[1]{(\ref{#1})}

\newcommand{\nn}{\nonumber}

\newcommand{\tr}{\mathrm{Tr}}

\newcommand{\BR}{\mathbb{R}}

\newcommand{\BZ}{\mathbb{Z}}

\newcommand{\comment}[1]{}

\newcommand{\CS}{{\cal S}}

\newcommand{\CT}{{\cal T}}
\newcommand{\CD}{{\cal D}}
\newcommand{\CM}{{\cal M}}
\newcommand{\CO}{{\cal O}}

\newcommand{\CN}{{\cal N}}

\newcommand{\CP}{{\cal P}}

\newcommand{\CZ}{{\cal Z}}
\newcommand{\CR}{{\cal R}}

\newcommand{\CU}{{\cal U}}

\newcommand{\tm}{\widetilde{m}}

\newcommand{\wt}{\widetilde}
\newcommand{\wh}{\widehat}

\newcommand{\ch}{\cosh \pi}
\newcommand{\s}{\sigma}
\newcommand{\bs}{\overline{\sigma}}

\newcommand{\Secref}[1]{Section~\ref{#1}}

\newcommand{\figref}[1]{Fig.~\ref{#1}}
\renewcommand{\eqref}[1]{(\ref{#1})}

\title{Higgs Branches of Argyres-Douglas theories as Quiver Varieties}
\author{Anindya Dey}
\affiliation{Department of Physics and Astronomy, Johns Hopkins University, 3400 North Charles Street,
Baltimore, MD 21218, USA}
\emailAdd{anindya.hepth@gmail.com}
\abstract{We present a general prescription for constructing 3d $\CN=4$ Lagrangians for the IR
SCFTs that arise from the circle reduction of a large class of Argyres-Douglas theories. The resultant Lagrangian 
gives a realization of the Higgs branch of the 4d SCFT as a quiver variety, up to a set of decoupled interacting SCFTs 
with empty Higgs branches. As representative examples, we focus on the 
families $(A_{p-N-1}, A_{N-1})$ and $D_p(SU(N))$. The Lagrangian in question is generically a non-ADE-type 
quiver gauge theory involving only unitary gauge nodes 
with fundamental and bifundamental hypermultiplets, as well as hypermultiplets which are only charged 
under the $U(1)$ subgroups of certain gauge nodes. 
Our starting point is the Lagrangian 3d mirror of the circle-reduced Argyres-Douglas theory, 
which can be read off from the class $\CS$ construction. 
Using the toolkit of the $S$-type operations, developed in \cite{Dey:2020hfe}, we show that the mirror of 
the 3d mirror for any Argyres-Douglas theory in the aforementioned families is guaranteed to be a 
Lagrangian theory of the above type, up to some decoupled free sectors. 
We comment on the extension of this procedure to 
other families of Argyres-Douglas theories. In addition, for the case of $D_p(SU(N))$ theories, we 
compare these 3d Lagrangians to the ones found in  
\cite{Closset:2020afy} and propose that the two are related by an IR duality. 
We check the proposed IR duality at the level of the three-sphere partition 
function for specific examples. In contrast to the 3d Lagrangians in \cite{Closset:2020afy}, which are linear chains involving 
unitary-special unitary nodes, we observe that the Coulomb branch global symmetries are manifest in the 3d Lagrangians 
that we find.}

\begin{document}
\maketitle

\section{Introduction and Summary}

\subsection{Overview and the basic idea of the paper}

\subsubsection{AD theories and 3d mirrors from the class $\CS$ construction}\label{Intro-classS}

Argyres-Douglas (AD) theories constitute a large class of 4d $\CN=2$ SCFTs for which the Coulomb branch operators 
have fractional scaling dimensions. The first examples of such theories were realized as special points on the Coulomb 
branch of Lagrangian theories in 4d \cite{Argyres:1995jj, Argyres:1995xn}. 
Over the years, various important subclasses of such theories have been constructed 
via the geometric engineering technique \cite{Cecotti:2010fi, Cecotti:2013lda} of compactifying Type IIB superstring theory on certain singular Calabi-Yau 
hypersurfaces, as well as by the class $\CS$ construction \cite{Gaiotto:2009hg, Xie:2012hs}. As is well known, the latter involves 
compactifying a 6d (2,0) theory (which has an ADE classification) on a Riemann sphere with either a single irregular puncture, 
or a single irregular puncture with a single regular puncture \cite{Xie:2012hs} \footnote{Certain AD theories can also be realized by the 
compactification of twisted ADE theories on a Riemann surface with only regular punctures, as pointed out in \cite{Beem:2020pry}.} . \\

One can compactify a given 4d $\CN=2$ SCFT to 3d by putting the theory on $\BR^3 \times S^1$, and flowing to 
the deep IR. The resultant 3d $\CN=4$ SCFT may or may not have a Lagrangian description, but is 
known to have a Lagrangian 3d mirror for a large subset of class $\CS$ theories.  
Given the class $\CS$ construction of a 4d $\CN=2$ SCFT, the basic idea is to show that the associated 
Hitchin moduli space (which is the Coulomb branch of the 
theory on $\BR^3 \times S^1$) in the IR limit is isomorphic to the Higgs branch of a 3d $\CN=4$ quiver gauge theory. 
The aforementioned 3d quiver gauge theory can be explicitly built from the class $\CS$ data in the following generic 
steps:
\begin{itemize}

\item To every puncture on the Riemann surface, one associates a quiver gauge theory, which can be read off 
from the singularity data of the puncture.

\item Given the set of quivers associated with the Riemann surface with punctures, one glues them together, 
following certain rules which depend on the specific punctures involved.

\end{itemize}

The 3d mirrors for 4d $\CN=2$ SCFTs, which arise from 6d $A_{N-1}$ (2,0) theories compactified on 
a Riemann surface with regular punctures, was found in \cite{boalch2007quivers, Benini:2010uu}. 
For an important subclass of irregular punctures, which appear 
in the realization of AD theories, the analogous result was first obtained in \cite{boalch2008irregular} and 
was extended in subsequent papers \cite{Xie:2012hs,Xie:2013jc,Wang:2018gvb}. 
For a recent review of the subject, we would like to refer the reader to \cite{Xie:2021ewm}.

In this work, we would be primarily interested in the 3d mirrors of the AD theories of the following families:  
$(A_{p-N-1}, D_{N-1})$ and $D_p(SU(N))$. The first family is realized by a single irregular puncture on a Riemann sphere, 
while the second is realized by a single irregular puncture and a single regular puncture. The 3d mirror for each family 
is discussed in detail in \Secref{AAgen} and \Secref{DpSUNgen}.

\subsubsection{$S$-type operations and 3d Lagrangians for AD theories}\label{Intro-SOp}

In \cite{Dey:2020hfe}, a systematic field theory construction for engineering new 3d mirror dualities was presented, 
starting from a well-defined set of basic dualities. Let us briefly summarize the construction. We consider a 
class of 3d SCFTs (which we refer to as class $\CU$) with UV Lagrangians for which the Higgs branch 
global symmetry has a subgroup $G^{\rm sub}_{\rm global}=\prod_\gamma U(M_\gamma) \subset G_H$.

Given a UV Lagrangian $X$ in class $\CU$, one can define a map which acts on $X$ to give a generically 
different theory $X'$, i.e.
\be \label{Smap}
\CO^\alpha_{\vec \CP}: X[\vec{\wh{A}}] \mapsto X'[\vec{\wh{B}}],
\ee
where the phrase ``different theory" implies that $X$ and $X'$ flow to different SCFTs in the IR. The precise definition 
of the map $\CO^\alpha_{\vec \CP}$ is reviewed in \Secref{SOp-rev}. In \eref{Smap}, $\wh{\vec{A}}$ and $\wh{\vec B}$ 
collectively represent the background fields associated with the global symmetries of the theories $X$ and $X'$ respectively. 
We refer to the above map as an ``elementary $S$-type operation". The name arises from the fact that in 
the special case, where $G^{\rm sub}_{\rm global}=U(1)$, and $\CO^\alpha_{\vec \CP}$ 
is a gauging operation of the said $U(1)$, the above map coincides with the $S$ generator of 
Witten's $SL(2,\BZ)$ action \cite{Witten:2003ya} on a 3d CFT.\\ 

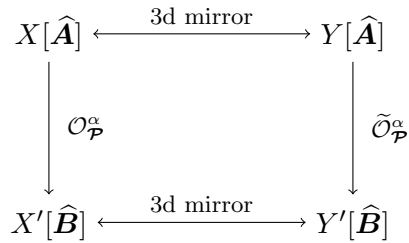
\begin{figure}[htbp]
\begin{center}
\begin{tikzpicture}
  \node (X) at (2,1.5) {$X[\wh{\vec A}]$};
  \node (Y) at (6,1.5) {$Y[\wh{\vec A}]$};
  \node (D) at (2,-1) {$X'[\wh{\vec B}]$};
  \node (E) at (6,-1) {$Y'[\wh{\vec B}]$};
   \draw[->] (X) -- (D) node [midway,  right=+3pt] {\footnotesize $\CO^\alpha_{\vec \CP}$};
  \draw[<->] (X) -- (Y)node [midway,above  ] {\footnotesize 3d mirror};
    \draw[->] (Y) -- (E) node [midway,  right=+3pt] {\footnotesize $\wt{\CO}^\alpha_{\vec \CP}$};
        \draw[<->] (D) -- (E)node  [midway,above  ] {\footnotesize 3d mirror};
\end{tikzpicture}
\end{center}
\caption{\footnotesize{Generating new dual pairs using an elementary $S$-type operation.} }
\label{S-OP-duality}
\end{figure}

Suppose the theory $X[\wh{\vec A}]$ has a 3d mirror $Y[\wh{\vec A}]$, such that both theories 
have Lagrangian realizations. For a given elementary $S$-type operation $\CO^\alpha_{\vec \CP}$ on $X[\wh{\vec A}]$, one can 
define a dual operation $\wt{\CO}^\alpha_{\vec \CP}$ on $Y[\wh{\vec A}]$: 
\be
\wt{\CO}^\alpha_{\vec \CP}: Y[\wh{\vec A}] \mapsto Y'[\wh{\vec B}],
\ee
which maps it to a different theory $Y'[\wh{\vec B}]$, such that the pair of theories $(X'[\wh{\vec B}],Y'[\wh{\vec B}])$ are 3d mirrors. 
The four theories $X[\wh{\vec A}], Y[\wh{\vec A}]$, $X'[\wh{\vec B}]$, and $Y'[\wh{\vec B}]$, are therefore related as shown in 
\figref{S-OP-duality}. 

The dual operation $\wt{\CO}^\alpha_{\vec \CP}$ can be determined explicitly, with the help of RG-invariant supersymmetric observables 
computed using localization, as discussed in \cite{Dey:2020hfe}. The procedure is reviewed in \Secref{SOp-rev}. Given the pair 
$(X[\wh{\vec A}],Y[\wh{\vec A}])$ and the map $\CO^\alpha_{\vec \CP}$, one can then read off the Lagrangian 
for the theory $Y'[\wh{\vec B}]$, assuming a weakly coupled description of the theory exists.
Given the recipe for finding the dual of the $S$-type operation, the strategy for engineering new mirror pairs can be stated as follows. 
We first define a conveniently small subset of mirror pairs, which are well understood from the String Theory
and/or the QFT perspective, and refer to it as the set of ``basic dualities". Starting with a dual pair $(X,Y)$ in this set of 
basic dualities, one can then implement the prescription of \figref{S-OP-duality} sequentially to generate new mirror pairs. 
We will pick the set of basic dualities to be the set of mirror pairs involving linear (A-type)
quiver gauge theories with unitary gauge groups. Additionally, we will demand that the quivers are good theories in the 
Gaiotto-Witten sense \cite{Gaiotto:2008ak}. Mirror symmetry for this class of theories has been studied in a lot of detail using the Type IIB 
Hanany-Witten brane construction \cite{Hanany:1996ie} on the one hand and QFT approaches on the other. 
In particular, it was shown in \cite{Gaiotto:2008ak} that the mirror of a good linear quiver is also a good linear quiver. 
Using the construction outlined above, one can then construct mirror pairs 
involving quivers of arbitrary shapes.\\

In this paper, we will use this construction to determine 3d $\CN=4$ Lagrangians for families of AD theories which have Lagrangian 
3d mirrors. Let this Lagrangian 3d mirror be $X'$ (see \figref{S-OP-duality} above). The strategy is to show that $X'$ can be engineered  
from a good linear quiver $X$ by a sequence of $S$-type operations. Since $X$ has a 3d mirror $Y$, which is also a good linear 
quiver, the 3d mirror of $X'$ can be obtained by a sequence of dual $S$-type operations on the theory $Y$. If the resultant theory 
$Y'$ is Lagrangian, then it gives a Lagrangian realization for the 3d SCFT obtained from the circle-reduction of the AD theory. 
As mentioned above, for a completely generic quiver gauge theory $X'$, the theory $Y'$ is not Lagrangian. However, for the AD theories under 
consideration, $X'$ can be engineered from a linear quiver $X$ by a very special class of $S$-type operations, for which 
the theory $Y'$ is guaranteed to have a Lagrangian description.

\subsubsection{Another 3d Lagrangian for a $D_p(SU(N))$ theory}\label{Intro-Closset}

In \cite{Closset:2020afy, Giacomelli:2020ryy}, the authors derived 3d Lagrangians for $D_p(SU(N))$ theories, by making use of a partially weakly-coupled description of the 4d $\CN=2$ SCFTs. We briefly summarize the results relevant for our paper below. \\

The $D_p(SU(N))$ theories are specified by the SW curve:
\be \label{SW-DpN}
x^N + z^p + \sum_{i,j \geq 0} u_{ij}\, z^i\, x^j =0, \quad \lambda_{\rm SW}= \frac{x\,dz}{z},
\ee
where the scaling dimension of the SW differential is $[\lambda_{\rm SW}]=1$, and this implies 
that the other scaling dimensions are given as
\begin{align}
[x]=1, \quad [z]=\frac{N}{p}, \quad [u_{ij}]= N-j - \frac{N}{p}\,i.
\end{align}

If $p,N$ are coprime, then $D_p(SU(N))$ is an isolated SCFT, i.e. it has no conformal manifold. There is 
another isolated SCFT that will be important in this story: $\CD_q(N, N-n)$, which has the following SW curve
\begin{align}
& x^N + x^{N-n}\, z^q +  \sum_{i,j \geq 0} u_{ij}\, z^i\, x^j =0, \quad \lambda_{\rm SW}= \frac{x\,dz}{z}, \nn \\
& \implies \, \frac{x^N}{z^q} +  x^{N-n} + \sum_{i,j \geq 0} v_{ij}\, z^i\, x^j =0.
\end{align}
If $n,q$ are chosen such that $n= N/{\rm GCD}(p,N)$ and $q= p/{\rm GCD}(p,N)$, then $\CD_q(N, N-n)$ is an 
isolated SCFT, as can be seen by counting the number of deformations of scaling dimension zero. These theories 
have a global symmetry $SU(N) \times U(1) \times SU(N-n)$.\\

Given an SCFT $D_p(SU(N))$, the monomial $x^{N-n}z^q$ in the SW curve corresponds to a marginal deformation. Also, the 
monomials $x^j z^q$ with $j=0,1,\ldots,N-n-1$ correspond to integer scaling dimensions 
$\Delta_{qj}=N-n, N-n-1, \ldots, 1$. Leaving out the mass deformation, the remaining deformations correspond 
to CB operators of a conformally gauged $SU(N-n)$ factor of the 4d SCFT. Together with the marginal deformation 
term, these terms in the SW curve can be organized as follows (after appropriate rescaling):
\be
x^{N-n}z^q + \sum^{N-n-2}_{j=0}\,u_{qj}x^j z^q =: (x^{N-n} + \sum^{N-n}_{j=2}\,u_{j}x^{N-n-j} )\,z^q
=: P_{N-n}(x)\,z^q.
\ee
The SW curve can then be written as
\begin{align}
x^N + \ldots + P_{N-n}(x)\,z^q + \ldots + z^p=0, \implies \frac{x^N}{z^q} + \ldots + P_{N-n}(x) + \ldots + z^{p-q}=0.
\end{align}
In this form, the above SW curve represents a $\CD_q(N, N-n)$ system and a $D_{p-q}(SU(N-n))$ system with the diagonal 
$SU(N-n)$ flavor symmetry being conformally gauged. This allows us to write a partially weakly-coupled description of the 
SCFT as follows:
\be
D_p(SU(N)) = \CD_q(N, N-n) \longleftarrow SU(N-n) \longrightarrow D_{p-q}(SU(N-n)).
\ee
Applying this argument iteratively, one can write the following partially weakly coupled description:
\begin{align} \label{DpN-wc}
& D_p(SU(N)) =  \CD_q(N, N-n) \longleftarrow {SU(N-n)} \longrightarrow \CD_q(N-n, N-2n)  \longleftarrow { SU(N-2n)}  \nn \\
& \longrightarrow \CD_q(N-2n, N-3n) \longleftarrow { SU(N-3n)} \longrightarrow  \CD_q(N-3n, N-4n) \longleftarrow {SU(N-4n)} \nn \\
& \longrightarrow \CD_q(N-4n, N-5n)  \longleftarrow \ldots \longleftarrow { SU(n)} \longrightarrow \CD_q(n,0),
\end{align}
where $\CD_q(n,0)= D_q(SU(n))$. This allows one to write the SCFT $D_p(SU(N))$ in terms of a set of isolated SCFTs with 
conformally-gauged special unitary flavor subgroups.\\

The isolated SCFT $\CD_q(N, K)$ can be realized as the low energy 
effective field theory at the origin of the Coulomb branch of a 4d Lagrangian theory, with appropriately tuned flavor 
masses. The 4d Lagrangian theory in question is a linear chain of special unitary gauge groups with $N$ fundamental flavors at one end,
and $K$ fundamental flavors at the other. For example, with $K=N-1$, i.e. $n=N/{\rm GCD}(p,N)=1$, the 4d Lagrangian is given as
\be \label{CDq-1}
\CD_q(N, N-1) = [N] - SU(N)_1 - SU(N)_2 - \ldots - SU(N)_{q-1} - [N-1].
\ee
The case for general $(N,K)$ can be found in \cite{Closset:2020afy}. Given the description in \eref{DpN-wc} and the above realization of the SCFTs 
$\CD_q(N, K)$, the 3d Lagrangian can be obtained as follows:

\begin{enumerate}

\item  The 3d limit of a given isolated SCFT $\CD_q(N, K)$ is obtained by studying the low energy effective theory of 
the dimensionally reduced linear quiver at the origin of the Coulomb branch, with the flavor masses turned on.
Let us call these 3d theories $\CT_i$, where $i=1,\ldots,m-1$. The theory $\CT_i$ associated with an isolated SCFT $\CD_q(N, K)$ 
turns out to be a linear quiver with unitary gauge nodes and Higgs branch global symmetry $SU(N) \times U(1) \times U(K)$.

\item The 4d vector multiplets of the conformally-gauged $SU$ factors become 3d vector multiplets under 
dimensional reduction. 

\item Given the set of 3d theories $\{ \CT_i \}$, the 3d limit of $D_p(SU(N))$ is then given by a linear chain of unitary and 
special unitary nodes with bifundamental and fundamental hypers:
\be \label{3dLag-gen}
[D_p(SU(N))]_{3d}: \CT_1- SU(N-n) - \CT_2 - SU(N-2n) - \ldots - SU(n) - \CT_{m-1}.
\ee
In certain cases, the 3d theory can have ugly nodes, but one can always reduce the theory to a quiver of the form \eref{3dLag-gen} 
and a set of free hypermultiplets. 
\end{enumerate}

In \Secref{DpSUNgen}, we note that the 3d Lagrangian in \eref{3dLag-gen} and the 3d Lagrangian constructed using $S$-type operations
are different quiver gauge theories, i.e. they have different gauge groups and matter content. We propose that the two 3d Lagrangians are 
related by an IR duality, and perform certain checks of the proposed duality for specific examples in \Secref{U2-USU}. 

\subsection{Summary of the main results}

The main results of the paper can be summarized as follows:

\begin{itemize}

\item \textbf{Existence of 3d Lagrangians for AD theories on circle reduction:} We present a general proof of the fact that the 
3d SCFTs which arise in the IR limit of the AD theories $(A_{p-N-1}, A_{N-1})$ ($p > N$) and $D_p(SU(N))$ ($p \gtreqless N$), 
compactified on a circle, always have an $\CN=4$ UV Lagrangian description (up to a free sector). 
Using the strategy outlined in \Secref{Intro-SOp}, the basic idea
is to show that the 3d mirrors of the circle-reduced AD theories, which can 
be constructed from the class $\CS$ construction (which we will refer to as class $\CS$ mirrors), themselves have a Lagrangian mirror. 
The latter then gives a 3d Lagrangian description of the IR SCFT that arises from the circle reduction of the AD theory in question.\\

For the two families of AD theories mentioned above, we show that the class $\CS$ mirror can be constructed from an $A$-type quiver gauge 
theory by a sequence of Abelian $S$-type operations. Using the properties of Abelian $S$-type operations, reviewed in \Secref{SOp-rev}, 
we then show that the 3d mirror of the class $\CS$ mirror is manifestly Lagrangian.
\bigskip
\item \textbf{General features of the 3d Lagrangian and examples:} The construction outlined above gives a systematic procedure for 
constructing the 3d Lagrangians. In \Secref{AAgen} and \Secref{DpSUNgen}, we outline some of the general features of the 3d Lagrangian for different 
families of AD theories under consideration. We present concrete examples in \Secref{AAgen}, \Secref{pltNEx}-\Secref{pgtNEx} where we explicitly write down 
the 3d Lagrangian. The results can be summarized as follows:

\begin{enumerate}

\item \textbf{$(A_{p-N-1}, A_{N-1})$ theories:} In this case, the 3d Lagrangian is a generically non-linear quiver built out of $U(1)$ gauge nodes with 
fundamental and bifundamental matter, along with a set of free twisted hypermultiplets. 
After presenting the general procedure for obtaining the 3d quiver in \Secref{AAgen}, we work out an 
example for which $m=GCD(p,N)=3$ and $m_G=\frac{N(p-N)}{9}$. The 3d Lagrangian for this example is given as:
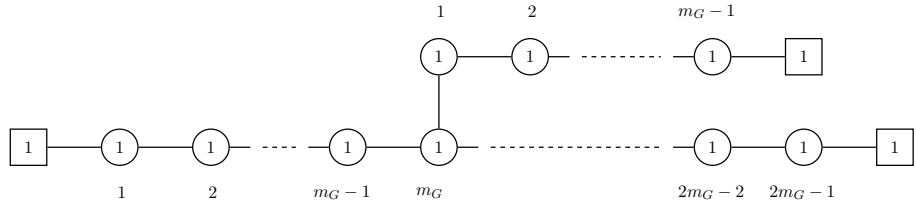
\begin{center}
\scalebox{.6}{\begin{tikzpicture}[node distance=2cm,cnode/.style={circle,draw,thick,minimum size=8mm},snode/.style={rectangle,draw,thick,minimum size=8mm},pnode/.style={rectangle,red,draw,thick,minimum size=8mm}]
\node[snode] (1) at (-2,0) {$1$};
\node[cnode] (2) at (0,0) {$1$};
\node[cnode] (3) at (2,0) {$1$};
\node[] (4) at (3,0) {};
\node[] (5) at (4,0) {};
\node[cnode] (6) at (5,0) {$1$};
\node[cnode] (7) at (7,0) {$1$};
\node[] (8) at (8,0) {};
\node[] (9) at (12,0) {};
\node[cnode] (10) at (13,0) {$1$};
\node[cnode] (11) at (15,0) {$1$};
\node[snode] (12) at (17,0) {$1$};
\node[cnode] (13) at (7,2) {$1$};
\node[cnode] (14) at (9,2) {$1$};
\node[] (15) at (10,2) {};
\node[] (16) at (12,2) {};
\node[cnode] (17) at (13,2) {$1$};
\node[snode] (30) at (15,2) {$1$};
\draw[thick] (1) -- (2);
\draw[thick] (2) -- (3);
\draw[thick] (3) -- (4);
\draw[thick,dashed] (4) -- (5);
\draw[thick] (5) -- (6);
\draw[thick] (6) -- (7);
\draw[thick] (7) -- (8);
\draw[thick,dashed] (8) -- (9);
\draw[thick] (9) -- (10);
\draw[thick] (10) -- (11);
\draw[thick] (11) -- (12);
\draw[thick] (7) -- (13);
\draw[thick] (13) -- (14);
\draw[thick] (14) -- (15);
\draw[thick,dashed] (15) -- (16);
\draw[thick] (16) -- (17);
\draw[thick] (17) -- (30);
\node[text width=0.1cm](20) at (0,-1) {$1$};
\node[text width=0.1cm](21) at (2,-1) {$2$};
\node[text width= 1.5 cm](22) at (5,-1) {$m_G-1$};
\node[text width=1 cm](23) at (7,-1) {$m_G$};
\node[text width=1.5 cm](24) at (13,-1) {$2m_G-2$};
\node[text width=1.5 cm](25) at (15,-1) {$2m_G-1$};
\node[text width=0.1 cm](26) at (7,3) {$1$};
\node[text width=0.1 cm](27) at (9,3) {$2$};
\node[text width=1.5 cm](28) at (13,3) {$m_G-1$};
\end{tikzpicture}}
\end{center}
and the number of free twisted hypermultiplets is $H=\frac{(N-3)(p-N-3)}{6}$. 

\bigskip

\item \textbf{$D_p(SU(N))$ theories:} In this case, the 3d Lagrangian is generically a non-linear quiver built out of unitary gauge nodes with 
fundamental and bifundamental hypermultiplets, as well as certain hypermultiplets charged only under certain $U(1)$ subgroups of the unitary 
groups. We will refer the latter as ``Abelian hypermultiplets". After presenting the general procedure for obtaining the 3d quiver in \Secref{DpSUNgen}, 
we present an example each for the cases $p < N$ and $p> N$ in \Secref{pltNEx} and \Secref{pgtNEx} respectively. The 3d Lagrangians in these 
cases are given as:

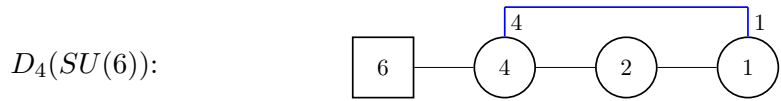
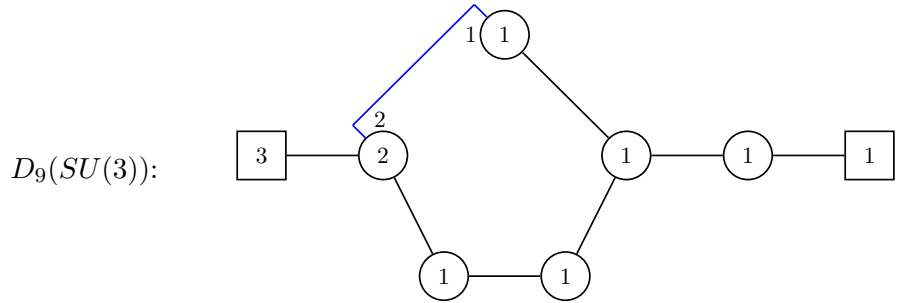
\begin{center}
\begin{tabular}{ccc}
{\begin{tikzpicture}
\node[text width=2cm] (1) at (0,0){$D_4(SU(6))$:}; 
\node[] (2) at (0,-1.6) {};
\end{tikzpicture}}
& \quad
& \scalebox{.8}{\begin{tikzpicture}[node distance=2cm, nnode/.style={circle,draw,thick, fill, inner sep=1 pt},cnode/.style={circle,draw,thick,minimum size=1.0 cm},snode/.style={rectangle,draw,thick,minimum size=1.0 cm}]
\node[snode] (1) at (0,0){6};
\node[cnode] (2) at (2,0){4};
\node[cnode] (3) at (4,0){2};
\node[cnode] (4) at (6,0){1};
\node[](5) at (2,-2){};
\node[](6) at (4,-2){};
\draw[-] (1) -- (2);
\draw[-] (2)-- (3);
\draw[-] (3)-- (4);
\draw[-, thick, blue] (2)-- (2,1);
\draw[-, thick, blue] (2,1)-- (6,1);
\draw[-, thick, blue] (6,1)-- (4);
\node[] (9) at (0,-0.75){};
\node[text width=.2cm](10) at (2.2, 0.75){4};
\node[text width=.2cm](11) at (6.2, 0.75){1};
\end{tikzpicture}}\\
{\begin{tikzpicture}
\node[text width=2cm] (1) at (0,0){$D_9(SU(3))$:}; 
\node[] (2) at (0,-1.6) {};
\end{tikzpicture}} 
& \quad
&\scalebox{.8}{\begin{tikzpicture}[node distance=2cm,cnode/.style={circle,draw,thick,minimum size=8mm},snode/.style={rectangle,draw,thick,minimum size=8mm},pnode/.style={rectangle,red,draw,thick,minimum size=8mm}]
\node[cnode] (1) at (-3,0) {$2$};
\node[snode] (2) at (-5,0) {$3$};
\node[cnode] (3) at (-1,2) {$1$};
\node[cnode] (4) at (-2,-2) {$1$};
\node[cnode] (5) at (0,-2) {$1$};
\node[cnode] (6) at (1,0) {$1$};
\node[cnode] (7) at (3,0) {$1$};
\node[snode] (8) at (5,0) {$1$};
\draw[thick] (1) -- (2);
\draw[thick, blue] (1) -- (-3.5,0.5);
\draw[thick, blue] (3) -- (-1.5, 2.5);
\draw[thick, blue] (-3.5,0.5) -- (-1.5, 2.5);
\draw[thick] (1) -- (4);
\draw[thick] (4) -- (5);
\draw[thick] (5) -- (6);
\draw[thick] (6) -- (7);
\draw[thick] (7) -- (8);
\draw[thick] (3) -- (6);
\node[text width=0.1cm](31) at (-3.1,0.6){2};
\node[text width=0.1cm](32) at (-1.6,2){1};
\end{tikzpicture}}
\end{tabular}
\end{center}

The blue lines in the quiver diagrams denote the Abelian hypermultiplets. The quiver notation is explained in detail in \Secref{Q-Not}.

\end{enumerate}

\bigskip

\item \textbf{Relation to 3d Lagrangians of Closset et al for $D_p(SU(N))$ theories and IR duality:} The authors of \cite{Closset:2020afy} showed that the circle reduction of the $D_p(SU(N))$ theories admit a Lagrangian realization in terms a linear chain of unitary and special unitary gauge nodes with fundamental 
and bifundamental hypermultiplets (reviewed in \Secref{Intro-Closset}). Therefore, comparing the 3d Lagrangian of \cite{Closset:2020afy} 
with that obtained by the construction above, one expects that for a given $D_p(SU(N))$ AD theory, there exists an IR duality (not mirror symmetry) between two 
theories belonging to the following distinct classes:

\begin{itemize}

\item Theory $\CT$: A linear chain of unitary and special unitary gauge groups with fundamentals and bifundamentals.

\item Theory $\CT^\vee$: A generically non-linear quiver involving only unitary gauge groups, decorated with Abelian hypermultiplets 
in addition to fundamental/bifundamental hypers.

\end{itemize}

In \Secref{IRdual-Dp-gen}, we present a general prescription for showing that the three-sphere partition function of the theory $\CT^\vee$ 
agrees with that of the theory $\CT$, associated with the circle reduction of a given $D_p(SU(N))$. Generically, the partition functions  
of the proposed duals agree for generic mass deformations, but only for a subspace of FI deformations for the theory $\CT^\vee$.  
The following IR duality between an $SU(N)$ gauge theory and a $U(N-1)$ gauge theory plays an important role in the prescription 
(see the quiver notation in \Secref{Q-Not}) :

\begin{center}
\begin{tabular}{ccc}
\scalebox{.8}{\begin{tikzpicture}[node distance=2cm, nnode/.style={circle,draw,thick, fill, inner sep=1 pt},cnode/.style={circle,draw,thick,minimum size=1.0 cm},snode/.style={rectangle,draw,thick,minimum size=1.0 cm}, pnode/.style={circle,double,draw,thick, minimum size=1.0cm}]
\node[pnode] (1) at (0,0){$N$};
\node[snode] (2) at (0,-2){$2N-1$};
\draw[-] (1) -- (2);
\node[text width=0.1cm](20)[below=0.5 cm of 2]{$(\CT)$};
\end{tikzpicture}}
& \qquad   \qquad
&\scalebox{.8}{\begin{tikzpicture}[node distance=2cm, nnode/.style={circle,draw,thick, fill, inner sep=1 pt},cnode/.style={circle,draw,thick,minimum size=1.0 cm},snode/.style={rectangle,draw,thick,minimum size=1.0 cm}]
\node[cnode] (1) at (0,0){$N-1$};
\node[snode] (2) at (0,-2){$2N-1$};
\node[snode,blue] (3) at (3,0){$1$};
\draw[-] (1) -- (2);
\draw[-, thick, blue] (1)-- (3);
\node[text width=1cm](10) at (1.2, 0.2){$N-1$};
\node[text width=1cm](11) at (0, 1.1){$\eta=0$};
\node[text width=0.1cm](20)[below=0.5 cm of 2]{$(\CT^\vee)$};
\end{tikzpicture}}
\end{tabular}
\end{center}

As explained in \Secref{IRdual-Dp-gen}, one can arrive at the theory $\CT^\vee$ by using this duality multiple times in the theory $\CT$, 
at the level of the partition function. In \Secref{IRdual-Dp-Ex1}-\Secref{IRdual-Dp-Ex2}, we present explicit examples of this IR duality 
associated with the $D_4(SU(6))$ and the $D_9(SU(3))$ respectively, and check the duality by showing that the three-sphere partition 
functions of the theories agree. Our computations show that the partition 
functions agree only if certain linear combinations of FI parameters in the $\CT^\vee$ theory are set to zero 
\footnote{These correspond to certain emergent $U(1)$ Coulomb branch global symmetries that appear in the IR for unitary-special unitary 
quivers. For the $SU(N)$ theory with $N_f=2N -1$, this was verified in \cite{Giacomelli:2020ryy}.}
The proposed IR dualities, 
for the examples studied in \Secref{IRdual-Dp}, can be summarized as follows:

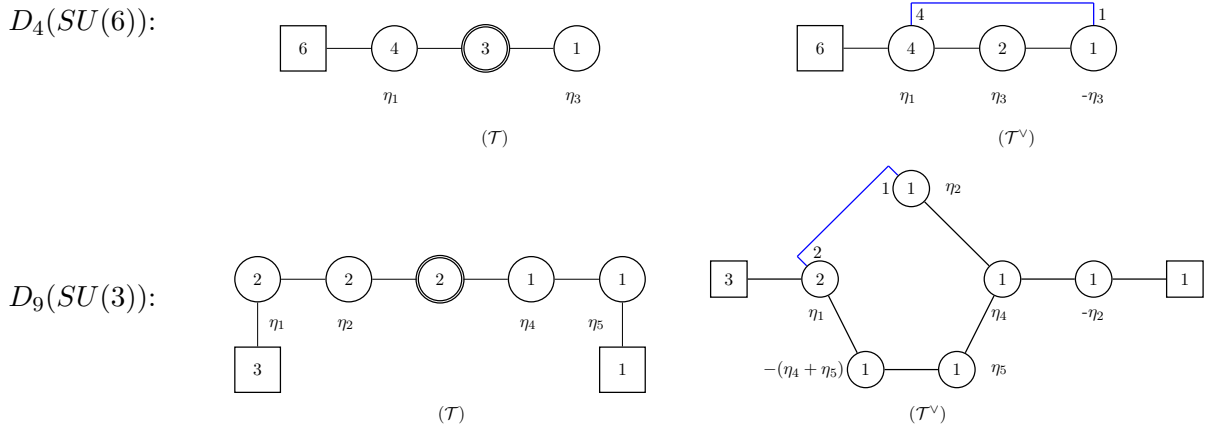
\begin{center}
\begin{tabular}{ccccc}
{\begin{tikzpicture}
\node[text width=2cm] (1) at (0,0){$D_4(SU(6))$:}; 
\node[] (2) at (0,-1.6) {};
\end{tikzpicture}}
& \qquad
&\scalebox{.6}{\begin{tikzpicture}[node distance=2cm, nnode/.style={circle,draw,thick, fill, inner sep=1 pt},cnode/.style={circle,draw,thick,minimum size=1.0 cm},snode/.style={rectangle,draw,thick,minimum size=1.0 cm}, pnode/.style={circle,double,draw,thick, minimum size=1.0cm}]
\node[snode] (1) at (0,0){6};
\node[cnode] (2) at (2,0){4};
\node[pnode] (3) at (4,0){3};
\node[cnode] (4) at (6,0){1};
\draw[-] (1) -- (2);
\draw[-] (2)-- (3);
\draw[-] (3)-- (4);
\node[text width=.2cm](12) at (4,-2){$(\CT)$};
\node[text width=0.5cm](11) at (2, -1.1){$\eta_1$};
\node[text width=0.5cm](12) at (6, -1.1){$\eta_3$};
\end{tikzpicture}}
& \qquad   
& \scalebox{.6}{\begin{tikzpicture}[node distance=2cm, nnode/.style={circle,draw,thick, fill, inner sep=1 pt},cnode/.style={circle,draw,thick,minimum size=1.0 cm},snode/.style={rectangle,draw,thick,minimum size=1.0 cm}]
\node[snode] (1) at (0,0){6};
\node[cnode] (2) at (2,0){4};
\node[cnode] (3) at (4,0){2};
\node[cnode] (4) at (6,0){1};
\draw[-] (1) -- (2);
\draw[-] (2)-- (3);
\draw[-] (3)-- (4);
\draw[-, thick, blue] (2)-- (2,1);
\draw[-, thick, blue] (2,1)-- (6,1);
\draw[-, thick, blue] (6,1)-- (4);
\node[text width=.2cm](10) at (2.2, 0.75){4};
\node[text width=.2cm](11) at (6.2, 0.75){1};
\node[text width=.2cm](12) at (4,-2){$(\CT^\vee)$};
\node[text width=0.5cm](12) at (2, -1.1){$\eta_1$};
\node[text width=0.5cm](13) at (4, -1.1){$\eta_3$};
\node[text width=0.5cm](14) at (6, -1.1){-$\eta_3$};
\end{tikzpicture}}\\
{\begin{tikzpicture}
\node[text width=2cm] (1) at (0,0){$D_9(SU(3))$:}; 
\node[] (2) at (0,-1.6) {};
\end{tikzpicture}}
& \qquad
&\scalebox{.6}{\begin{tikzpicture}[
cnode/.style={circle,draw,thick, minimum size=1.0cm},snode/.style={rectangle,draw,thick,minimum size=1cm},pnode/.style={circle,double,draw,thick, minimum size=1.0cm}]
\node[cnode] (1) at (0,0){2};
\node[snode] (2) at (0,-2){3};
\node[cnode] (3) at (2,0){2};
\node[pnode] (4) at (4,0){2};
\node[cnode] (5) at (6,0){1};
\node[cnode] (6) at (8,0){1};
\node[snode] (7) at (8,-2){1};
\draw[-] (1) -- (2);
\draw[-] (1)-- (3);
\draw[-] (3)-- (4);
\draw[-] (4)-- (5);
\draw[-] (5)-- (6);
\draw[-] (6)-- (7);
\node[text width=0.1cm](30) at (4,-3){$(\CT)$};
\node[text width=0.5cm](11) at (0.5, -1){$\eta_1$};
\node[text width=0.5cm](12) at (2, -1){$\eta_2$};
\node[text width=0.5cm](13) at (6, -1){$\eta_4$};
\node[text width=0.5cm](14) at (7.5, -1){$\eta_5$};
\end{tikzpicture}}
&\qquad
&\scalebox{.6}{\begin{tikzpicture}[node distance=2cm,cnode/.style={circle,draw,thick,minimum size=8mm},snode/.style={rectangle,draw,thick,minimum size=8mm},pnode/.style={rectangle,red,draw,thick,minimum size=8mm}]
\node[cnode] (1) at (-3,0) {$2$};
\node[snode] (2) at (-5,0) {$3$};
\node[cnode] (3) at (-1,2) {$1$};
\node[cnode] (4) at (-2,-2) {$1$};
\node[cnode] (5) at (0,-2) {$1$};
\node[cnode] (6) at (1,0) {$1$};
\node[cnode] (7) at (3,0) {$1$};
\node[snode] (8) at (5,0) {$1$};
\draw[thick] (1) -- (2);
\draw[thick, blue] (1) -- (-3.5,0.5);
\draw[thick, blue] (3) -- (-1.5, 2.5);
\draw[thick, blue] (-3.5,0.5) -- (-1.5, 2.5);
\draw[thick] (1) -- (4);
\draw[thick] (4) -- (5);
\draw[thick] (5) -- (6);
\draw[thick] (6) -- (7);
\draw[thick] (7) -- (8);
\draw[thick] (3) -- (6);
\node[text width=0.1cm](31) at (-3.1,0.6){2};
\node[text width=0.1cm](32) at (-1.6,2){1};
\node[text width=0.1cm](30) at (-1,-3){$(\CT^\vee)$};
\node[text width=0.5cm](11) at (-3, -0.8){$\eta_1$};
\node[text width=2.5 cm](12) at (-3, -2){$-(\eta_4 +\eta_5)$};
\node[text width=0.5cm](13) at (1, -2){$\eta_5$};
\node[text width=0.5cm](14) at (0, 2){$\eta_2$};
\node[text width=0.5cm](15) at (1, -0.8){$\eta_4$};
\node[text width=0.5cm](16) at (3, -0.8){-$\eta_2$};
\end{tikzpicture}}
\end{tabular}
\end{center}
For the above dual pairs, we have indicated how the FI parameters map across the duality.\\

For the class of Lagrangians $\CT^\vee$, the Coulomb branch symmetry can be read off from the quiver gauge theory. 
This is, however, not the case for the class of Lagrangians $\CT$, where the Coulomb branch symmetry (or some subgroup 
of it) appears as an emergent symmetry in the IR \cite{Giacomelli:2020ryy}. 

\bigskip

\item \textbf{3d Lagrangians for other families of Argyres-Douglas theories:} The construction using the $S$-type operation, as summarized 
in \figref{S-OP-duality}, can be deployed whenever the 3d mirror of the AD theory reduced on a circle has a Lagrangian description. 
In the generic case, where the 3d mirror can only be engineered from a linear quiver with unitary gauge groups by a non-Abelian 
$S$-type operation, there is no guarantee that the dual operation will produce a Lagrangian theory. 
However, for families of AD theories for which the aforementioned 3d mirror can be engineered from a linear quiver 
by a sequence of elementary Abelian $S$-type operations, the mirror of the 3d mirror is guaranteed to be Lagrangian. 
This in turn implies that the 3d SCFT obtained by the circle-reduction of the associated AD theory is guaranteed to have a 
Lagrangian description. Aside from the two infinite families -- $(A_{p-N-1}, A_{N-1})$ and $D_p(SU(N)$ -- that we study in 
the present work, there are other infinite families for which the 3d mirrors have the same property and can therefore be treated 
in an analogous fashion. These include the families $(A_s, D_{(s+1)p+2})$ and $D^{N-1}_p(SU(N))$ (for $p \gtreqless N-1$).
In the first case, the 3d Lagrangians can be worked out in a fashion very similar to the $(A_{p-N-1}, A_{N-1})$ theories, 
while in the second case, the construction closely mimics the $D_p(SU(N))$ theories. 
It would be interesting to have a complete classification of the AD theories for which the 3d mirrors have 
the above property, and construct the associated 3d Lagrangians. We leave this to a future work.

\end{itemize}

\section{Review of $S$-type operations and a result for Abelian operations}\label{SOp-rev}

In this section, we review the $S$-type operations introduced in \cite{Dey:2020hfe} and discuss a result on Abelian $S$-type operations 
which will be central to our construction of the 3d Lagrangians. 

\subsection{Quiver Notation}\label{Q-Not}

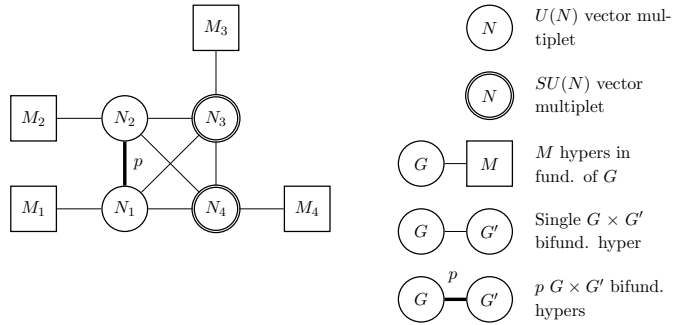
\begin{figure}[htbp]
\begin{center}
\scalebox{0.6}{\begin{tikzpicture}[node distance=2cm,
cnode/.style={circle,draw,thick, minimum size=1.0cm},snode/.style={rectangle,draw,thick,minimum size=1cm}, pnode/.style={circle,double,draw,thick, minimum size=1.0cm}]
\node[cnode] (1) at (-2,0) {$N_1$};
\node[cnode] (2) at (-2,2) {$N_2$};
\node[pnode] (3) at (0,2) {$N_3$};
\node[pnode] (4) at (0,0) {$N_4$};
\node[snode] (5) at (-4,0) {$M_1$};
\node[snode] (6) at (-4,2) {$M_2$};
\node[snode] (7) at (0,4) {$M_3$};
\node[snode] (8) at (2,0) {$M_4$};
\draw[line width=0.75mm, black] (1) -- (2);
\draw[-] (2) -- (3);
\draw[-] (3) -- (4);
\draw[-] (1) -- (4);
\draw[-] (1) -- (5);
\draw[-] (2) -- (6);
\draw[-] (3) -- (7);
\draw[-] (4) -- (8);
\draw[-] (2) -- (4);
\draw[-] (1) -- (3);
\node[text width=0.1](40) at (-1.8, 1){$p$};
\node[cnode] (9) at (6,4) {$N$};
\node[text width=3cm](10) at (8.5, 4){$U(N)$ vector multiplet};
\node[pnode] (11) at (6,2.5) {$N$};
\node[text width=3cm](12) at (8.5, 2.5){$SU(N)$ vector multiplet};
\node[snode] (13) at (6,1) {$M$};
\node[cnode] (14) at (4.5,1) {$G$};
\draw[-] (13)--(14);
\node[text width=3cm](15) at (8.5, 1){$M$ hypers in fund. of $G$};
\node[cnode] (16) at (6, -0.5) {$G'$};
\node[cnode] (17) at (4.5, -0.5) {$G$};
\draw[-] (16)--(17);
\node[text width=3cm](18) at (8.5, - 0.5){Single $G \times G'$ bifund. hyper};
\node[cnode] (31) at (6,-2) {$G'$};
\node[cnode] (32) at (4.5,-2) {$G$};
\draw[line width=0.75mm, black] (31)--(32);
\node[text width=0.2cm](33) at (5.2, -1.5){$p$};
\node[text width=3cm](34) at (8.5, -2){$p$ $G \times G'$ bifund. hypers};
\end{tikzpicture}}
\end{center}
\caption{\footnotesize{A quiver gauge theory with $G=U(N_1) \times U(N_2) \times SU(N_3) \times SU(N_4)$ and hypermultiplets in bifundamental representations, with the various conventions listed on the RHS.}}
\label{fig: QuivConv}
\end{figure}

To begin with, we set up the 3d $\CN=4$ quiver gauge theory notation that we will use for the rest of the paper
\footnote{For a short recent review on 3d $\CN=4$ Lagrangian theories, we refer the reader to the paper \cite{Bullimore:2015lsa}.}. 
The quiver notation, relevant for this paper, is summarized in \figref{fig: QuivConv}.
 
In this work, we will also encounter hypermultiplets charged only under the $U(1)$ subgroup of certain 
unitary factors in the quiver gauge group\footnote{For the examples treated in this paper, it is more appropriate to think of these hypermultiplets as transforming in determinant/anti-determinant representation of the unitary gauge group.}. Given their importance in the current discussion, we will distinguish these hypermultiplets in the 
quiver diagram with the following notation:
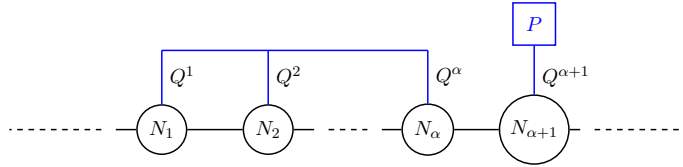
\begin{center}
\scalebox{0.7}{\begin{tikzpicture}[node distance=2cm,cnode/.style={circle,draw,thick,minimum size=8mm},snode/.style={rectangle,draw,thick,minimum size=8mm},pnode/.style={rectangle,red,draw,thick,minimum size=8mm}]
\node[] (10) at (-3,0) {};
\node[] (1) at (-1,0) {};
\node[cnode] (2) at (0,0) {$N_1$};
\node[cnode] (3) at (2,0) {$N_2$};
\node[] (4) at (3,0) {};
\node[] (5) at (4,0) {};
\node[cnode] (6) at (5,0) {$N_{\alpha}$};
\node[cnode] (7) at (7,0) {$N_{\alpha+1}$};
\node[] (8) at (8,0) {};
\node[] (9) at (10,0) {};
\node[snode, blue] (12) at (7,2) {$P$};
\draw[thick] (1) -- (2);
\draw[thick] (2) -- (3);
\draw[thick] (3) -- (4);
\draw[thick,dashed] (4) -- (5);
\draw[thick] (5) -- (6);
\draw[thick] (6) -- (7);
\draw[thick] (7) -- (8);
\draw[thick,blue] (7) -- (12);
\draw[thick,dashed] (8) -- (9);
\draw[thick,dashed] (1) -- (10);
\draw[thick, blue] (2) -- (0,1.5);
\draw[thick, blue] (0,1.5) -- (5,1.5);
\draw[thick, blue] (2,1.5) -- (3);
\draw[thick, blue] (5,1.5) -- (6);
\node[text width=.2cm](15) at (0.25,1){$Q^1$};
\node[text width=.2cm](16) at (2.25,1){$Q^2$};
\node[text width=.2cm](17) at (5.25,1){$Q^\alpha$};
\node[text width=.2cm](18) at (7.25,1){$Q^{\alpha+1}$};
\end{tikzpicture}}
\end{center}
The blue line in the above quiver diagram, connecting the gauge nodes $U(N_1)$, $U(N_2)$ and $U(N_\alpha)$, 
denotes a \textit{single} hypermultiplet charged under the gauge subgroup $U(1)_1 \times U(1)_2 \times U(1)_\alpha$, 
where $U(1)_i \subset U(N_i)$, with the charges $(Q^1, Q^2, Q^\alpha)$. The blue line connected to the flavor node 
denotes $P$ copies of a hypermultiplet with charge $Q^{\alpha+1}$ under $U(1)_{\alpha +1} \subset U(N_{\alpha +1})$.
More generally, we will denote the charges of these hypermultiplets by the matrix $Q^a_i$, where the index $a$ runs over 
the unitary gauge nodes in the quiver and $i$ runs over the hypermultiplets (i.e. a flavor index) in question.

We will occasionally use another notation for quiver gauge theories, which is particularly useful for representing linear chains of unitary and special unitary gauge 
groups with fundamental and bifundamental matter:
\be
\CT = [N] - SU(N_1) - SU(N_2) - \ldots - N_k - N_{k+1} - [M],
\ee
where the unitary gauge group factor $U(N)$ is denoted by the integer $N$, the special unitary factors are explicitly written, $[M],[N]$ denote 
fundamental hypers for the associated gauge group factor, while consecutive gauge group factors are connected by a single 
bifundamental hyper.

\subsection{Review of $S$-type operations}

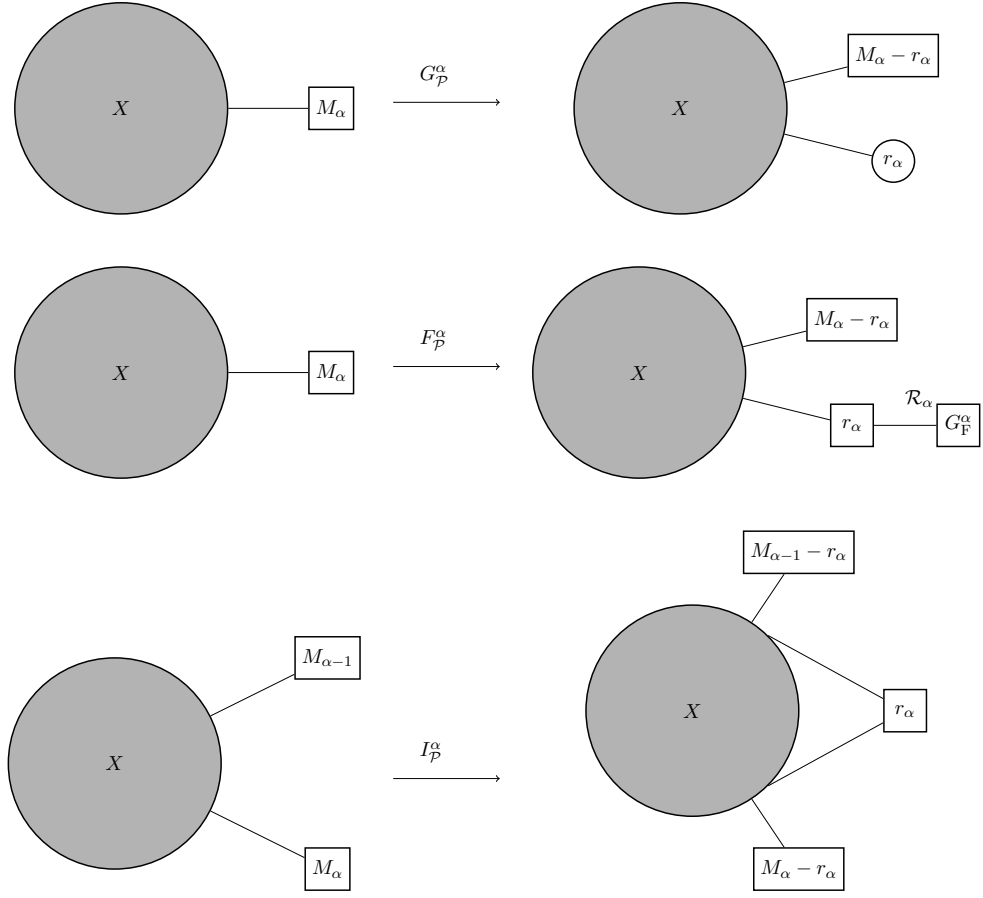
\begin{figure}[htbp]
\begin{center}
\begin{tabular}{ccc}
\scalebox{0.7}{\begin{tikzpicture}[
cnode/.style={circle,draw,thick,minimum size=4mm},snode/.style={rectangle,draw,thick,minimum size=8mm},pnode/.style={rectangle,red,draw,thick,minimum size=1.0cm}, bnode/.style={circle,draw, thick, fill=black!30,minimum size=4cm}]
\node[bnode] (1) at (0,0){$X$} ;
\node[snode] (2) [right=1.5cm  of 1]{$M_\alpha$} ;
\draw[-] (1)--(2);
\end{tikzpicture}}
&\scalebox{0.7}{\begin{tikzpicture}
\draw[->] (5,0) -- (7,0); 
\node[text width=1cm](6) at (6,0.5){$G^\alpha_\CP$};
\node[] at (6, -2){};
\end{tikzpicture}}
& \scalebox{0.7}{\begin{tikzpicture}[
cnode/.style={circle,draw,thick,minimum size=4mm},snode/.style={rectangle,draw,thick,minimum size=8mm},pnode/.style={rectangle,red,draw,thick,minimum size=1.0cm}, bnode/.style={circle,draw, thick, fill=black!30,minimum size=4cm}]\node[bnode] (3) at (10,0){$X$} ;
\node[snode] (4) at (14, 1){$M_\alpha- r_\alpha$} ;
\node[cnode] (5) at (14,-1){$r_\alpha$} ;
\draw[-] (3)--(4);
\draw[-] (3)--(5);
\end{tikzpicture}} \\
\qquad & \qquad & \qquad \\
\scalebox{0.7}{\begin{tikzpicture}[
cnode/.style={circle,draw,thick,minimum size=4mm},snode/.style={rectangle,draw,thick,minimum size=8mm},pnode/.style={rectangle,red,draw,thick,minimum size=1.0cm}, bnode/.style={circle,draw, thick, fill=black!30,minimum size=4cm}]
\node[bnode] (1) at (0,0){$X$} ;
\node[snode] (2) [right=1.5cm  of 1]{$M_\alpha$} ;
\draw[-] (1)--(2);
\end{tikzpicture}}
& \scalebox{0.7}{\begin{tikzpicture}
\draw[->] (5,0) -- (7,0); 
\node[text width=1cm](10) at (6,0.5){$F^\alpha_\CP$};
\node[] at (6, -2){};
\end{tikzpicture}}
&\scalebox{0.7}{\begin{tikzpicture}[
cnode/.style={circle,draw,thick,minimum size=4mm},snode/.style={rectangle,draw,thick,minimum size=8mm},pnode/.style={rectangle,red,draw,thick,minimum size=1.0cm}, bnode/.style={circle,draw, thick, fill=black!30,minimum size=4cm}]
\node[bnode] (3) at (10,0){$X$} ;
\node[snode] (4) at (14, 1){$M_\alpha- r_\alpha$} ;
\node[snode] (5) at (14,-1){$r_\alpha$} ;
\node[snode] (6) at (16,-1){$G^\alpha_{\rm F}$} ;
\draw[-] (3)--(4);
\draw[-] (3)--(5);
\draw[-] (5)--(6);
\node[text width=1cm](11) at (15.5, - 0.5){$\CR_\alpha$};
\end{tikzpicture}}\\
\qquad & \qquad & \qquad \\
\scalebox{0.7}{\begin{tikzpicture}[
cnode/.style={circle,draw,thick,minimum size=4mm},snode/.style={rectangle,draw,thick,minimum size=8mm},pnode/.style={rectangle,red,draw,thick,minimum size=1.0cm}, bnode/.style={circle,draw, thick, fill=black!30,minimum size=4cm}]
\node[bnode] (1) at (0,0){$X$} ;
\node[snode] (2) at (4,-2) {$M_\alpha$} ;
\node[snode] (3) at (4, 2) {$M_{\alpha -1}$} ;
\draw[-] (1)--(2);
\draw[-] (1)--(3);
\end{tikzpicture}}
& \scalebox{0.7}{\begin{tikzpicture}
\draw[->] (5,0) -- (7,0);
\node[text width=1cm](6) at (6,0.5){$I^\alpha_\CP$};
\node[] at (6, -2){};
\end{tikzpicture}}
& \scalebox{0.7}{\begin{tikzpicture}[
cnode/.style={circle,draw,thick,minimum size=4mm},snode/.style={rectangle,draw,thick,minimum size=8mm},pnode/.style={rectangle,red,draw,thick,minimum size=1.0cm}, bnode/.style={circle,draw, thick, fill=black!30,minimum size=4cm}]
\node[bnode] (3) at (10,0){$X$} ;
\node[snode] (4) at (12, 3){$M_{\alpha-1}- r_\alpha$} ;
\node[snode] (5) at (14,0){$r_\alpha$} ;
\node[snode] (6) at (12,-3){$M_\alpha- r_\alpha$} ;
\draw[-] (3)--(4);
\draw[-] (3.north east) to  (5);
\draw[-] (3.south east) to (5);
\draw[-] (3)--(6);
\end{tikzpicture}}
\end{tabular}
\end{center}
\caption{\footnotesize{The gauging, flavoring and identification quiver operations on a generic quiver $X$ of class $\CU$. Each operation involves two steps. 
In the first step, one splits the flavor node $U(M_\alpha)$ into two, corresponding to $U(r_\alpha) \times U(M_\alpha - r_\alpha)$ flavor nodes.
In the second step, one gauges, adds hypermultiplets to, or identifies the $U(r_\alpha)$ flavor node respectively. The identification operation may 
involve two or more $U(r_\alpha)$ flavor nodes.}}
\label{fig: QuivOps}
\end{figure}

Let us now review the $S$-type operations \cite{Dey:2020hfe}. Consider a class of 3d quiver gauge theories for which the Higgs branch 
global symmetry has a subgroup $G^{\rm sub}_{\rm global}=\prod_\gamma U(M_\gamma) \subset G_H$, and are also good theories in 
the Gaiotto-Witten sense \cite{Gaiotto:2008ak}.
We will refer to them as class $\CU$. Let $X[\wh{\vec A}]$ be a generic theory in this class, where $\wh{\vec A}$ collectively denotes the 
background vector multiplets (including twisted ones) for the global symmetries. 
Given such a quiver $X[\wh{\vec A}]$, one can define a 
set of three basic quiver operations\footnote{One can also include a defect operation in this list, but we will not need it for this work.
See \cite{Dey:2021jbf} for details on the defect operation.} (see \cite{Dey:2020hfe} for details), which are shown in \figref{fig: QuivOps}:
\begin{enumerate}
\item {\bf{Gauging ($G^\alpha_\CP$)}}
\item {\bf{Flavoring ($F^\alpha_\CP$)}}
\item \textbf{Identification($I^\alpha_{\vec \CP}$)}
\end{enumerate}

The superscript $\alpha$ denotes the flavor node at which the operation is being performed. Note that each operation involves splitting the flavor node 
$U(M_\alpha)$ into two, corresponding to $U(r_\alpha) \times U(M_\alpha - r_\alpha)$ flavor nodes. The $U(1)^{M_\alpha}$ masses are 
related to the $U(1)^{r_\alpha} \times U(1)^{M_\alpha-r_\alpha}$ masses by the following map:
\be \label{uvdef0}
\overrightarrow{m^\alpha}_{i_\alpha} = \CP_{i_\alpha i} \, \overrightarrow{u}^\alpha_i + \CP_{i_\alpha \, r_\alpha + j} \, \overrightarrow{v}^\alpha_j, \quad( i_\alpha=1,\ldots, M_\alpha,\quad i=1,\ldots, r_\alpha, \quad j=1,\ldots, M_\alpha - r_\alpha),
\ee
where $\CP$ is a permutation matrix of order $M_\alpha$. The resultant theory deformed by the $U(r_\alpha) \times U(M_\alpha - r_\alpha)$ mass parameters is
denoted as $(X, \CP)$. For the identification operation, this procedure is carried out at multiple flavor nodes, involving a permutation matrix for each. Therefore, 
in the most general case, we will refer the above theory as $(X, \{\CP_\beta \})$ or $(X, \vec \CP)$.\\ 

\begin{definition}
\textbf{An elementary $S$-type operation} $\CO^\alpha_{\vec \CP}$ on $X$ at a flavor node $\alpha$, is defined as any possible combination of the 
identification ($I^\alpha_{\vec \CP}$) and the  flavoring ($F^\alpha_{\CP}$) quiver operations followed by a single gauging operation $G^\alpha_{\CP}$, 
i.e.
\be \label{Sbasic-def}
\boxed{\CO^\alpha_{\vec \CP}(X) := (G^\alpha_{\vec \CP}) \circ (F^\alpha_{\vec \CP})^{n_2} \circ (I^\alpha_{\vec \CP})^{n_1}(X), 
\quad (n_i=0,1, \,\, \forall i).}
\ee
The operation $\CO^\alpha_{\vec \CP}$ generically maps a quiver gauge theory $X[\wh{\vec A}]$ to a new quiver gauge theory $X'[\wh{\vec B}]$:
\be 
\CO^\alpha_{\vec \CP} : X[\wh{\vec A}] \mapsto X'[\wh{\vec B}].
\ee
\end{definition}
The order of the different constituent operations in the above definition is important to emphasize. The composition of the identification and the flavoring operations is commutative, while none of the operations commute with the gauging operation. From the above definition, one finds that there are four distinct types of 
elementary $S$-type operations depending on the constituent quiver operations:
\begin{itemize}
\item Gauging.

\item Flavoring-gauging.

\item Identification-gauging.

\item Identification-flavoring-gauging.

\end{itemize}

\begin{definition}
\textbf{An elementary Abelian $S$-type operation} is defined as an elementary $S$-type operation where the gauging operation 
$G^\alpha_{\vec \CP}$ involves a $U(1)$ global symmetry.
\end{definition}
Note that the elementary Abelian gauging operation gives the action of the $SL(2,\BZ)$ $S$-generator \cite{Witten:2003ya} 
on a 3d SCFT with a $U(1)$ global symmetry.

 \begin{definition}
\textbf{ A generic $S$-type operation} is a combination of any number of elementary $S$-type operations: 
\be \label{Comp-OP}
\boxed{\CO^{(\alpha_1, \ldots, \alpha_l)}_{({\vec \CP_1},\ldots,{\vec \CP_l})}(X) :={\CO^{\alpha_l}_{\vec \CP_l}} \circ {\CO^{\alpha_{l-1}}_{\vec \CP_{l-1}}} \circ \ldots \circ {\CO^{\alpha_2}_{\vec \CP_2}} \circ  {\CO^{\alpha_1}_{\vec \CP_1}}(X).}
\ee
\end{definition}
The superscript $(\alpha_1, \ldots, \alpha_l)$ labels the flavor nodes at which the elementary $S$-type operations are performed, with 
$({\vec \CP_1},\ldots,{\vec \CP_l})$ denoting the associated permutation matrices.\\

One of the main applications of the $S$-type operations, as discussed in \cite{Dey:2020hfe}, is to construct new pairs of 3d mirror 
theories starting from a given pair of Lagrangian mirrors $(X[\wh{\vec A}],Y[\wh{\vec A}])$, with $X$ in class $\CU$. We now briefly 
review the construction. Given the operation
\be 
\CO^\alpha_{\vec \CP} : X[\wh{\vec A}] \mapsto X'[\wh{\vec B}],
\ee
one can then define a dual operation $\wt{\CO}^\alpha_{\vec \CP}$ which acts on the quiver gauge theory $Y$:
\be
\wt{\CO}^\alpha_{\vec \CP} : Y[\wh{\vec A}] \mapsto Y'[\wh{\vec B}], 
\ee
such that the pair $(X'[\wh{\vec B}], Y'[\wh{\vec B}])$ is IR dual. This can be summarized in \figref{S-OP-duality2}.
\begin{figure}[htbp]
\begin{center}
\begin{tikzpicture}
  \node (X) at (2,1.5) {$X[\wh{\vec A}]$};
  \node (Y) at (6,1.5) {$Y[\wh{\vec A}]$};
  \node (D) at (2,-1) {$X'[\wh{\vec B}]$};
  \node (E) at (6,-1) {$Y'[\wh{\vec B}]$};
   \draw[->] (X) -- (D) node [midway,  right=+3pt] {\footnotesize $\CO^\alpha_{\vec \CP}$};
  \draw[<->] (X) -- (Y)node [midway,above  ] {\footnotesize 3d Mirror };
    \draw[->] (Y) -- (E) node [midway,  right=+3pt] {\footnotesize $\wt{\CO}^\alpha_{\vec \CP}$};
        \draw[<->] (D) -- (E)node  [midway,above  ] {\footnotesize 3d Mirror};
\end{tikzpicture}
\end{center}
\caption{\footnotesize{Generating new dual pairs using an elementary $S$-type operation.} }
\label{S-OP-duality2}
\end{figure}

The goal now is to precisely determine the dual operation $\wt{\CO}^\alpha_{\vec \CP}$ on the theory $Y$ and 
therefore determine the theory $Y'$. This is achieved using supersymmetric observables computed using 
localization. Below, we illustrate the construction using the three-sphere partition function. For 
a similar construction using superconformal index, we refer the reader to Appendix C of \cite{Dey:2020hfe}. \\

The three-sphere partition function of a 3d $\CN=4$ Lagrangian theory with gauge group $G$ is given by a matrix integral of the following form\cite{Kapustin:2010xq}: 
\begin{align}\label{PF-main}
Z^{(\CT)}(\vec m; \vec \eta)=  \int  \Big[d\vec s\Big] \, Z_{\rm int}(\vec s, \vec m, \vec \eta) = : \int  \Big[d\vec s\Big] \, Z_{\rm FI}(\vec s, \vec \eta)\, Z^{\rm vector}_{\rm{1-loop}}(\vec s)\,Z^{\rm hyper}_{\rm{1-loop}}(\vec s, \vec m),
\end{align}
where the measure $\Big[d\vec s\Big]=\frac{d^k \vec{s}}{|{W}(G)|}$, and $|{W}(G)|$ is the order of the Weyl group of $G$. 
The partition function is a function of the real masses $\vec m$ and real FI parameters $\vec \eta$, 
which are the real scalar components inside background 3d $\CN=4$ vector multiplets and twisted vector multiplets respectively. 
The individual terms in the integrand on the RHS consist of classical and one-loop contributions, and are given as follows. 
\begin{itemize}

\item The classical contribution to the matrix integral comes from the $l$ $U(1)$ factors in the gauge group $G$:
\begin{equation} \label{PF-main-FI}
Z_{\rm FI}(\vec s, \vec \eta) = \prod^l_{\gamma=1} e^{2\pi i \eta_\gamma \, \text{Tr}(\vec{s}^\gamma)}\,,
\end{equation}
where $\gamma$ runs over the $l$ unitary gauge nodes of the theory and $\eta_\gamma$ is the FI parameter for a given $\gamma$.

\item The  one-loop contribution of the $\mathcal{N}=4$ vector multiplet is given as:
\begin{equation}\label{PF-main-vec}
Z^{\rm vector}_{\text{1-loop}}(\vec s)=\prod_{\alpha} \sinh{\pi \alpha(\vec s)}\,,
\end{equation}
where the product extends over the roots of the Lie algebra of $G$.

\item The one-loop contribution of an $\mathcal{N}=4$ hypermultiplet transforming in a representation  $\CR$ of $G \times G_{H}$ is given as:
\begin{equation} \label{PF-main-hyper}
Z^{\rm hyper}_{\text{1-loop}}(\vec s, \vec m)=\prod_{\rho(\CR)} \frac{1}{\cosh{\pi \rho(\vec s, \vec m)}}\,,
\end{equation}
where the product extends over the weights of the representation $\CR$.
\end{itemize}

An elementary $S$-type operation can be implemented in terms of the three-sphere partition function as follows:
\begin{align}\label{PF-OP-woD}
Z^{\CO^\alpha_{\vec \CP}(X)} (\vec{m}^{\CO^\alpha_{\vec \CP}}, \ldots; \vec \eta, \eta_\alpha)= \int \Big[d\vec{u}^\alpha\Big] \, \CZ_{\CO^\alpha_{\vec \CP}(X)}(\vec u^{\alpha}, \{\vec{u}^\beta\}_{\beta \neq \alpha}, \eta_\alpha, \vec{m}^{\CO^\alpha_{\vec \CP}})
\cdot Z^{(X,\{ P_\beta\})} (\{\vec{u}^\beta\},\ldots; \vec \eta),
\end{align}
where $Z^{\CO^\alpha_{\vec \CP}(X)}$ denotes the partition function of the theory $X'=\CO^\alpha_{\vec \CP}(X)$, 
and $\CZ_{\CO^\alpha_{\vec \CP}(X)}$ is an operator that we will momentarily specify.
The parameter $\eta_\alpha$ is an FI parameter associated with the gauging operation, and the parameters $\vec{m}^{\CO^\alpha_{\vec \CP}}$ 
collectively denote the hypermultiplet masses associated 
with the flavoring and/or the identification operations. The operator $\CZ_{\CO^\alpha_{\vec \CP}(X)}$ can be built out of the partition function contributions 
of the constituent gauging ($G^\alpha_\CP$) , flavoring ($F^\alpha_\CP$), and identification ($I^\alpha_\CP$) operations following 
the definition \eref{Sbasic-def}:
\be \label{CZ-OP}
\CZ_{\CO^\alpha_{\vec \CP}(X)}(\vec u^{\alpha}, \{\vec{u}^\beta\}_{\beta \neq \alpha}, \eta_\alpha, \vec{m}^{\CO^\alpha_{\vec \CP}}) =\CZ_{G^\alpha_\CP(X)} \cdot \Big(\CZ_{F^\alpha_\CP(X)}\Big)^{n_2} \cdot \Big(\CZ_{I^\alpha_\CP(X)}\Big)^{n_1}, \quad (n_1,n_2=0,1),
\ee
where the individual constituents are:
\begin{align}
& \CZ_{G^\alpha_\CP(X)}(\vec{u}^\alpha,\eta_\alpha) = Z_{\rm FI} (\vec{u}^\alpha,\eta_\alpha) \,Z_{\rm 1-loop} ^{\rm vector} (\vec{u}^\alpha)=e^{2\pi i \eta_\alpha \sum_i u^\alpha_i} \, \prod_{i < j} \sinh^2{\pi (u^\alpha_i - u^\alpha_j)}, \label{CZ-gauging}\\
& \CZ_{F^\alpha_\CP(X)}(\vec{u}^\alpha, \vec{m}^\alpha_F)= Z_{\rm 1-loop} ^{\rm hyper} (\vec{u}^\alpha, \vec{m}^\alpha_F)=\prod_{\rho(\CR_\alpha)}\frac{1}{ 
\ch{\rho(\vec{u}^\alpha, \vec{m}^\alpha_F)}}, \label{CZ-flavoring}\\
&\CZ_{I^\alpha_\CP(X)}(\vec u^{\alpha}, \{\vec{u}^\beta\}, \vec \mu) = \int \prod^{p}_{j=1} \prod^{r_\alpha}_{i=1} d {u^{\gamma_j}}_i \, \prod^{p}_{j=1} \delta^{(r_\alpha)}\Big(\vec{u}^{\alpha} - \vec{u}^{\gamma_{j}} + {\mu}^{\gamma_j} \Big). \label{CZ-identification}
\end{align}
In \eref{CZ-flavoring}, the parameters $\vec{m}^\alpha_F$ denote the masses for the added flavors. In \eref{CZ-identification}, the identified nodes (we assume that 
there are $p$ of them) are labelled as $\{ \gamma_j \}_{j=1,\ldots, p}$, while $\{ \mu^{\gamma_j} \}_{j=1,\ldots, p}$ denote the masses introduced by the identification 
operation. The set of masses $(\vec{m}^\alpha_F, \{ \mu^{\gamma_j} \}_{j=1,\ldots, p})$ are collectively denoted as $\vec{m}^{\CO^\alpha_{\vec \CP}}$ in the 
subsequent formulas.\\

We can now proceed to work out the dual operation in terms of the partition function.
Mirror symmetry implies that the partition functions of $X$ and $Y$ are related as
\begin{align} \label{MS-XY}
Z^{(X,\{ P_\beta\})} (\{\vec{u}^\beta\}, \{\vec{v}^\beta\},\ldots;\vec{\eta}) 
 = e^{2\pi i \sum_{i,l,\beta}\,b^{il}_\beta u^\beta_i \eta_l} \cdot  C_{XY}(\{\vec u^\beta=0\},\ldots, \vec \eta) \cdot Z^{(Y, \{\CP_\beta\})}(\vec{m}^Y; \vec{\eta}^Y).
\end{align}
The partition function of the theory $Y$ has the following form
\begin{align} \label{Z-int-Y}
Z^{(Y, \{\CP_\beta\})}= & \int \prod_{\gamma'} \Big[d\vec{\s}^{\gamma'} \Big]\, Z^{(Y,\{\CP_\beta\})}_{\rm int}(\{\vec \s^{\gamma'} \}, \vec{m}^Y(\vec{\eta}), \vec{\eta}^Y(\{\vec{u}^\beta\},\ldots)) \nn \\
= & \int \prod_{\gamma'} \Big[d\vec{\s}^{\gamma'} \Big]\,e^{2\pi i \,\sum_{i,\beta}g^i_\beta (\{\vec \s^{\gamma'} \}, \CP_\beta)\,u^{\beta}_i}\,Z^{(Y,\{\CP_\beta\})}_{\rm int}(\{\vec \s^{\gamma'} \}, \vec{m}^Y(\vec{\eta}), \vec{\eta}^Y(\{\vec{u}^\beta =0 \},\ldots)),
\end{align}
where $\gamma'$ labels the gauge nodes of the quiver $Y$, and in the second equality, we have isolated the $\vec u$-dependent terms in the integrand of the 
matrix model.

Following the logic of \figref{S-OP-duality}, the partition function of the dual theory $Y'= \wt{\CO}^\alpha_{\vec \CP}(Y)$ can be written in the  
following form:
\begin{empheq}[box=\widefbox]{align}\label{PF-wtOPgen}
Z^{\wt{\CO}^\alpha_{\vec \CP}(Y)}(\vec{m}'; \vec \eta')
= \int \prod_{\gamma'} \Big[d\vec{\s}^{\gamma'} \Big]\, & \CZ_{\wt{\CO}^\alpha_{\vec \CP}(Y)}(\{\s^{\gamma'}\},\vec{m}^{{\CO}^\alpha_{\vec\CP}}, \eta_{\alpha},\vec \eta)\,\cdot C_{XY}(\{\vec u^\beta=0\},\ldots, \vec \eta) \nn \\
& \times Z^{(Y,\{\CP_\beta\})}_{\rm int}(\{\vec \s^{\gamma'} \}, \vec{m}^Y(\vec{\eta}), \vec{\eta}^Y(\{\vec{u}^\beta =0 \},\ldots)), 
\end{empheq}
where $\vec m'=\vec{m}'(\vec{\eta},\eta_{\alpha}), \vec \eta'=\vec \eta'(\vec{m}^{{\CO}^\alpha_{\vec\CP}},\ldots)$ are linear functions of their 
arguments. The function $\CZ_{\wt{\CO}^\alpha_{\vec \CP}(Y)}$ can be formally written as a Fourier-transform of the operator 
$\CZ_{\CO^\alpha_{\vec \CP}(X)}$ defined in \eref{CZ-OP}: 
\begin{align} \label{CZ-wtOP}
\CZ_{\wt{\CO}^\alpha_{\vec \CP}(Y)}
= \int \Big[d\vec{u}^\alpha\Big] \, \CZ_{\CO^\alpha_{\vec \CP}(X)}(\vec u^{\alpha}, \{\vec{u}^\beta\}_{\beta \neq \alpha}, \eta_\alpha, \vec{m}^{\CO^\alpha_{\vec \CP}})\, \cdot \,e^{2\pi i \,\sum_{i,\beta}(g^i_\beta (\{\vec \s^{\gamma'} \}, \CP_\beta) + \sum_l b^{il}_\beta \eta_l)\,u^{\beta}_i},
\end{align}
where $g^i_\beta$ is the set of functions defined in \eref{Z-int-Y}, and $b^{il}_\beta$ are a set of integers defined in \eref{MS-XY}. 
The equations \eref{PF-wtOPgen}-\eref{CZ-wtOP} gives a working definition of the dual $S$-type operation on the 
theory $Y$. Although the RHS of \eref{PF-wtOPgen} is written as an operation on a Lagrangian theory $Y$, the 
theory $Y'$ may not turn out to be Lagrangian. However, if the theory $Y'$ is indeed Lagrangian, one can rewrite the RHS of 
\eref{PF-wtOPgen} in the standard form of \eref{PF-main}, after some manipulations of the matrix integral. The 
gauge group and matter content of the theory $Y'$ can then be read off from \eref{PF-wtOPgen}, 
as well as the contact term $C_{X'Y'}$.

\subsection{Dual operations for the Abelian case}\label{SOp-Ab}

The dual operation greatly simplifies for the special case where $\CO^\alpha_{\vec \CP}$ is an Abelian $S$-type operation. The 
dual theory for the four types of elementary operations can be explicitly determined from the general expression in \eref{PF-wtOPgen}
and turn out to be Lagrangian. The details of the computation can be found in Section 4.1 of \cite{Dey:2020hfe}, and we summarize the results 
below. The general features of the dual theory in each case are as follows:

\begin{itemize}

\item \textbf{Gauging and Identification-gauging:} The dual partition function for the gauging operation is given as
\begin{align}\label{dual-AbGgen}
& Z^{\wt{G}^\alpha_{\CP}(Y)}
=\int \prod_{\gamma'} \Big[d\vec{\s}^{\gamma'} \Big]\, \delta \Big(\eta_\alpha + \sum_l b^{l}_\alpha \eta_l + g_\alpha \Big(\{\vec \s^{\gamma'} \}, \CP \Big) \Big)
 \, Z^{(Y, \CP)}_{\rm int}(\{\vec \s^{\gamma'} \}, \vec{m}^Y(\vec{\eta}), \vec{\eta}^Y({u}^\alpha =0, v^\alpha, \ldots)),
\end{align}
while that for the identification-gauging operation is given as
\begin{align}\label{dual-AbGIgen}
Z^{\wt{(G^\alpha_{\vec\CP} \circ I^\alpha_{\vec\CP})}(Y)} 
=\int \prod_{\gamma'} \Big[d \vec\s^{\gamma'}\Big]\, & \delta \Big(\eta_{\alpha} + \sum_{\beta,l} b^{l}_\beta \eta_l + \sum_{\beta} g_\beta \Big(\{\vec \s^{\gamma'} \}, \CP_\beta \Big) \Big)\nn \\ &\times \, Z^{(Y,\{\CP_\beta\})}_{\rm int}(\{\vec \s^{\gamma'} \},\vec{m}^Y(\vec{\eta}), \vec{\eta}^Y(\{\vec{u}^\beta = \mu^\beta \}, \{\vec{v}^\beta\},\ldots)).
\end{align}
In both cases, the dual operation amounts to gauging a $U(1)_J$ topological symmetry 
of the theory $Y$, which is equivalent to ungauging a $U(1)$ subgroup of the gauge group of $Y$. 
The precise $U(1)$ subgroup being ungauged can be read off from the argument of the 
delta function in the matrix model on the RHS of \eref{dual-AbGgen} and \eref{dual-AbGIgen} 
respectively. 

\item \textbf{Flavoring-gauging and Identification-flavoring-gauging:} Consider the flavoring-gauging operation first. For the sake of concreteness 
(and relevance to theories studied in this paper), we will assume that the flavoring involves adding hypermultiplets with charge 1 under the
$U(1)$ flavor symmetry to be gauged. In this case, the dual partition function can be written in the following form:
\begin{align}\label{dual-AbGFgen}
Z^{\wt{(G^\alpha_{\CP} \circ F^\alpha_\CP)}(Y)}
=  \int \prod_{\gamma'} \Big[d\vec{\s}^{\gamma'} \Big]\, \prod^{N^{\alpha}_F-1}_{k=1} d\tau^{\alpha}_{k} \, & Z^{(\CT[N^{\alpha}_F -1])}_{\rm int}(\vec{\tau}^{\alpha}, 0,-\vec m^{\alpha}_F)\cdot Z^{\rm hyper}_{\wt{(G^\alpha_{\CP} \circ F^\alpha_\CP)}(Y)}\Big(\{\vec \s^{\gamma'} \}, \tau^\alpha_1, \eta_\alpha, \vec \eta \Big)\nn \\ & \times Z^{(Y, \CP)}_{\rm int}(\{\vec \s^{\gamma'} \}, \vec{m}^Y(\vec{\eta}), \vec{\eta}^Y(\vec{u}^\alpha =m^\alpha_{1\,F}, \vec{v}^\alpha,\ldots)),
\end{align}
with $Z^{(\CT[N^{\alpha}_F -1])}_{\rm int}(\vec{\tau}^{\alpha}, 0, \vec m^{\alpha}_F)$ being the partition function of the theory $\CT[N^{\alpha}_F -1]$:
\begin{center}
\scalebox{0.8}{\begin{tikzpicture}[
cnode/.style={circle,draw,thick,minimum size=4mm},snode/.style={rectangle,draw,thick,minimum size=8mm},pnode/.style={rectangle,red,draw,thick,minimum size=8mm},nnode/.style={circle,red,draw,thick,minimum size=8mm}]
\node[text width=2cm](1) at (-2,0) {$\CT[N^{\alpha}_F-1]$:};
\node[cnode] (25) at (8,0){1};
\node[snode] (26) [right=1cm of 25]{1};
\node[cnode] (27) [left=1.5 cm of 25]{1};
\node[cnode] (28) [left=1.5 cm of 27]{1};
\node[cnode] (29) [left=1.5 cm of 28]{1};
\draw[-] (25) -- (26);
\draw[-] (25) -- (27);
\draw[thick, dashed] (27) -- (28);
\draw[-] (28) -- (29);
\node[text width=2cm](31) [above=0.25 of 25] {$N^{\alpha}_F-1$};
\node[text width=2 cm](32) [above=0.25 of 27] {$N^{\alpha}_F-2$};
\node[text width=1cm](33) [above=0.25 of 28] {$2$};
\node[text width=1cm](34) [above=0.25 of 29] {$1$};
\end{tikzpicture}}
\end{center}
where the mass parameter for the single fundamental hyper is set to zero, the FI parameters (from left to right) are given by 
the parameters $\{ m^{\alpha}_{i\,F} \}_{i=1,\ldots, N^{\alpha}_F}$, and the matrix model integration variables (from left to right)
are labelled as $\{ \tau^{\alpha}_{k\,F} \}_{k=1,\ldots, N^{\alpha}_F-1}$. Finally, $Z^{\rm hyper}_{\wt{(G^\alpha_{\CP} \circ F^\alpha_\CP)}(Y)}$ 
is the contribution of a single hypermultiplet which is charged under the $U(1)_1$ node of $\CT[N^{\alpha}_F -1]$ and under $U(1)$ 
subgroups of certain unitary gauge nodes of the quiver $Y$. Explicitly, the function $Z^{\rm hyper}_{\wt{(G^\alpha_{\CP} \circ F^\alpha_\CP)}(Y)}$ 
is given as:
\be
Z^{\rm hyper}_{\wt{(G^\alpha_{\CP} \circ F^\alpha_\CP)}(Y)}\Big(\{\vec \s^{\gamma'} \}, \tau^\alpha_1, \eta_\alpha, \vec \eta \Big)= \frac{1}{\ch{(g_\alpha (\{\vec \s^{\gamma'} \}, \CP)+ \tau^\alpha_1+ \eta_\alpha + \sum_l b^{l}_\alpha \eta_l)}}, \label{HyperGFgen}
\ee
where the various $U(1)$ charges can be read off from the argument of the cosh-function.\\

In a similar fashion, the dual partition function for the identification-flavoring-gauging operation is given as:
\begin{align}\label{dual-AbGFIgen}
Z^{\wt{(G^\alpha_{\vec\CP} \circ F^\alpha_{\vec\CP} \circ I^\alpha_{\vec\CP})}(Y)}= \int \prod_{\gamma'} \Big[d\vec{\s}^{\gamma'} \Big]\, & \prod^{N^{\alpha}_F-1}_{k=1} d\tau^{\alpha}_{k} \,  Z^{(\CT[N^{\alpha}_F -1])}_{\rm int}(\vec{\tau}^{\alpha},0, -\vec m^{\alpha}_F)\cdot Z^{\rm hyper}_{\wt{(G^\alpha_{\vec\CP} \circ F^\alpha_{\vec\CP} \circ I^\alpha_{\vec\CP})}(Y)}\Big(\{\vec \s^{\gamma'} \}, \tau^\alpha_1, \eta_\alpha, \vec\eta \Big) \nn \\
& \times Z^{(Y,\{\CP_\beta\})}_{\rm int}(\{\vec \s^{\gamma'} \}, \vec{m}^Y(\vec{\eta}), \vec{\eta}^Y(\{\vec{u}^\beta =\mu^\beta + m^\alpha_{1\,F}\}, \{\vec{v}^\beta \},\ldots)),
\end{align}
where $Z^{(\CT[N^{\alpha}_F -1])}_{\rm int}$ is defined as above, and $Z^{\rm hyper}_{\wt{(G^\alpha_{\vec\CP} \circ F^\alpha_{\vec\CP} \circ I^\alpha_{\vec\CP})}(Y)}$ is the contribution of a single hypermultiplet:
\be \label{HyperGFIgen}
Z^{\rm hyper}_{\wt{(G^\alpha_{\vec\CP} \circ F^\alpha_{\vec\CP} \circ I^\alpha_{\vec\CP})}(Y)}\Big(\{\vec \s^{\gamma'} \}, \tau^\alpha_1, \eta_\alpha, \vec\eta \Big)= \frac{1}{\ch{(\sum_\beta g_\beta (\{\vec \s^{\gamma'} \}, \CP_\beta)+ \tau^\alpha_1+ \eta_\alpha + \sum_{\beta,l} b^{l}_\beta \eta_l)}}.
\ee

Therefore, for both types of operations, one can write down the general form of 
the quiver $Y'$. Assuming that the flavoring operation involves adding $N^{\alpha}_F$ hypermultiplets of gauge charge $1$, 
$Y'$ has the following form:
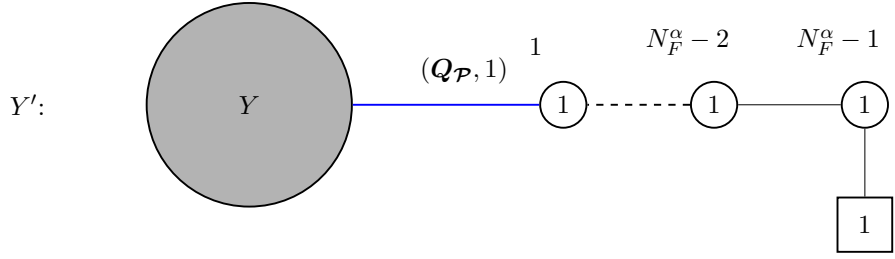
\begin{center}
\scalebox{0.9}{\begin{tikzpicture}[
cnode/.style={circle,draw,thick,minimum size=4mm},snode/.style={rectangle,draw,thick,minimum size=8mm},pnode/.style={rectangle,red,draw,thick,minimum size=1.0cm}, bnode/.style={circle,draw, thick, fill=black!30,minimum size=3cm}]
\node[text width=1 cm](34) at (-3,0) {$Y'$:};
\node[bnode] (1) at (0,0){$Y$} ;
\node[cnode] (25) at (9,0){1};
\node[snode] (26) [below=1cm of 25]{1};
\node[cnode] (27) [left= 1.5 cm of 25]{1};
\node[cnode] (28) [left=1.5 cm of 27]{1};
\draw[-] (25) -- (26);
\draw[-] (25) -- (27);
\draw[thick, dashed] (27) -- (28);
\draw[thick,blue] (28) -- (1);
\node[text width=2cm](31) [above=0.25 of 25] {$N^{\alpha}_F-1$};
\node[text width=2 cm](32) [above=0.25 of 27] {$N^{\alpha}_F-2$};
\node[text width=1cm](33) [above=0.25 of 28] {$1$};
\node[text width=1 cm](34) at (3,0.5) {$(\vec Q_{\vec \CP}, 1)$};
\end{tikzpicture}} 
\end{center}
The quiver $Y'$ can be built from the two subquivers - $Y$ and the linear quiver $\CT[N^{\alpha}_F -1]$, which are
connected by a single hypermultiplet which has charge $1$ under $U(1)_1$ gauge node of $\CT[N^{\alpha}_F -1]$, and 
is also charged under the $U(1)$ subgroups of certain unitary factors in the gauge group of $Y$. The charge vector corresponding to the latter 
is denoted by $\vec Q_{\vec \CP}$, so that the full charge vector is given by $(\vec Q_{\vec \CP},1)$. The precise charges can be read off 
from the formula \eref{HyperGFgen} and \eref{HyperGFIgen} respectively.

\end{itemize} 

This leads directly to the following result, which will be the main tool in this paper for constructing 3d Lagrangians for the AD theories:

\begin{result}
 If a quiver gauge theory $X'$ is related to another quiver $X$ by a sequence of elementary
Abelian $S$-type operations, and $X$ has a Lagrangian mirror $Y$, then $X'$ is guaranteed to have a Lagrangian mirror $Y'$.
\end{result}

The proof of this statement is straightforward. Starting from a dual pair of Lagrangian theories $(X,Y)$, the action of a single 
elementary Abelian $S$-type operation always gives a new pair of Lagrangian dual theories, as we discussed above. 
Therefore, implementing any number of elementary Abelian 
$S$-type operations on the dual pair $(X,Y)$ will obviously lead to a Lagrangian dual pair $(X',Y')$.\\

A important corollary of the above result is as follows. Assume that the pair of dual theories $(X,Y)$ both have unitary gauge groups. Then a necessary 
(but not sufficient) condition for $Y'$ to have a special unitary factor is that the generic Abelian $S$-type operation 
$\CO^{(\alpha_1, \ldots, \alpha_l)}_{({\vec \CP_1},\ldots,{\vec \CP_l})}$ consists of at least 
one gauging or identification-gauging elementary operation. In other words, if $\CO^{(\alpha_1, \ldots, \alpha_l)}_{({\vec \CP_1},\ldots,{\vec \CP_l})}$ consists 
only of flavoring-gauging and identification-flavoring-gauging operations, and the dual pair $(X,Y)$ have only unitary gauge groups, then $(X',Y')$
are guaranteed to have only unitary gauge groups. We will use this corollary to infer certain general features of 3d Lagrangians for AD theories in 
\Secref{3d-AD-Lag}.

\section{Unitary 3d Lagrangians from $S$-type operations}\label{3d-AD-Lag}

In this section, we discuss the general strategy for writing down the 3d Lagrangians for the $(A_{p-N-1}, A_{N-1})$ 
and the $D_p(SU(N))$ theories using $S$-type operations, and work out a few examples explicitly. 

\subsection{Constructing the 3d Lagrangians for the $(A_{p-N-1}, A_{N-1})$ theories}\label{AAgen}
In this subsection, we discuss the procedure for constructing the 3d Lagrangians for $(A_{p-N-1}, A_{N-1})$ on circle reduction. 
We give a systematic prescription for writing down the 3d Lagrangian for the generic case, 
and then present a specific example.

The basic idea is to study the 3d mirror of $(A_{p-N-1}, A_{N-1})$, which can be explicitly written down from the 
class $\CS$ construction of the 4d theory \cite{Xie:2012hs}. Given these 3d mirrors, one observes that they can be engineered 
from a $U(1)$ gauge theory with fundamental hypermultiplets, by a sequence of elementary Abelian $S$-type operations. 
Since the dual operation, following the discussion in \Secref{SOp-rev}, is guaranteed to produce a Lagrangian theory, 
we can explicitly write down a Lagrangian for the mirror of the 3d mirror i.e. a Lagrangian for the AD theory 
reduced on a circle.\\

\begin{figure}[htbp]
\begin{center}
\begin{tabular}{ccc}
\scalebox{0.9}{\begin{tikzpicture}[node distance=2cm,cnode/.style={circle,draw,thick,minimum size=8mm},snode/.style={rectangle,draw,thick,minimum size=8mm},pnode/.style={rectangle,red,draw,thick,minimum size=8mm}]
\node[cnode] (1) at (-2,0) {$1$};
\node[cnode] (2) at (2,0) {$1$};
\node[cnode] (3) at (0,2) {$1$};
\draw[line width=0.75mm, black] (1) -- (2);
\draw[line width=0.75mm, black] (2) -- (3);
\draw[line width=0.75mm, black] (1) -- (3);
\node[text width=0.1cm](15) at (0,0.25){$m_G$};
\node[text width=0.1cm](16) at (-1.4,1.25){$m_G$};
\node[text width=0.1cm](17) at (1,1.25){$m_G$};
\node[text width=1cm](4) at (0,-0.5){$(X')$};
\end{tikzpicture}}
& \qquad  \qquad
& \scalebox{0.9}{\begin{tikzpicture}[node distance=2cm,cnode/.style={circle,draw,thick,minimum size=8mm},snode/.style={rectangle,draw,thick,minimum size=8mm},pnode/.style={rectangle,red,draw,thick,minimum size=8mm}]
\node[cnode] (1) at (-2,0) {$1$};
\node[cnode] (2) at (2,0) {$1$};
\node[snode] (3) at (-2,-2) {$m_G$};
\node[snode] (4) at (2,-2) {$m_G$};
\draw[thick] (1) -- (3);
\draw[thick] (2) -- (4);
\draw[line width=0.75mm, black] (1) to (2);
\node[text width=0.1cm](15) at (0,0.25){$m_G$};
\node[text width=0.1cm](16) at (0,-3){$(X')$};
\end{tikzpicture}}
\end{tabular}
\caption{\footnotesize{The quiver gauge theory corresponding to the case with $m=3$ and generic $m_G$. The quiver on the right is obtained by decoupling a $U(1)$ vector multiplet associated with one of the vertices of the complete graph on the left.}}
\label{3dmirr-AA}
\end{center}
\end{figure}
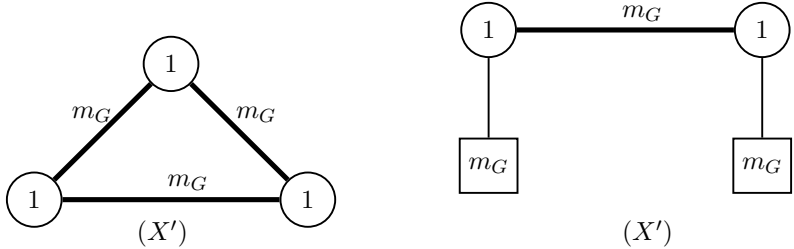

Given the $(A_{p-N-1}, A_{N-1})$ theory, one can define the following parameters:
\be
m=GCD(p,N) , \quad n=\frac{N}{m}, \quad q=\frac{p}{m}, \quad m_G= n(q-n).
\ee
The 3d mirror of the $(A_{p-N-1}, A_{N-1})$ theory is given by a quiver gauge theory with a set of free hypermultiplets.
The former is given by a complete graph of $m$ vertices and edge multiplicity $m_G$,
where each vertex denotes a $U(1)$ gauge node and an edge of multiplicity $m_G$ denotes $m_G$ bifundamental hypermultiplets, as 
shown in \figref{3dmirr-AA}. The number of free multiplets is $H_{\rm free} = \frac{(N-m)(p-N -m)}{2m}$.

We now observe that a generic quiver gauge theory of the form given in \figref{3dmirr-AA} can be constructed from a $U(1)$ theory 
with fundamental hypermultiplets by a sequence of elementary $S$-type operations. This can be done using the following steps:
\begin{enumerate}

\item The starting point is the quiver gauge theory $X$:
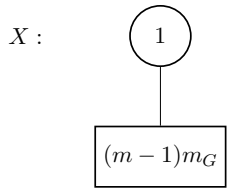
\begin{center}
\scalebox{.8}{\begin{tikzpicture}[node distance=2cm, cnode/.style={circle,draw,thick,minimum size=1.0 cm},snode/.style={rectangle,draw,thick,minimum size=1.0 cm}, bnode/.style={circle,draw, thick, fill=black!30,minimum size=1cm}]
\node[text width=1.0 cm](10) at (-2,0){$X:$};
\node[cnode] (1) at (0,0){$1$};
\node[snode] (2) at (0,-2){$(m-1)m_G$};
\draw[-] (1) -- (2);
\end{tikzpicture}}
\end{center}

\item Consider splitting the flavor node $U((m-1)m_G) \to U((m-2)m_G) \times U(m_G)$, and the resultant $U(m_G)$ flavor node into $m_G$ $U(1)$ 
flavor nodes, as shown below. Now, we implement the following elementary $S$-type operation -- the  $m_G$ $U(1)$ flavor nodes are identified to a single $U(1)$ 
flavor node, which is then flavored by $(m-2)m_G$ fundamental hypermultiplets, followed by gauging the identified $U(1)$. This identification-flavoring-gauging 
operation $\CO^{(1)}_{\CP}$ leads to the following quiver gauge theory:
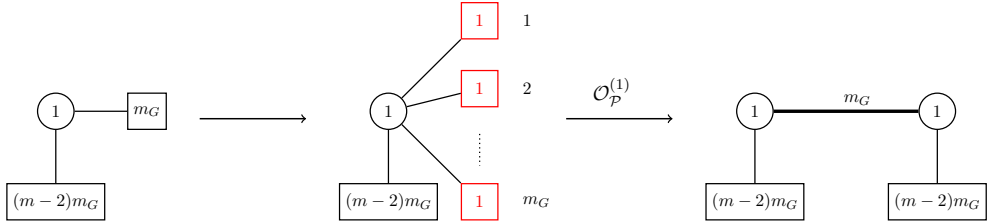
\begin{center}
\begin{tabular}{ccccc}
\scalebox{.6}{\begin{tikzpicture}[node distance=2cm,cnode/.style={circle,draw,thick,minimum size=8mm},snode/.style={rectangle,draw,thick,minimum size=8mm},pnode/.style={rectangle,red,draw,thick,minimum size=8mm}]
\node[cnode] (1) at (0,0) {$1$};
\node[snode] (2) at (0,-2) {$(m-2)m_G$};
\node[snode] (3) at (2,0) {$m_G$};
\draw[thick] (1) -- (2);
\draw[thick] (1) -- (3);
\end{tikzpicture}}
& \scalebox{.7}{\begin{tikzpicture} \draw[thick, ->] (0,0) -- (2,0); 
\node[] at (1,-1.8) {};
\end{tikzpicture}}
&\scalebox{.6}{\begin{tikzpicture}[node distance=2cm,cnode/.style={circle,draw,thick,minimum size=8mm},snode/.style={rectangle,draw,thick,minimum size=8mm},pnode/.style={rectangle,red,draw,thick,minimum size=8mm}]
\node[cnode] (1) at (0,0) {$1$};
\node[snode] (2) at (0,-2) {$(m-2)m_G$};
\node[pnode] (3) at (2,2) {$1$};
\node[pnode] (4) at (2,0.5) {$1$};
\node[pnode] (5) at (2,-2) {$1$};
\draw[thick] (1) -- (2);
\draw[thick] (1) -- (3);
\draw[thick] (1) -- (4);
\draw[thick] (1) -- (5);
\draw[thick, dotted, black] (2, -0.5) -- (2,-1.2);
\node[text width=0.1cm](15) at (3, 2){$1$};
\node[text width=0.1cm](16) at (3, 0.5){$2$};
\node[text width=0.1cm](17) at (3, -2){$m_G$};
\end{tikzpicture}}
& \scalebox{.7}{\begin{tikzpicture} \draw[thick, ->] (0,0) -- (2,0); 
\node[] at (1,-1.8) {};
\node[text width=1cm](1) at (1,0.5) {$\CO^{(1)}_{\CP}$};
\end{tikzpicture}}
&\scalebox{.6}{\begin{tikzpicture}[node distance=2cm,cnode/.style={circle,draw,thick,minimum size=8mm},snode/.style={rectangle,draw,thick,minimum size=8mm}]
\node[cnode] (1) at (-2,0) {$1$};
\node[cnode] (2) at (2,0) {$1$};
\node[snode] (3) at (-2,-2) {$(m-2)m_G$};
\node[snode] (4) at (2,-2) {$(m-2)m_G$};
\draw[thick] (1) -- (3);
\draw[thick] (2) -- (4);
\draw[line width=0.75mm, black] (1) to (2);
\node[text width=0.1cm](15) at (0,0.25){$m_G$};
\end{tikzpicture}}
\end{tabular}
\end{center}

\item In the next step, we split the two flavor nodes as $U((m-2)m_G)_1 \to U((m-3)m_G)_1 \times U(m_G)_1$ and 
$U((m-2)m_G)_2 \to U((m-3)m_G)_2 \times U(m_G)_2$. In addition, we split each of the two $U(m_G)$ flavor nodes 
into $m_G$ $U(1)$ nodes. Now, we implement the following elementary $S$-type operation -- the  $2m_G$ $U(1)$ 
flavor nodes are identified to a single $U(1)$ flavor node, which is then flavored by $(m-3)m_G$ fundamental hypermultiplets, 
followed by gauging the identified $U(1)$. This identification-flavoring-gauging operation leads to the following quiver gauge theory:
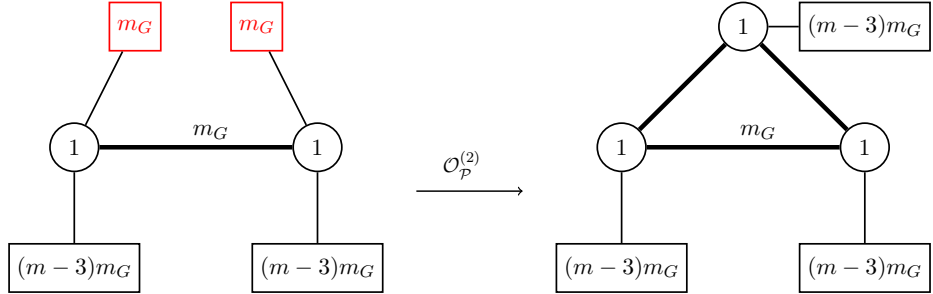
\begin{center}
\begin{tabular}{ccc}
\scalebox{.8}{\begin{tikzpicture}[node distance=2cm,cnode/.style={circle,draw,thick,minimum size=8mm},snode/.style={rectangle,draw,thick,minimum size=8mm}]
\node[cnode] (1) at (-2,0) {$1$};
\node[cnode] (2) at (2,0) {$1$};
\node[snode] (3) at (-2,-2) {$(m-3)m_G$};
\node[snode] (4) at (2,-2) {$(m-3)m_G$};
\node[snode, red] (5) at (-1, 2) {$m_G$};
\node[snode, red] (6) at (1, 2) {$m_G$};
\draw[thick] (1) -- (3);
\draw[thick] (2) -- (4);
\draw[thick] (1) -- (5);
\draw[thick] (2) -- (6);
\draw[line width=0.75mm, black] (1) to (2);
\node[text width=0.1cm](15) at (0,0.25){$m_G$};
\end{tikzpicture}}
& \scalebox{.7}{\begin{tikzpicture} \draw[thick, ->] (0,0) -- (2,0); 
\node[] at (1,-1.8) {};
\node[text width=1cm](1) at (1,0.5) {$\CO^{(2)}_{\CP}$};
\end{tikzpicture}}
&\scalebox{.8}{\begin{tikzpicture}[node distance=2cm,cnode/.style={circle,draw,thick,minimum size=8mm},snode/.style={rectangle,draw,thick,minimum size=8mm}]
\node[cnode] (1) at (-2,0) {$1$};
\node[cnode] (2) at (2,0) {$1$};
\node[snode] (3) at (-2,-2) {$(m-3)m_G$};
\node[snode] (4) at (2,-2) {$(m-3)m_G$};
\node[cnode] (5) at (0,2) {$1$};
\node[snode] (6) at (2,2) {$(m-3)m_G$};
\draw[thick] (1) -- (3);
\draw[thick] (2) -- (4);
\draw[thick] (5) -- (6);
\draw[line width=0.75mm, black] (1) to (5);
\draw[line width=0.75mm, black] (2) to (5);
\draw[line width=0.75mm, black] (1) to (2);
\node[text width=0.1cm](15) at (0,0.25){$m_G$};
\end{tikzpicture}}
\end{tabular}
\end{center}

\item Proceeding in the same fashion, we implement $(m-2)$ identification-flavoring-gauging operation to engineer the quiver 
gauge theory $X'$ from the quiver $X$, i.e.
\be \label{SOp-AA}
X' = \CO^{(m-2)}_{\vec \CP_{m-2}} \circ \CO^{(m-3)}_{\vec \CP_{m-3}} \circ \ldots \circ \CO^{(2)}_{\vec \CP_{2}} \circ \CO^{(1)}_{\vec \CP_{1}}(X),
\ee
where $X'$ is the quiver gauge theory given in \figref{3dmirr-AA}, for generic $m$ and $m_G$.  

\end{enumerate}

Using the general prescription  in \Secref{SOp-Ab} for finding out the dual of a given elementary Abelian $S$-type operation, 
one can explicitly determine the 3d mirror of $X'$. Since $X'$ can be engineered from the linear quiver $X$ by a sequence 
of elementary Abelian $S$-type operations, the dual quiver $Y'$ is guaranteed to be a Lagrangian theory. The quiver $Y'$, 
for generic $p$ and $N$, will be a non-linear quiver involving $U(1)$ gauge groups. As discussed earlier, the quiver $Y'$, 
along with $H$ free twisted hypermultiplets, then gives a Lagrangian description for the $(A_{p-N-1}, A_{N-1})$ 
AD theory reduced on a circle. We will now explicitly find the quiver $Y'$ in a specific example.\\

\begin{figure}[htbp]
\begin{center}
\scalebox{0.6}{\begin{tikzpicture}[node distance=2cm,cnode/.style={circle,draw,thick,minimum size=8mm},snode/.style={rectangle,draw,thick,minimum size=8mm},pnode/.style={rectangle,red,draw,thick,minimum size=8mm}]
\node[cnode] (1) at (-2,0) {$1$};
\node[cnode] (2) at (2,0) {$1$};
\node[snode] (3) at (-2,-2) {$m_G$};
\node[snode] (4) at (2,-2) {$m_G$};
\draw[thick] (1) -- (3);
\draw[thick] (2) -- (4);
\draw[line width=0.75mm, black] (1) to (2);
\node[text width=0.1cm](15) at (0,0.25){$m_G$};
\node[text width=0.1cm](16) at (0,-3){$(X')$};
\end{tikzpicture}}
\qquad \qquad
\scalebox{0.5}{\begin{tikzpicture}[node distance=2cm,cnode/.style={circle,draw,thick,minimum size=8mm},snode/.style={rectangle,draw,thick,minimum size=8mm},pnode/.style={rectangle,red,draw,thick,minimum size=8mm}]
\node[snode] (1) at (-2,0) {$1$};
\node[cnode] (2) at (0,0) {$1$};
\node[cnode] (3) at (2,0) {$1$};
\node[] (4) at (3,0) {};
\node[] (5) at (4,0) {};
\node[cnode] (6) at (5,0) {$1$};
\node[cnode] (7) at (7,0) {$1$};
\node[] (8) at (8,0) {};
\node[] (9) at (12,0) {};
\node[cnode] (10) at (13,0) {$1$};
\node[cnode] (11) at (15,0) {$1$};
\node[snode] (12) at (17,0) {$1$};
\node[cnode] (13) at (7,2) {$1$};
\node[cnode] (14) at (9,2) {$1$};
\node[] (15) at (10,2) {};
\node[] (16) at (12,2) {};
\node[cnode] (17) at (13,2) {$1$};
\node[snode] (30) at (15,2) {$1$};
\draw[thick] (1) -- (2);
\draw[thick] (2) -- (3);
\draw[thick] (3) -- (4);
\draw[thick,dashed] (4) -- (5);
\draw[thick] (5) -- (6);
\draw[thick] (6) -- (7);
\draw[thick] (7) -- (8);
\draw[thick,dashed] (8) -- (9);
\draw[thick] (9) -- (10);
\draw[thick] (10) -- (11);
\draw[thick] (11) -- (12);
\draw[thick] (7) -- (13);
\draw[thick] (13) -- (14);
\draw[thick] (14) -- (15);
\draw[thick,dashed] (15) -- (16);
\draw[thick] (16) -- (17);
\draw[thick] (17) -- (30);
\node[text width=0.1cm](20) at (0,-1) {$1$};
\node[text width=0.1cm](21) at (2,-1) {$2$};
\node[text width= 1.5 cm](22) at (5,-1) {$m_G-1$};
\node[text width=1 cm](23) at (7,-1) {$m_G$};
\node[text width=1.5 cm](24) at (13,-1) {$2m_G-2$};
\node[text width=1.5 cm](25) at (15,-1) {$2m_G-1$};
\node[text width=0.1 cm](26) at (7,3) {$1$};
\node[text width=0.1 cm](27) at (9,3) {$2$};
\node[text width=1.5 cm](28) at (13,3) {$m_G-1$};
\node[text width=0.1cm](30) at (7,-3){$(Y')$};
\end{tikzpicture}}
\caption{\footnotesize{The quiver $X'$ is the 3d mirror of $(A_{p-N-1}, A_{N-1})$ (with $m=2$ and $m_G$ arbitrary) reduced on a circle, 
constructed from the class $\CS$ prescription. The quiver $Y'$ is the 3d mirror of $X'$, i.e. a Lagrangian realization of the SCFT obtained 
by the reduction of $(A_{p-N-1}, A_{N-1})$ (with $m=2$ and $m_G$ arbitrary) on a circle.}}
\label{AbEx1gen}
\end{center}
\end{figure}
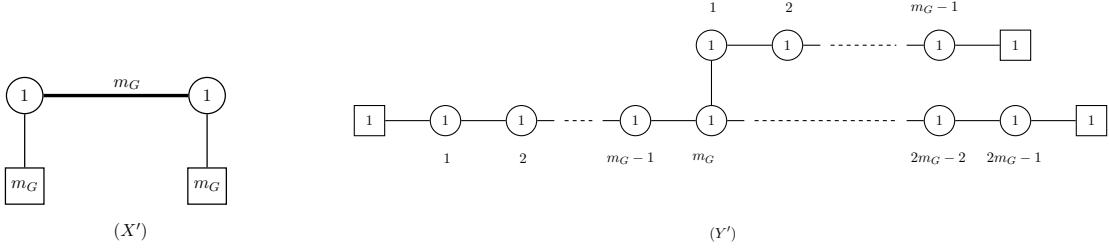

Consider the case where $m=3$ and $m_G$ is arbitrary. The 3d mirror of the associated AD theory is then given by the 
theory $(X')$ in \figref{3dmirr-AA}. One can now construct the quiver $X'$ starting from a simple linear quiver using 
the $S$-type operation technique, and determine its mirror dual $Y'$. The result is shown in \figref{AbEx1gen}, 
where the dual theory $Y'$ has the shape of a star-shaped quiver consisting of three linear quivers glued at a 
common $U(1)$ gauge node. The dimensions of the respective Higgs and Coulomb branches, and the associated 
global symmetries are shown in Table \ref{Tab:AbEx1gen}.

\begin{center}
\begin{table}[htbp]
\resizebox{\textwidth}{!}{%
\begin{tabular}{|c|c|c|}
\hline
Moduli space data & Theory $X'$ & Theory $Y'$ \\
\hline \hline 
dim\,$\CM_H$ & $3m_G-2$ & $2$\\
\hline
dim\,$\CM_C$ & $2$ & $3m_G-2$\\
\hline
$G_H$ & $SU(m_G) \times SU(m_G) \times SU(m_G) \times U(1) $  & $U(1) \times U(1)$\\
\hline
$G_C$ & $U(1) \times U(1)$ & $SU(m_G) \times SU(m_G) \times SU(m_G) \times U(1) $ \\
\hline
\end{tabular}}
\caption{Summary table for the moduli space dimensions and global symmetries for the mirror pair in \figref{AbEx1gen}.}
\label{Tab:AbEx1gen}
\end{table}
\end{center}

\begin{figure}[htbp]
\begin{center}
\scalebox{0.6}{\begin{tikzpicture}[node distance=2cm,cnode/.style={circle,draw,thick,minimum size=8mm},snode/.style={rectangle,draw,thick,minimum size=8mm},pnode/.style={rectangle,red,draw,thick,minimum size=8mm}]
\node[cnode] (1) at (0,0) {$1$};
\node[snode] (2) at (0,-2) {$2m_G$};
\draw[thick] (1) -- (2);
\node[text width=0.1cm](3) at (0,-3){$(X)$};
\end{tikzpicture}}
\qquad \qquad
\scalebox{0.6}{\begin{tikzpicture}[node distance=2cm,cnode/.style={circle,draw,thick,minimum size=8mm},snode/.style={rectangle,draw,thick,minimum size=8mm},pnode/.style={rectangle,red,draw,thick,minimum size=8mm}]
\node[cnode] (1) at (0,0) {$1$};
\node[cnode] (2) at (2,0) {$1$};
\node[] (3) at (3,0) {};
\node[] (4) at (4,0) {};
\node[cnode] (5) at (5,0) {$1$};
\node[cnode] (6) at (7,0) {$1$};
\node[snode] (7) at (0,-2) {$1$};
\node[snode] (8) at (7, -2) {$1$};
\draw[thick] (1) -- (2);
\draw[thick] (2) -- (3);
\draw[thick, dashed] (3) -- (4);
\draw[thick] (4) -- (5);
\draw[thick] (5) -- (6);
\draw[thick] (6) -- (8);
\draw[thick] (1) -- (7);
\node[text width=0.1cm](20) at (0,1) {$1$};
\node[text width=0.1cm](21) at (2,1) {$2$};
\node[text width= 1.5 cm](22) at (5,1) {$2m_G-2$};
\node[text width=1.5 cm](23) at (7,1) {$2m_G-1$};
\node[text width=0.1cm](20) at (3.5,-3){$(Y)$};
\end{tikzpicture}}
\caption{\footnotesize{A pair of mirror dual Abelian linear quivers.}}
\label{AbEx1LQ}
\end{center}
\end{figure}
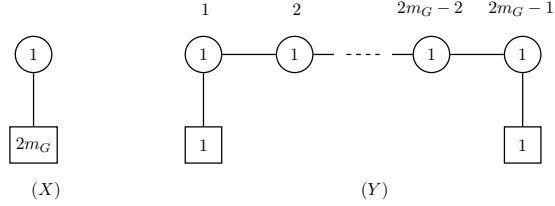

Let us now construct the quiver $Y'$ explicitly, using the $S$-type operations.
The starting point is the mirror pair $(X,Y)$, as shown in \figref{AbEx1LQ}. 
The partition functions of the theories $(X,Y)$ are given as
\begin{align}
 Z^{(X)}(\vec m, \vec t) =& \int\, ds \, \frac{e^{2\pi i \,(t_1- t_2)\,s}}{\prod^{2m_G}_{i=1}\, \ch{(s-m_i)}}=: \int\, ds \, Z^{(X)}_{\rm int} (s, \vec m, \vec t), \\
Z^{(Y)}(\vec t, \vec m) = &\int\, \prod^{2m_G-1}_{k=1}d\s_k \, \frac{\prod^{2m_G-1}_{k=1}\, e^{2\pi i \,(m_k- m_{k+1})\,\s_k}}{\ch{(\s_1 -t_1)}\,\prod^{2m_G-2}_{k=1}\ch{(\s_k -\s_{k+1})} \,\ch{(\s_{2m_G-1} -t_2)}}\\
=: & \int\, \prod^{2m_G-1}_{k=1}d\s_k \, Z^{(Y)}_{\rm int} (\{\s_k \}, \vec m, \vec t),
\end{align}
where $Z^{(X)}_{\rm int}$ and $Z^{(Y)}_{\rm int}$ are the respective matrix model integrands for $X$ and $Y$, and the parameters $\vec m, \vec t$ are unrestricted.
Mirror symmetry of $X$ and $Y$ implies that
\begin{align}
& Z^{(X)}(\vec m, \vec t)  =C_{XY}(\vec m, \vec t)\, Z^{(Y)}(\vec t, -\vec m), \\
& C_{XY}(\vec m, \vec t)=  e^{2\pi i (m_1\,t_1- m_{2m_G}\,t_2)},
\end{align}
where $C_{XY}$ is a contact term, which reminds us that the duality holds for unrestricted $\vec m$ and $\vec t$, only when we turn on a certain background 
Chern-Simons sector. 

\begin{figure}[htbp]
\begin{center}
\begin{tabular}{ccc}
\scalebox{.5}{\begin{tikzpicture}[node distance=2cm,cnode/.style={circle,draw,thick,minimum size=8mm},snode/.style={rectangle,draw,thick,minimum size=8mm},pnode/.style={rectangle,red,draw,thick,minimum size=8mm}]
\node[cnode] (1) at (0,0) {$1$};
\node[snode] (2) at (0,-2) {$m_G$};
\node[pnode] (3) at (2,2) {$1$};
\node[pnode] (4) at (2,0.5) {$1$};
\node[pnode] (5) at (2,-2) {$1$};
\draw[thick] (1) -- (2);
\draw[thick] (1) -- (3);
\draw[thick] (1) -- (4);
\draw[thick] (1) -- (5);
\draw[thick, dotted, black] (2, -0.5) -- (2,-1.2);
\node[text width=0.1cm](3) at (0,-3){$(X)$};
\node[text width=0.1cm](15) at (3, 2){$1$};
\node[text width=0.1cm](16) at (3, 0.5){$2$};
\node[text width=0.1cm](17) at (3, -2){$m_G$};
\end{tikzpicture}}
& \qquad \qquad 
& \scalebox{.5}{\begin{tikzpicture}[node distance=2cm,cnode/.style={circle,draw,thick,minimum size=8mm},snode/.style={rectangle,draw,thick,minimum size=8mm},pnode/.style={rectangle,red,draw,thick,minimum size=8mm}]
\node[cnode] (1) at (0,0) {$1$};
\node[cnode] (2) at (2,0) {$1$};
\node[] (3) at (3,0) {};
\node[] (4) at (4,0) {};
\node[cnode] (5) at (5,0) {$1$};
\node[cnode] (6) at (7,0) {$1$};
\node[snode] (7) at (0,-2) {$1$};
\node[snode] (8) at (7, -2) {$1$};
\draw[thick] (1) -- (2);
\draw[thick] (2) -- (3);
\draw[thick, dashed] (3) -- (4);
\draw[thick] (4) -- (5);
\draw[thick] (5) -- (6);
\draw[thick] (6) -- (8);
\draw[thick] (1) -- (7);
\node[text width=0.1cm](20) at (0,1) {$1$};
\node[text width=0.1cm](21) at (2,1) {$2$};
\node[text width= 1.5 cm](22) at (5,1) {$2m_G-2$};
\node[text width= 1.5 cm](23) at (7,1) {$2m_G-1$};
\node[text width=0.1cm](20) at (3.5,-3){$(Y)$};
\end{tikzpicture}}\\
 \scalebox{.7}{\begin{tikzpicture}
\draw[->] (15,-3) -- (15,-5);
\node[text width=0.1cm](20) at (14.0, -4) {$\CO^\alpha_{\CP}$};
\end{tikzpicture}}
&\qquad \qquad 
& \scalebox{.7}{\begin{tikzpicture}
\draw[->] (15,-3) -- (15,-5);
\node[text width=0.1cm](29) at (15.5, -4) {$\wt{\CO}^\alpha_{\CP}$};
\end{tikzpicture}}\\
\scalebox{.6}{\begin{tikzpicture}[node distance=2cm,cnode/.style={circle,draw,thick,minimum size=8mm},snode/.style={rectangle,draw,thick,minimum size=8mm},pnode/.style={rectangle,red,draw,thick,minimum size=8mm}]
\node[cnode] (1) at (-2,0) {$1$};
\node[cnode] (2) at (2,0) {$1$};
\node[snode] (3) at (-2,-2) {$m_G$};
\node[snode] (4) at (2,-2) {$m_G$};
\draw[thick] (1) -- (3);
\draw[thick] (2) -- (4);
\draw[line width=0.75mm, black] (1) to (2);
\node[text width=0.1cm](15) at (0,0.25){$m_G$};
\node[text width=0.1cm](16) at (0,-3){$(X')$};
\end{tikzpicture}}
&\qquad \qquad 
& \scalebox{.5}{\begin{tikzpicture}[node distance=2cm,cnode/.style={circle,draw,thick,minimum size=8mm},snode/.style={rectangle,draw,thick,minimum size=8mm},pnode/.style={rectangle,red,draw,thick,minimum size=8mm}]
\node[snode] (1) at (-2,0) {$1$};
\node[cnode] (2) at (0,0) {$1$};
\node[cnode] (3) at (2,0) {$1$};
\node[] (4) at (3,0) {};
\node[] (5) at (4,0) {};
\node[cnode] (6) at (5,0) {$1$};
\node[cnode] (7) at (7,0) {$1$};
\node[] (8) at (8,0) {};
\node[] (9) at (12,0) {};
\node[cnode] (10) at (13,0) {$1$};
\node[cnode] (11) at (15,0) {$1$};
\node[snode] (12) at (17,0) {$1$};
\node[cnode] (13) at (7,2) {$1$};
\node[cnode] (14) at (9,2) {$1$};
\node[] (15) at (10,2) {};
\node[] (16) at (12,2) {};
\node[cnode] (17) at (13,2) {$1$};
\node[snode] (30) at (15,2) {$1$};
\draw[thick] (1) -- (2);
\draw[thick] (2) -- (3);
\draw[thick] (3) -- (4);
\draw[thick,dashed] (4) -- (5);
\draw[thick] (5) -- (6);
\draw[thick] (6) -- (7);
\draw[thick] (7) -- (8);
\draw[thick,dashed] (8) -- (9);
\draw[thick] (9) -- (10);
\draw[thick] (10) -- (11);
\draw[thick] (11) -- (12);
\draw[thick] (7) -- (13);
\draw[thick] (13) -- (14);
\draw[thick] (14) -- (15);
\draw[thick,dashed] (15) -- (16);
\draw[thick] (16) -- (17);
\draw[thick] (17) -- (30);
\node[text width=0.1cm](20) at (0,-1) {$1$};
\node[text width=0.1cm](21) at (2,-1) {$2$};
\node[text width= 1.5 cm](22) at (5,-1) {$m_G-1$};
\node[text width=1 cm](23) at (7,-1) {$m_G$};
\node[text width=1.5 cm](24) at (13,-1) {$2m_G-2$};
\node[text width=1.5 cm](25) at (15,-1) {$2m_G-1$};
\node[text width=0.1 cm](26) at (7,3) {$1$};
\node[text width=0.1 cm](27) at (9,3) {$2$};
\node[text width=1.5 cm](28) at (13,3) {$m_G-1$};
\node[text width=0.1cm](30) at (7,-3){$(Y')$};
\end{tikzpicture}}
\end{tabular}
\caption{\footnotesize{Construction of the pair of mirror dual theories $(X',Y')$, by a single elementary Abelian $S$-type operation 
of the identification-flavoring-gauging type. The dual operation involves attaching a balanced linear chain of $U(1)$ gauge nodes 
to the $U(1)_{m_G}$ gauge node of $Y$.}}
\label{SimpAbEx1GFI}
\end{center}
\end{figure}
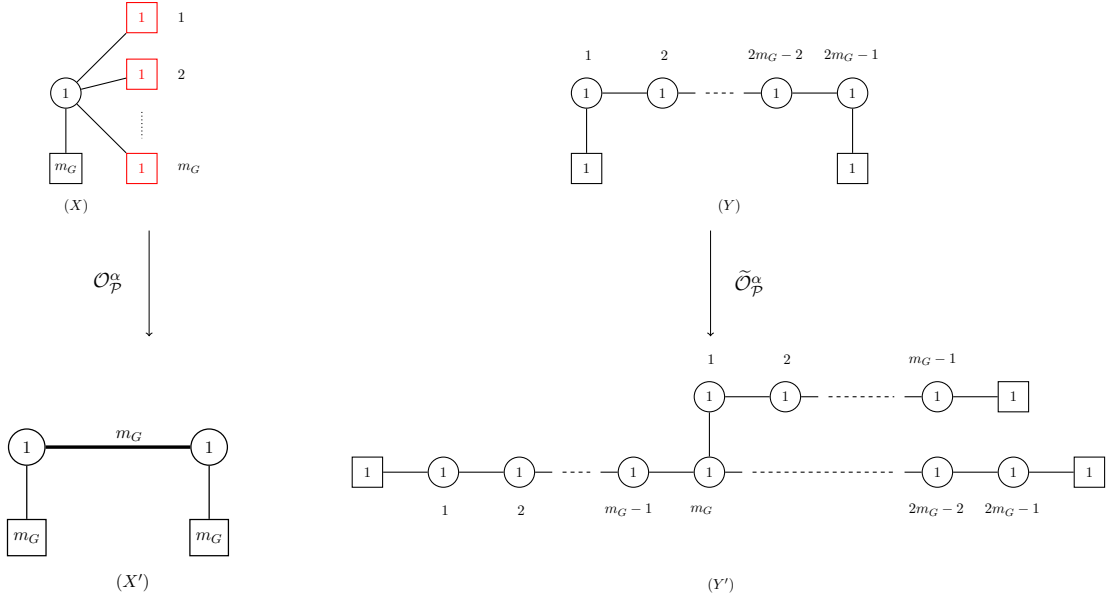

To obtain the quiver $X'$ from the linear quiver $X$, one implements an elementary Abelian identification-flavoring-gauging operation $\CO^\alpha_{\CP}$ 
on $X$ (as stated in Step 2 of the general procedure above) :
\begin{align}
\CO^\alpha_{\CP}(X)= G^\alpha_\CP \circ F^\alpha_\CP \circ I^\alpha_\CP(X) , \label{SOp-1a}
\end{align}
where $\CO^\alpha_{\CP}$ is shown explicitly in \figref{SimpAbEx1GFI}. The $m_G$ $U(1)$ nodes being identified are shown in red. 
The mass parameters associated with the $U(1)^{m_G}$ global symmetry being identified are chosen as :
\begin{align}
u_j=  m_{m_G+j}, \qquad j=1,\ldots, m_G, \label{SOp-1b}
\end{align}
which corresponds to a specific choice of the permutation matrix $\CP$.
The partition function of the theory $X'= \CO^\alpha_{\CP}(X)$ is then given as
\begin{align}
Z^{\CO^\alpha_{\CP}(X)}= \int\, d{u}^{\alpha}\, \CZ_{\CO^\alpha_{\CP}(X)}(u^\alpha, \vec u, \eta_\alpha, \vec m^f, \vec \mu) \cdot Z^{(X,\CP)}(\vec u, \ldots; \vec t), 
\end{align}
where the operator $\CZ_{\CO^\alpha_{\CP}(X)}$ can be constructed from the general prescription of \eref{CZ-OP} and has the following form:
\begin{align}
\CZ_{\CO^\alpha_{\CP}(X)}= & Z_{\rm FI} ({u}^{\alpha},\eta_{\alpha})\, Z_{\rm 1-loop}^{\rm hyper} ({u}^{\alpha}, \vec m^{\alpha}_F)\, \int \prod^{m_G}_{j=1} \, d {u_{j}} \, \delta \Big({u}^{\alpha} - {u}_{j} + {\mu}_{j} \Big) \\
= & \frac{e^{2\pi i \eta_\alpha u^\alpha}}{\prod^{m_G}_{a=1}\, \ch{(u^\alpha - m^f_a)}}\, \int \prod^{m_G}_{j=1} \, d {u_{j}} \, \delta \Big({u}^{\alpha} - {u}_{j} + {\mu}_{j} \Big).
\end{align}
Using the above expression, the partition function of the theory $X'$ can be explicitly written as
\begin{align}\label{PF-X1}
Z^{(X')}= \int\, ds\, du^\alpha\, \frac{e^{2\pi i \eta_\alpha u^\alpha}\, e^{2\pi i \,(t_1- t_2)\,s}}{\prod^{m_G}_{i=1} \ch{(s-m^{(1)}_i)}\, \prod^{m_G}_{a=1} \ch{(u^\alpha-m^{(2)}_a)} \, 
\prod^{m_G}_{j=1} \ch{(s- u^\alpha -m^{\rm bif}_j)}},
\end{align}
where the fundamental masses $\vec{m}^{(1)}, \vec{m}^{(2)}$ and the bifundamental masses $\vec m^{\rm bif}$ are given in terms of mass parameters introduced above 
in the following fashion:
\begin{align}
& m^{(1)}_i = m_{i}, \qquad i=1,\ldots, m_G,\\
& m^{(2)}_a= m^f_a, \qquad a=1,\ldots, m_G,\\
& m^{\rm bif}_j= \mu_j,  \qquad j=1,\ldots, m_G.
\end{align}
The matrix integral on the RHS of \eref{PF-X1} manifestly corresponds to the quiver $X'$ in \figref{AbEx1gen}. 
By shifting the integration variables, one can check that the matrix integral can be written in terms 
of $3m_G-2$ independent real masses, which live in the Cartan subalgebra of the Higgs branch global symmetry group of $X'$, i.e. 
$G^{X'}_H=SU(m_G)^3 \times U(1)$.
The Coulomb branch global symmetry can be read off from the quiver itself -- it has two unbalanced nodes (unless $m_G=1$, in which case 
$X'$ is a linear quiver) leading to a $G^{X'}_C=U(1)^2$.

The dual theory can be read off from the general construction in \eref{PF-wtOPgen}-\eref{CZ-wtOP}, with the function $ \CZ_{\wt{\CO}^\alpha_{\vec \CP}(Y)}$ being 
given as
\begin{align}
& \CZ_{\wt{\CO}^\alpha_{\vec \CP}(Y)} = \int \, du^\alpha \, \CZ_{\CO^\alpha_{\vec \CP}(X)} \cdot \prod^{m_G}_{j=1}\, e^{2\pi i (g^j(\{\s_k\}, \CP) + \sum_l b^{jl}t_l)\,u_j},\\
&\prod^{m_G}_{j=1} e^{2\pi i g^j(\{\s_k\}, \CP) \,u_j}= \Big(\prod^{m_G-1}_{j=1}\, e^{2\pi i u_j \, (\s_{m_G+j-1} - \s_{m_G+j})}\Big)\,e^{2\pi i u_{m_G}\, \s_{2m_G-1}}, \\
& \prod^{m_G}_{j=1} e^{2\pi i u_j \sum_l b^{jl}t_l} = e^{-2\pi i u_{m_G}\, t_2}.
\end{align}
This leads to the following dual partition function
\begin{align}\label{Ab-DualPf}
Z^{\wt{\CO}^\alpha_{\vec \CP}(Y)}(\vec{m}'; \vec \eta')= C_{X'Y'}\cdot \int \, & \prod^{2m_G-1}_{k=1}\,  d\s_k\,\prod^{m_G-1}_{b=1}\, d\tau_b\,
Z^{\rm bif}_{\rm 1-loop}(\tau_1, \s_{m_G}, 0)\, Z^{(\CT[m_G-1])}_{\rm int}(\{ \tau_b\}, \eta_\alpha -t_2, -\vec m^f) \nn \\
\times \, &  Z^{(Y)}_{\rm int}(\{ \s_k\}, \vec t, \{-m_1, \ldots, -m_{m_G}, -\mu'_1,\ldots, - \mu'_{m_G} \}), 
\end{align}
where the mass parameters $\mu'_j = \mu_j + m^f_1$ for $j=1,\ldots, m_G$. The function $C_{X'Y'}$ and the function 
$Z^{(\CT[m_G-1])}_{\rm int}$ (defined in \Secref{SOp-Ab}) are given as
\begin{align}
& C_{X'Y'}(\vec m^f, \vec \mu,\vec t)= e^{2\pi i m^f_{m_G}(\eta_\alpha -t_2)}\, e^{2\pi i (m_1 t_1 - \mu_{m_G} t_2)},\\
& Z^{(\CT[m_G-1])}_{\rm int}(\{ \tau_k\}, \eta_\alpha -t_2, \vec m^f) = 
\frac{\prod^{m_G-1}_{b=1} e^{2\pi i \tau_b(m^f_b - m^f_{b+1})}}{\prod^{m_G-2}_{b=1}\ch{(\tau_b -\tau_{b+1})}\, \ch{(\tau_{m_G-1} - \eta_\alpha +t_2)}}.
\end{align}
From the RHS of \eref{Ab-DualPf}, the gauge group and the matter content of the dual theory can be read off:
\begin{align}
Z^{\wt{\CO}^\alpha_{\vec \CP}(Y)}(\vec{m}'; \vec \eta')=:  C_{X'Y'}(\vec m^f, \vec \mu,\vec t) \cdot Z^{(Y')}\Big(\vec{m}'(\vec t, \eta_\alpha); \vec{\eta}'(\vec m, \vec m^f, \vec \mu) \Big),
\end{align}
where $Y'$ is the quiver gauge theory shown in \figref{SimpAbEx1GFI}, and $ C_{X'Y'}$ is a contact term associated with the new duality.
The mirror map for the masses and FI parameters can be read off from the RHS of \eref{Ab-DualPf}.  After a shift in the integration variables, 
the matrix integral can be written in terms of two independent real masses, associated with the Higgs branch global symmetry 
$G^{(Y')}_H = U(1)^2$. The Coulomb branch global symmetry can be read off from $Y'$ as follows. The quiver has three  
subquivers each consisting of a linear chain of $m_G-1$ gauge nodes and a single unbalanced gauge node at the center.
This leads to the Coulomb branch global symmetry $G^{(Y')}_H = SU(m_G)^3 \times U(1)$.

\subsection{Constructing the 3d Lagrangians for the $D_p(SU(N))$ theories}\label{DpSUNgen}

In this subsection, we discuss the procedure for constructing the 3d Lagrangians for $D_p(SU(N))$ on circle reduction, 
using $S$-type operations. We make a general argument that the resultant quiver gauge theory always has unitary 
gauge group factors with fundamentals and bifundamentals, as well as the Abelian hypermultiplets.

The starting point is the 3d mirror of $D_p(SU(N))$, which can be explicitly written down from the 
class $\CS$ construction of the 4d theory \cite{Wang:2015mra, Giacomelli:2020ryy}. Below, 
we treat the two cases $p<N$ and $p> N$ separately. 

\subsubsection{The case of $p<N$}\label{pltNGen}

Let us first consider the case of $p < N$, for which we define the parameters:
\begin{align}
& x=\lfloor \frac{N}{p} \rfloor, \quad M=N - (x+1), \quad m=GCD(p,N) , \quad n=\frac{N}{m}, \quad q=\frac{p}{m}, \nn \\
& m_1=q(1+x) -n, \quad m_2=n-qx. \label{pltN-pars}
\end{align}

The 3d mirrors for $D_p(SU(N))$ with $p<N$ can be constructed using the following steps.

\begin{enumerate}

\item  Consider three subquivers from which the 3d quiver will be built out:

\begin{itemize}

\item A linear chain of unitary gauge groups with bifundamental matter : $$ \CT_1 = (p-1) - 2(p-1) - \ldots - (x-1)(p-1) - x(p-1)$$

\item Another linear chain of unitary gauge groups with bifundamental matter : $$ \CT_2 = M - (M-1) - \ldots - 2 - 1$$

\item A complete graph $\CT_3$ of $m$ vertices, where each vertex is a $U(1)$ gauge node, and edge multiplicity $m_G= m_1\,m_2$.

\end{itemize}

\item Connect the $U(x(p-1))$ gauge node of $\CT_1$ and the $U(M)$ gauge node of $\CT_2$ with an edge of multiplicity 1, i.e.  single
bifundamental hyper. 

\item Connect the $U(x(p-1))$ gauge node of $\CT_1$ to each of the $U(1)$ gauge nodes of $\CT_3$ with an edge of multiplicity $m_1$.

\item Connect the $U(M)$ gauge node of $\CT_2$ to each of the $U(1)$ gauge nodes of $\CT_3$ with an edge of multiplicity $m_2$.

\item Finally, one needs to decouple an overall $U(1)$ factor from the quiver, and we will choose to implement this by ungauging one of 
the $U(1)$ gauge nodes of $\CT_3$.

\item In addition, one can have a free sector consisting of $H$ free hypers where $$H= \frac{(N-px -m)(p(1+x) - N -m)}{2m}.$$
In the special cases, where $N-px -m=0$, or $p(1+x) - N -m=0$, the 3d mirror is given by the quiver gauge theory constructed 
above.
\end{enumerate}
It is convenient to rewrite the linear chains $\CT_1$ and $\CT_2$ in the following fashion:
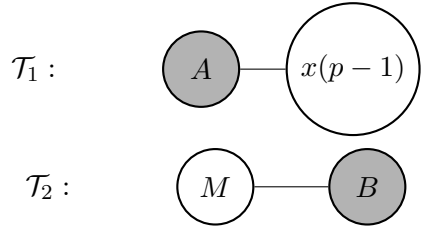
\begin{center}
\begin{tabular}{c}
{\begin{tikzpicture} 
\node[text width=1 cm] (5) at (-2,0){$\CT_1:$}; 
\node[circle,draw, thick, fill=black!30,minimum size=1cm] (1) at (0,0){$A$}; 
\node[circle,draw,thick,minimum size=1.0 cm] (2) at (2,0){$x(p-1)$};
\draw[-] (1) -- (2);
\end{tikzpicture}} \\
\bigskip
{\begin{tikzpicture} 
\node[text width=1 cm] (5) at (-2,0){$\CT_2:$}; 
\node[circle,draw, thick, fill=black!30,minimum size=1cm] (1) at (2,0){$B$};
\node[circle,draw,thick,minimum size=1.0 cm] (2) at (0,0){$M$};
\draw[-] (1) -- (2);
\end{tikzpicture}}
\end{tabular}
\end{center}

The grey nodes $A$ and $B$ are linear quivers that can be read off from the linear quivers $\CT_1$ and $\CT_2$ respectively, given above.
The 3d mirror for $D_p(SU(N))$ for the case $p<N$ can then be written down explicitly, as shown \figref{3dmirr-pltN}.

\begin{figure}[htbp]
\begin{center}
\begin{tabular}{ccc}
\scalebox{.8}{\begin{tikzpicture}[node distance=2cm, cnode/.style={circle,draw,thick,minimum size=1.0 cm},snode/.style={rectangle,draw,thick,minimum size=1.0 cm}, bnode/.style={circle,draw, thick, fill=black!30,minimum size=1cm}]
\node[bnode] (1) at (0,0){$A$};
\node[cnode] (2) at (2,0){$x(p-1)$};
\node[cnode] (3) at (4,0){$M$};
\node[bnode] (4) at (6,0){$B$};
\node[cnode] (5) at (1,-3){$1$};
\node[cnode] (6) at (5,-3){$1$};
\node[cnode] (7) at (3,-5){$1$};
\draw[-] (1) -- (2);
\draw[-] (2)-- (3);
\draw[-] (3)-- (4);
\draw[line width=0.75mm, black] (2) -- (5);
\draw[line width=0.75mm, black] (2) -- (6);
\draw[line width=0.75mm, black] (2) -- (7);
\draw[line width=0.75mm, black] (3) -- (5);
\draw[line width=0.75mm, black] (3) -- (6);
\draw[line width=0.75mm, black] (3) -- (7);
\draw[line width=0.75mm, gray] (6) -- (5);
\draw[line width=0.75mm, gray] (7) -- (6);
\draw[line width=0.75mm, gray] (5) -- (7);
\node[text width=0.1 cm] (15) at (1.8,-1.2){$m_1$};
\node[text width=0.1 cm] (16) at (2.4,-1.0){$m_1$};
\node[text width=0.1 cm] (17) at (3.8,-1.0){$m_2$};
\node[text width=0.1 cm] (18) at (3.5,-0.7){$m_2$};
\node[text width=0.1 cm] (19) at (2.9,-2.8){$m_G$};
\end{tikzpicture}}
& \qquad \qquad
& \scalebox{.8}{\begin{tikzpicture}[node distance=2cm, cnode/.style={circle,draw,thick,minimum size=1.0 cm},snode/.style={rectangle,draw,thick,minimum size=1.0 cm}, bnode/.style={circle,draw, thick, fill=black!30,minimum size=1cm}]
\node[bnode] (1) at (0,0){$A$};
\node[cnode] (2) at (2,0){$x(p-1)$};
\node[cnode] (3) at (4,0){$M$};
\node[bnode] (4) at (6,0){$B$};
\node[cnode] (5) at (1,-3){$1$};
\node[cnode] (6) at (5,-3){$1$};
\node[snode] (7) at (1,-5){$m_1m_2$};
\node[snode] (8) at (5,-5){$m_1m_2$};
\node[snode] (9) at (0,-2){$m_1$};
\node[snode] (10) at (6,-2){$m_2$};
\draw[-] (1) -- (2);
\draw[-] (2)-- (3);
\draw[-] (3)-- (4);
\draw[line width=0.75mm, black] (2) -- (5);
\draw[line width=0.75mm, black] (2) -- (6);
\draw[-] (2) -- (9);
\draw[line width=0.75mm, black] (3) -- (5);
\draw[line width=0.75mm, black] (3) -- (6);
\draw[-] (3) -- (10);
\draw[line width=0.75mm, gray] (6) -- (5);
\draw[-] (7) -- (5);
\draw[-] (8) -- (6);
\node[text width=0.1 cm] (15) at (1.8,-1.2){$m_1$};
\node[text width=0.1 cm] (16) at (2.4,-1.0){$m_1$};
\node[text width=0.1 cm] (17) at (3.8,-1.0){$m_2$};
\node[text width=0.1 cm] (18) at (3.5,-0.7){$m_2$};
\node[text width=0.1 cm] (19) at (3,-2.8){$m_G$};
\end{tikzpicture}}
\end{tabular}
\caption{\footnotesize{The 3d mirror for the $D_p(SU(N))$ ($p<N$) AD theory reduced on a circle, for the case $m=GCD(p,N)=3$.
The quiver on the right is obtained by decoupling a $U(1)$ vector multiplet associated with one of the vertices of the complete graph 
from the quiver on the left. The multiplicity of the edges connecting $U(x(p-1))$ with the $U(1)$ nodes is $m_1$, and that of the 
edges connecting $U(M)$ with the $U(1)$ nodes is $m_2$. The multiplicity of edges connecting two $U(1)$ nodes is $m_G=m_1m_2$.}}
\label{3dmirr-pltN} 
\end{center}
\end{figure}

We now observe that the quiver gauge theory in \figref{3dmirr-pltN} can be constructed from a linear quiver with unitary gauge groups 
by a sequence of elementary $S$-type operations. This can be done using the following steps:
\begin{enumerate}

\item The starting point is the linear quiver $X$:
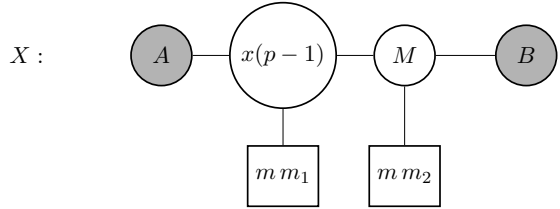
\begin{center}
\scalebox{.8}{\begin{tikzpicture}[node distance=2cm, cnode/.style={circle,draw,thick,minimum size=1.0 cm},snode/.style={rectangle,draw,thick,minimum size=1.0 cm}, bnode/.style={circle,draw, thick, fill=black!30,minimum size=1cm}]
\node[text width=1.0 cm](10) at (-2,0){$X:$};
\node[bnode] (1) at (0,0){$A$};
\node[cnode] (2) at (2,0){$x(p-1)$};
\node[cnode] (3) at (4,0){$M$};
\node[bnode] (4) at (6,0){$B$};
\node[snode] (5) at (2,-2){$m\,m_1$};
\node[snode] (6) at (4,-2){$m\,m_2$};
\draw[-] (1) -- (2);
\draw[-] (2) -- (3);
\draw[-] (3) -- (4);
\draw[-] (2) -- (5);
\draw[-] (3) -- (6);
\end{tikzpicture}}
\end{center}

\item Consider splitting the flavor nodes $U(mm_1) \to U((m-1)m_1) \times U(m_1)$ and $U(mm_2) \to U((m-1)m_2) \times U(m_2)$. In addition, 
let us split the new $U(m_1)$ and $U(m_2)$ flavor nodes into $m_1$ $U(1)$ nodes and $m_2$ $U(1)$ nodes respectively. Now, we implement the following 
elementary $S$-type operation -- the $m_1$ $U(1)$ nodes and $m_2$ $U(1)$ nodes are identified to a single $U(1)$ flavor node, which is then flavored 
by $(m-1)\,m_G$ fundamental hypermultiplets, followed by gauging the identified $U(1)$. This identification-flavoring-gauging operation 
$\CO^{1}_{\CP_1}$ leads to the following quiver gauge theory:
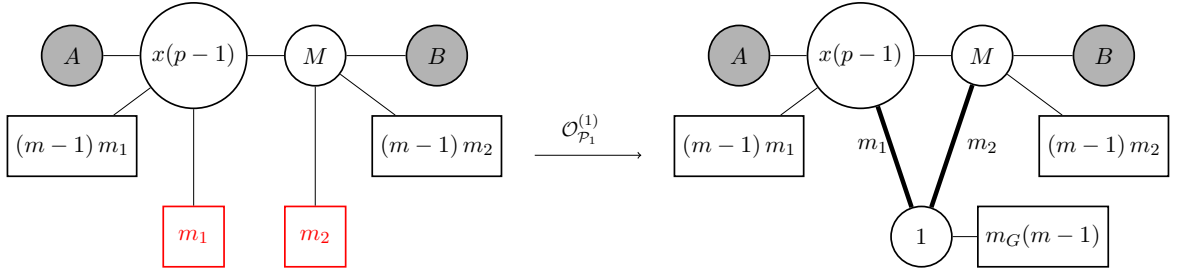
\begin{center}
\begin{tabular}{ccc}
\scalebox{.8}{\begin{tikzpicture}[node distance=2cm, cnode/.style={circle,draw,thick,minimum size=1.0 cm},snode/.style={rectangle,draw,thick,minimum size=1.0 cm}, bnode/.style={circle,draw, thick, fill=black!30,minimum size=1cm}, srnode/.style={red, rectangle,draw,thick,minimum size=1.0 cm}]
\node[bnode] (1) at (0,0){$A$};
\node[cnode] (2) at (2,0){$x(p-1)$};
\node[cnode] (3) at (4,0){$M$};
\node[bnode] (4) at (6,0){$B$};
\node[srnode] (5) at (2,-3){$m_1$};
\node[srnode] (6) at (4,-3){$m_2$};
\node[snode] (7) at (0,-1.5){$(m-1)\,m_1$};
\node[snode] (8) at (6,-1.5){$(m-1)\,m_2$};
\draw[-] (1) -- (2);
\draw[-] (2) -- (3);
\draw[-] (3) -- (4);
\draw[-] (2) -- (5);
\draw[-] (3) -- (6);
\draw[-] (2) -- (7);
\draw[-] (3) -- (8);
\end{tikzpicture}}
& \scalebox{0.7}{\begin{tikzpicture}
\draw[->] (5,0) -- (7,0);
\node[text width=1cm](6) at (6,0.5){$\CO^{(1)}_{\CP_1}$};
\node[] at (6, -2){};
\end{tikzpicture}} & \scalebox{.8}{\begin{tikzpicture}[node distance=2cm, cnode/.style={circle,draw,thick,minimum size=1.0 cm},snode/.style={rectangle,draw,thick,minimum size=1.0 cm}, bnode/.style={circle,draw, thick, fill=black!30,minimum size=1cm}, srnode/.style={red, rectangle,draw,thick,minimum size=1.0 cm}]
\node[bnode] (1) at (0,0){$A$};
\node[cnode] (2) at (2,0){$x(p-1)$};
\node[cnode] (3) at (4,0){$M$};
\node[bnode] (4) at (6,0){$B$};
\node[cnode] (5) at (3,-3){$1$};
\node[snode] (6) at (5,-3){$m_G(m-1)$};
\node[snode] (7) at (0,-1.5){$(m-1)\,m_1$};
\node[snode] (8) at (6,-1.5){$(m-1)\,m_2$};
\draw[-] (1) -- (2);
\draw[-] (2) -- (3);
\draw[-] (3) -- (4);
\draw[line width=0.75mm, black] (2) -- (5);
\draw[line width=0.75mm, black] (3) -- (5);
\draw[-] (5) -- (6);
\draw[-] (2) -- (7);
\draw[-] (3) -- (8);
\node[text width=0.1 cm] (9) at (2,-1.5){$m_1$};
\node[text width=0.1 cm] (10) at (3.8,-1.5){$m_2$};
\end{tikzpicture}}
\end{tabular}
\end{center}

\item In the next step, we split the flavor nodes $U((m-1)m_1) \to U((m-2)m_1) \times U(m_1)$, $U((m-1)m_2) \to U((m-2)m_2) \times U(m_2)$, 
and $U(m_G(m-1)) \to U(m_G(m-2)) \times U(m_G)$. In addition, we split the new $U(m_1)$, $U(m_2)$, and $U(m_G)$ flavor nodes into 
$m_1$ $U(1)$ nodes, $m_2$ $U(1)$ nodes, and $m_G$ $U(1)$ nodes respectively. 
Now, we implement the second elementary $S$-type operation -- the $m_1$ $U(1)$ nodes, $m_2$ $U(1)$ nodes, and $m_G$ $U(1)$ nodes 
are identified to a single $U(1)$ flavor node, which is then flavored 
by $(m-2)\,m_G$ fundamental hypermultiplets, followed by gauging the identified $U(1)$. This identification-flavoring-gauging operation 
$\CO^{2}_{\CP_2}$ leads to the following quiver gauge theory:

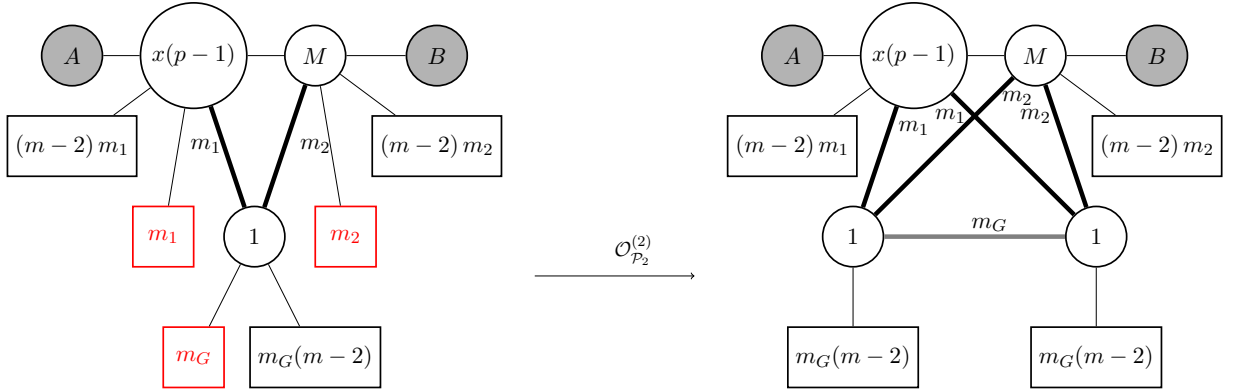
\begin{center}
\begin{tabular}{ccc}
\scalebox{.8}{\begin{tikzpicture}[node distance=2cm, cnode/.style={circle,draw,thick,minimum size=1.0 cm},snode/.style={rectangle,draw,thick,minimum size=1.0 cm}, bnode/.style={circle,draw, thick, fill=black!30,minimum size=1cm}, srnode/.style={red, rectangle,draw,thick,minimum size=1.0 cm}]
\node[bnode] (1) at (0,0){$A$};
\node[cnode] (2) at (2,0){$x(p-1)$};
\node[cnode] (3) at (4,0){$M$};
\node[bnode] (4) at (6,0){$B$};
\node[cnode] (5) at (3,-3){$1$};
\node[snode] (6) at (4,-5){$m_G(m-2)$};
\node[srnode] (7) at (2,-5){$m_G$};
\node[snode] (8) at (0,-1.5){$(m-2)\,m_1$};
\node[snode] (9) at (6,-1.5){$(m-2)\,m_2$};
\node[srnode] (10) at (1.5,-3){$m_1$};
\node[srnode] (11) at (4.5,-3){$m_2$};
\draw[-] (1) -- (2);
\draw[-] (2) -- (3);
\draw[-] (3) -- (4);
\draw[line width=0.75mm, black] (2) -- (5);
\draw[line width=0.75mm, black] (3) -- (5);
\draw[-] (5) -- (6);
\draw[-] (5) -- (7);
\draw[-] (2) -- (8);
\draw[-] (2) -- (10);
\draw[-] (3) -- (9);
\draw[-] (3) -- (11);
\node[text width=0.1 cm] (9) at (2,-1.5){$m_1$};
\node[text width=0.1 cm] (10) at (3.8,-1.5){$m_2$};
\end{tikzpicture}}
& \scalebox{0.7}{\begin{tikzpicture}
\draw[->] (5,0) -- (8,0);
\node[text width=1cm](6) at (7,0.5){$\CO^{(2)}_{\CP_2}$};
\node[] at (6, -2){};
\end{tikzpicture}}
& \scalebox{.8}{\begin{tikzpicture}[node distance=2cm, cnode/.style={circle,draw,thick,minimum size=1.0 cm},snode/.style={rectangle,draw,thick,minimum size=1.0 cm}, bnode/.style={circle,draw, thick, fill=black!30,minimum size=1cm}, srnode/.style={red, rectangle,draw,thick,minimum size=1.0 cm}]
\node[bnode] (1) at (0,0){$A$};
\node[cnode] (2) at (2,0){$x(p-1)$};
\node[cnode] (3) at (4,0){$M$};
\node[bnode] (4) at (6,0){$B$};
\node[cnode] (5) at (5,-3){$1$};
\node[cnode] (6) at (1,-3){$1$};
\node[snode] (7) at (5,-5){$m_G(m-2)$};
\node[snode] (8) at (1,-5){$m_G(m-2)$};
\node[snode] (9) at (0,-1.5){$(m-2)\,m_1$};
\node[snode] (10) at (6,-1.5){$(m-2)\,m_2$};
\draw[-] (1) -- (2);
\draw[-] (2) -- (3);
\draw[-] (3) -- (4);
\draw[line width=0.75mm, black] (2) -- (5);
\draw[line width=0.75mm, black] (2) -- (6);
\draw[line width=0.75mm, black] (3) -- (5);
\draw[line width=0.75mm, black] (3) -- (6);
\draw[line width=0.75mm, gray] (5) -- (6);
\draw[-] (2) -- (9);
\draw[-] (3) -- (10);
\draw[-] (5) -- (7);
\draw[-] (6) -- (8);
\node[text width=0.1 cm] (15) at (1.8,-1.2){$m_1$};
\node[text width=0.1 cm] (16) at (2.4,-1.0){$m_1$};
\node[text width=0.1 cm] (17) at (3.8,-1.0){$m_2$};
\node[text width=0.1 cm] (18) at (3.5,-0.7){$m_2$};
\node[text width=0.1 cm] (19) at (3,-2.8){$m_G$};
\end{tikzpicture}}
\end{tabular}
\end{center}

Note that the quiver on the right is the 3d mirror for $D_p(SU(N))$ ($p<N$) for the special case $m=3$.

\item Proceeding in the same fashion, we implement a total of $(m-1)$ identification-flavoring-gauging operations on the linear 
quiver $X$ leading to the quiver gauge theory $X'$, i.e.
\be \label{SOp-pltN}
X' = \CO^{(m-1)}_{\vec \CP_{m-1}} \circ \CO^{(m-2)}_{\vec \CP_{m-2}} \circ \ldots \circ \CO^{(2)}_{\vec \CP_{2}} \circ \CO^{(1)}_{\vec \CP_{1}}(X),
\ee
where $X'$ is the quiver gauge theory given in \figref{3dmirr-pltN}. 
\end{enumerate}

Using the general prescription  in \Secref{SOp-rev} for finding out the dual of a given elementary $S$-type operation, 
one can explicitly determine the 3d mirror of $X'$. One can make some general statements about the mirror $Y'$:

\begin{itemize}

\item Since $X$ is a good linear quiver (a balanced quiver) with unitary gauge groups, its mirror dual $Y$ is also a Lagrangian theory. 
Since the quiver $X'$ is obtained from $X$ by a sequence of Abelian elementary $S$-type operations, we conclude that the 3d mirror 
$Y'$ must be a Lagrangian theory, using the main result of \Secref{SOp-rev}. 

\item In addition, we note that the elementary $S$-type operations in \eref{SOp-pltN} are of the identification-flavoring-gauging type.
Using the corollary to the main result, this implies that the quiver $Y'$ will only have unitary gauge groups.

\item In addition to fundamental and bifundamental matter, the quiver will involve several hypermultiplets which are only charged under 
the $U(1)$ subgroups of certain unitary factors, since this is a general feature of the dual of an identification-flavoring-gauging operation 
on a theory with unitary gauge group.

\end{itemize}

As discussed earlier, the quiver $Y'$, along with $H$ free twisted hypermultiplets, gives a Lagrangian description for the $D_p(SU(N))$ 
AD theory reduced on a circle. We will write down the quiver $Y'$ for an explicit example in \Secref{pltNEx}. 

\subsubsection{The case of $p>N$}\label{pgtNGen}

Now let us consider the case $p > N$, for which we define the following parameters:
\begin{align}\label{pgtN-pars}
& m=GCD(p,N) , \quad n=\frac{N}{m}, \quad q=\frac{p}{m}.
\end{align}
The 3d mirror for $D_p(SU(N))$ with $p> N$ can now be constructed in the following fashion:
\begin{enumerate}

\item Consider the following subquivers from which the 3d quiver can be built:

\begin{itemize}

\item A linear chain of unitary gauge groups with bifundamental matter : $$ \CT_1 = (N-1) - (N-2) - \ldots - 2 - 1$$

\item A complete graph $\CT_2$ of $m$ vertices, where each vertex is a $U(1)$ gauge node, and edge multiplicity $m_G= n(q-n)$.

\end{itemize}

\item Connect the $U(N-1)$ gauge node of $\CT_1$ to each of the $U(1)$ gauge nodes of $\CT_2$ with an edge of multiplicity $n$.

\item Finally, one needs to decouple an overall $U(1)$ factor from the quiver, and we will choose to implement this by ungauging one of 
the $U(1)$ gauge nodes of $\CT_2$.

\item In addition, one can have a free sector consisting of $H$ free hypers where $$H= \frac{(N-m)(p - N -m)}{2m}.$$
In the special cases, where $N-m=0$, or $p - N -m=0$, the 3d mirror is given by the quiver gauge theory constructed 
above.

\end{enumerate}

\begin{figure}[htbp]
\begin{center}
\begin{tabular}{ccc}
\scalebox{.8}{\begin{tikzpicture}[node distance=2cm, cnode/.style={circle,draw,thick,minimum size=1.0 cm},snode/.style={rectangle,draw,thick,minimum size=1.0 cm}, bnode/.style={circle,draw, thick, fill=black!30,minimum size=1cm}]
\node[bnode] (1) at (0,0){$C$};
\node[cnode] (2) at (2,0){$N-1$};
\node[cnode] (3) at (4,2){1};
\node[cnode] (4) at (4,-2){1};
\node[cnode] (5) at (7,2){1};
\node[cnode] (6) at (7,-2){1};
\draw[-] (1) -- (2);
\draw[line width=0.75mm, black] (2) -- (3);
\draw[line width=0.75mm, black] (2) -- (4);
\draw[line width=0.75mm, black] (2) -- (5);
\draw[line width=0.75mm, black] (2) -- (6);
\draw[line width=0.75mm, gray] (3) -- (4);
\draw[line width=0.75mm, gray] (3) -- (5);
\draw[line width=0.75mm, gray] (3) -- (6);
\draw[line width=0.75mm, gray] (4) -- (5);
\draw[line width=0.75mm, gray] (4) -- (6);
\draw[line width=0.75mm, gray] (5) -- (6);
\node[text width=0.1 cm] (15) at (3,0){$n$};
\node[text width=0.1 cm] (16) at (7.2,0){$m_G$};
\end{tikzpicture}}
& \qquad \qquad
&\scalebox{.8}{\begin{tikzpicture}[node distance=2cm, cnode/.style={circle,draw,thick,minimum size=1.0 cm},snode/.style={rectangle,draw,thick,minimum size=1.0 cm}, bnode/.style={circle,draw, thick, fill=black!30,minimum size=1cm}]
\node[bnode] (1) at (0,0){$C$};
\node[cnode] (2) at (2,0){$N-1$};
\node[cnode] (3) at (4,2){1};
\node[cnode] (4) at (4,-2){1};
\node[cnode] (5) at (7,2){1};
\node[snode] (6) at (7,-2){$m_G$};
\node[snode] (7) at (2,-2){$n$};
\node[snode] (8) at (9,2){$m_G$};
\node[snode] (9) at (7,0){$m_G$};
\draw[-] (1) -- (2);
\draw[line width=0.75mm, black] (2) -- (3);
\draw[line width=0.75mm, black] (2) -- (4);
\draw[line width=0.75mm, black] (2) -- (5);
\draw[line width=0.75mm, gray] (3) -- (4);
\draw[line width=0.75mm, gray] (3) -- (5);
\draw[line width=0.75mm, gray] (4) -- (5);
\draw[-] (4) -- (6);
\draw[-] (2) -- (7);
\draw[-] (5) -- (8);
\draw[-] (3) -- (9);
\end{tikzpicture}}
\end{tabular}
\caption{\footnotesize{The 3d mirror for the $D_p(SU(N))$ ($p> N$) AD theory reduced on a circle, for the case $m=GCD(p,N)=4$.
The quiver on the right is obtained by decoupling a $U(1)$ vector multiplet associated with one of the vertices of the complete graph 
from the quiver on the left. The multiplicity of the edges connecting $U(N-1)$ with the $U(1)$ nodes is $n$, and that of the 
edges connecting two $U(1)$ nodes is $m_G=n(q-n)$.}}
\label{3dmirr-pgtN}
\end{center}
\end{figure}
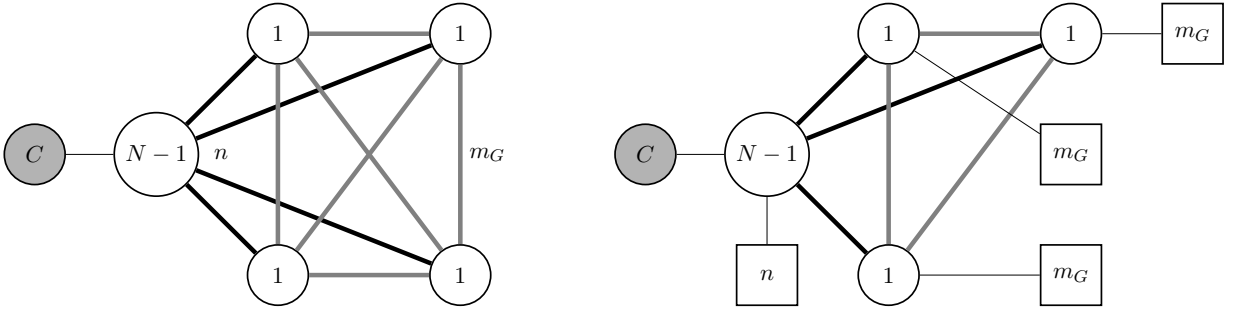

It is convenient to rewrite the linear chain $\CT_1$ in the following fashion:
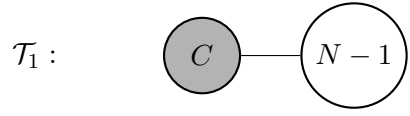
\begin{center}
\begin{tabular}{c}
{\begin{tikzpicture} 
\node[text width=1 cm] (5) at (-2,0){$\CT_1:$}; 
\node[circle,draw, thick, fill=black!30,minimum size=1cm] (1) at (0,0){$C$}; 
\node[circle,draw,thick,minimum size=1.0 cm] (2) at (2,0){$N-1$};
\draw[-] (1) -- (2);
\end{tikzpicture}}
\end{tabular}
\end{center}
The quiver $C$ is a linear chain $C : 1-2-\ldots-(N-2)$. The 3d mirror for $D_p(SU(N))$ for the case $p<N$ can then be written down explicitly, 
as shown \figref{3dmirr-pgtN}.

We observe that the quiver gauge theory in \figref{3dmirr-pgtN} can be constructed from a linear quiver with unitary gauge groups by a 
sequence of elementary Abelian $S$-type operations. This can be performed using the following steps:
\begin{enumerate}

\item The starting point is the following linear quiver $X$ (the $T(U(N))$ quiver with $N=mn$):

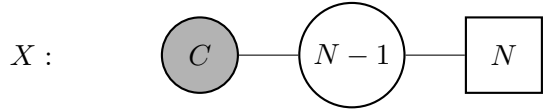
\begin{center}
{\begin{tikzpicture} 
\node[text width=1 cm] (5) at (-2,0){$X:$}; 
\node[circle,draw, thick, fill=black!30,minimum size=1cm] (1) at (0,0){$C$}; 
\node[circle,draw,thick,minimum size=1.0 cm] (2) at (2,0){$N-1$};
\node[rectangle,draw,thick,minimum size=1.0 cm] (3) at (4,0){$N$};
\draw[-] (1) -- (2);
\draw[-] (2) -- (3);
\end{tikzpicture}}
\end{center}

\item Consider splitting the $U(mn)$ flavor node as $U(mn) \to U((m-1)n) \times U(n)$, and further splitting the new $U(n)$ node as
$U(n) \to U(1)^n$. Now, we implement the following elementary $S$-type operation -- the $n$ $U(1)$ nodes are identified to a single 
$U(1)$, which is then flavored by $(m-1)m_G$ fundamental hypers, followed by gauging the identified $U(1)$. The identification-flavoring-gauging 
operation $\CO^{(1)}_{\vec\CP_1}$ leads to the following quiver gauge theory:
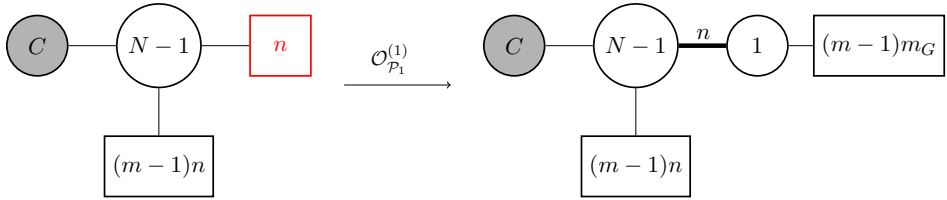
\begin{center}
\begin{tabular}{ccc}
\scalebox{.8}{\begin{tikzpicture}[node distance=2cm, cnode/.style={circle,draw,thick,minimum size=1.0 cm},snode/.style={rectangle,draw,thick,minimum size=1.0 cm}, bnode/.style={circle,draw, thick, fill=black!30,minimum size=1cm}, srnode/.style={red, rectangle,draw,thick,minimum size=1.0 cm}]
\node[bnode](1) at (0,0){$C$}; 
\node[cnode](2) at (2,0){$N-1$};
\node[srnode](3) at (4,0){$n$};
\node[snode](4) at (2,-2){$(m-1)n$};
\draw[-] (1) -- (2);
\draw[-] (2) -- (3);
\draw[-] (2) -- (4);
\end{tikzpicture}}
& \scalebox{0.7}{\begin{tikzpicture}
\draw[->] (5,0) -- (7,0);
\node[text width=1cm](6) at (6,0.5){$\CO^{(1)}_{\CP_1}$};
\node[] at (6, -2){};
\end{tikzpicture}}
&\scalebox{.8}{\begin{tikzpicture}[node distance=2cm, cnode/.style={circle,draw,thick,minimum size=1.0 cm},snode/.style={rectangle,draw,thick,minimum size=1.0 cm}, bnode/.style={circle,draw, thick, fill=black!30,minimum size=1cm}, srnode/.style={red, rectangle,draw,thick,minimum size=1.0 cm}]
\node[bnode](1) at (0,0){$C$}; 
\node[cnode](2) at (2,0){$N-1$};
\node[cnode](3) at (4,0){1};
\node[snode](4) at (2,-2){$(m-1)n$};
\node[snode](5) at (6,0){$(m-1)m_G$};
\draw[-] (1) -- (2);
\draw[line width=0.75mm, black] (2) -- (3);
\draw[-] (2) -- (4);
\draw[-] (3) -- (5);
\node[text width=1cm](6) at (3.5,0.2){$n$};
\end{tikzpicture}}
\end{tabular}
\end{center}

\item In the next step, we split the $U((m-1)m_G)$ flavor node as $U((m-1)m_G) \to U((m-2)m_G)\times U(m_G)$, and the $U((m-1)n)$ flavor 
node as $U((m-1)n) \to U((m-2)n) \times U(n)$. In addition, we split the new $U(m_G)$ and $U(n)$ flavor nodes into $m_G$ $U(1)$ nodes 
and $n$ $U(1)$ nodes respectively. Then, we implement the following elementary $S$-type operation -- the $m_G$ $U(1)$ nodes and 
$n$ $U(1)$ nodes are identified to a single $U(1)$, which is then flavored by $(m-2)m_G$ fundamental hypers, followed by gauging the 
identified $U(1)$. The identification-flavoring-gauging operation $\CO^{(2)}_{\vec\CP_2}$ leads to the following quiver gauge theory:
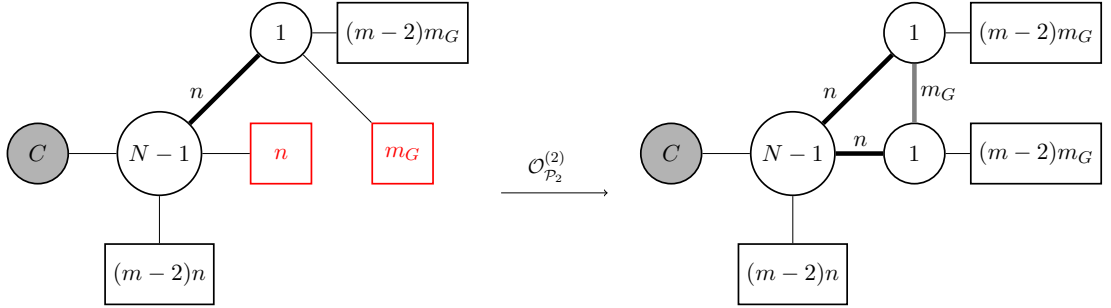
\begin{center}
\begin{tabular}{ccc}
\scalebox{.8}{\begin{tikzpicture}[node distance=2cm, cnode/.style={circle,draw,thick,minimum size=1.0 cm},snode/.style={rectangle,draw,thick,minimum size=1.0 cm}, bnode/.style={circle,draw, thick, fill=black!30,minimum size=1cm}, srnode/.style={red, rectangle,draw,thick,minimum size=1.0 cm}]
\node[bnode](1) at (0,0){$C$}; 
\node[cnode](2) at (2,0){$N-1$};
\node[cnode](3) at (4,2){1};
\node[snode](4) at (2,-2){$(m-2)n$};
\node[snode](5) at (6,2){$(m-2)m_G$};
\node[srnode](6) at (4,0){$n$};
\node[srnode](7) at (6,0){$m_G$};
\draw[-] (1) -- (2);
\draw[line width=0.75mm, black] (2) -- (3);
\draw[-] (2) -- (4);
\draw[-] (3) -- (5);
\draw[-] (2) -- (6);
\draw[-] (3) -- (7);
\node[text width=1cm](6) at (3, 1.0){$n$};
\end{tikzpicture}}
& \scalebox{0.7}{\begin{tikzpicture}
\draw[->] (5,0) -- (7,0);
\node[text width=1cm](6) at (6,0.5){$\CO^{(2)}_{\CP_2}$};
\node[] at (6, -2){};
\end{tikzpicture}}
&\scalebox{.8}{\begin{tikzpicture}[node distance=2cm, cnode/.style={circle,draw,thick,minimum size=1.0 cm},snode/.style={rectangle,draw,thick,minimum size=1.0 cm}, bnode/.style={circle,draw, thick, fill=black!30,minimum size=1cm}, srnode/.style={red, rectangle,draw,thick,minimum size=1.0 cm}]
\node[bnode](1) at (0,0){$C$}; 
\node[cnode](2) at (2,0){$N-1$};
\node[cnode](3) at (4,2){1};
\node[snode](4) at (2,-2){$(m-2)n$};
\node[snode](5) at (6,2){$(m-2)m_G$};
\node[cnode](6) at (4,0){1};
\node[snode](7) at (6,0){$(m-2)m_G$};
\draw[-] (1) -- (2);
\draw[line width=0.75mm, black] (2) -- (3);
\draw[-] (2) -- (4);
\draw[-] (3) -- (5);
\draw[line width=0.75mm, gray] (3) -- (6);
\draw[line width=0.75mm, black] (2) -- (6);
\draw[-] (6) -- (7);
\node[text width=1cm](6) at (3, 1.0){$n$};
\node[text width=1cm](7) at (3.5, 0.2){$n$};
\node[text width=1cm](8) at (4.6, 1){$m_G$};
\end{tikzpicture}}
\end{tabular}
\end{center}

\item Proceeding in the same fashion, we implement a total of $(m-1)$ identification-flavoring-gauging operations on the linear 
quiver $X$ leading to the quiver gauge theory $X'$, i.e.
\be \label{SOp-pgtN}
X' = \CO^{(m-1)}_{\vec \CP_{m-1}} \circ \CO^{(m-2)}_{\vec \CP_{m-2}} \circ \ldots \circ \CO^{(2)}_{\vec \CP_{2}} \circ \CO^{(1)}_{\vec \CP_{1}}(X),
\ee
where $X'$ is the quiver gauge theory given in \figref{3dmirr-pgtN}. 

\end{enumerate}

Similar to the case of $p <N$, one can make some general statements about the 3d mirror of $X'$, i.e. $Y'$ for the $p> N$ as follows:

\begin{itemize}

\item The 3d mirror $Y'$ of the quiver gauge theory $X'$ must be a Lagrangian theory. 

\item The gauge group of $Y'$ will be a product of unitary factors.

\item In addition to fundamental and bifundamental matter, the quiver will involve several hypermultiplets which are only charged under 
$U(1)$ subgroups of certain unitary factors of the gauge group.

\end{itemize}

The resultant quiver gauge theory $Y'$, along with $H$ free twisted hypermultiplets, gives a Lagrangian realization of the 
$D_p(SU(N))$ ($p> N$) AD theory reduced on a circle. In \Secref{pgtNEx}, we work out the explicit form for the quiver $Y'$ in 
a specific example.

\subsection{An example with $p< N$ : $D_4(SU(6))$}\label{pltNEx}

In this section, we consider a specific example of the class of quiver gauge theories discussed in \Secref{pltNGen} and explicitly 
determine its 3d mirror following the general prescription given in \Secref{SOp-Ab}.

\subsubsection{The unitary Lagrangian}

Consider the quiver gauge theory $X'$ in \figref{3dmirr-pltN} for $p=4, N=6$. The various parameters that encode the quiver 
data are given as:
\begin{align}
& x=1, \quad M=4, \quad m=2 , \quad n=3, \quad q=2, \nn \\
& m_1=1, \quad m_2=1.
\end{align}

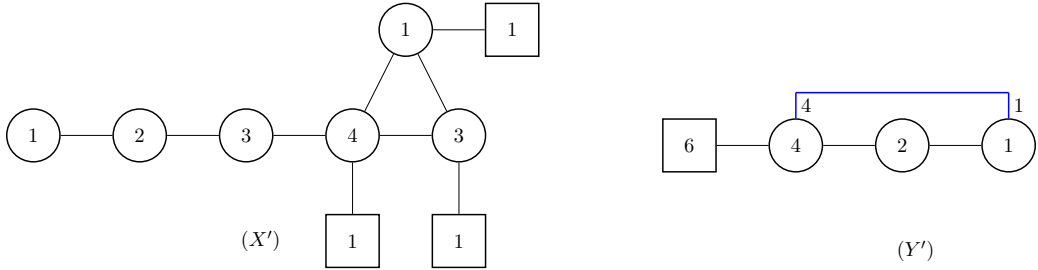
\begin{figure}[htbp]
\begin{center}
\begin{tabular}{ccc}
\scalebox{.7}{\begin{tikzpicture}[node distance=2cm, nnode/.style={circle,draw,thick, fill, inner sep=1 pt},cnode/.style={circle,draw,thick,minimum size=1.0 cm},snode/.style={rectangle,draw,thick,minimum size=1.0 cm}, pnode/.style={circle,double,draw,thick, minimum size=1.0cm}]
\node[cnode] (1) at (0,0){1};
\node[cnode] (2) at (2,0){2};
\node[cnode] (3) at (4,0){3};
\node[cnode] (4) at (6,0){4};
\node[cnode] (5) at (8,0){3};
\node[cnode] (6) at (7,2){1};
\node[snode] (7) at (6,-2){1};
\node[snode] (8) at (8,-2){1};
\node[snode] (9) at (9,2){1};
\draw[-] (1) -- (2);
\draw[-] (2)-- (3);
\draw[-] (3)-- (4);
\draw[-] (5)-- (4);
\draw[-] (6)-- (4);
\draw[-] (5)-- (6);
\draw[-] (9)-- (6);
\draw[-] (4)-- (7);
\draw[-] (5)-- (8);
\node[] (9) at (0,-0.75){};
\node[text width=.2cm](11) at (4,-2){$(X')$};
\end{tikzpicture}}
& \qquad   \qquad
& \scalebox{.7}{\begin{tikzpicture}[node distance=2cm, nnode/.style={circle,draw,thick, fill, inner sep=1 pt},cnode/.style={circle,draw,thick,minimum size=1.0 cm},snode/.style={rectangle,draw,thick,minimum size=1.0 cm}]
\node[snode] (1) at (0,0){6};
\node[cnode] (2) at (2,0){4};
\node[cnode] (3) at (4,0){2};
\node[cnode] (4) at (6,0){1};
\node[](5) at (2,-2){};
\node[](6) at (4,-2){};
\draw[-] (1) -- (2);
\draw[-] (2)-- (3);
\draw[-] (3)-- (4);
\draw[-, thick, blue] (2)-- (2,1);
\draw[-, thick, blue] (2,1)-- (6,1);
\draw[-, thick, blue] (6,1)-- (4);
\node[] (9) at (0,-0.75){};
\node[text width=.2cm](10) at (2.2, 0.75){4};
\node[text width=.2cm](11) at (6.2, 0.75){1};
\node[text width=.2cm](11) at (4,-2){$(Y')$};
\end{tikzpicture}}
\end{tabular}
\caption{\footnotesize{The quiver $X'$ is the 3d mirror of $D_4(SU(6))$ reduced on a circle, constructed from a 
class $\CS$ prescription. The quiver $Y'$ is the 3d mirror of $X'$, i.e. a Lagrangian realization of the SCFT obtained 
by the reduction of $D_4(SU(6))$ on a circle.}}
\label{fig: pltN-quivers}
\end{center}
\end{figure}

This corresponds to the quiver gauge theory:

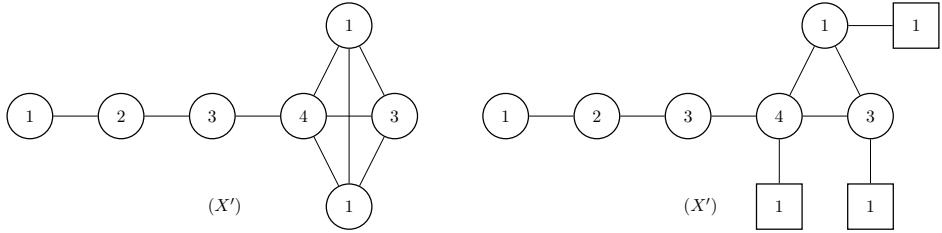
\begin{center}
\begin{tabular}{ccc}
\scalebox{.6}{\begin{tikzpicture}[node distance=2cm, nnode/.style={circle,draw,thick, fill, inner sep=1 pt},cnode/.style={circle,draw,thick,minimum size=1.0 cm},snode/.style={rectangle,draw,thick,minimum size=1.0 cm}, pnode/.style={circle,double,draw,thick, minimum size=1.0cm}]
\node[cnode] (1) at (0,0){1};
\node[cnode] (2) at (2,0){2};
\node[cnode] (3) at (4,0){3};
\node[cnode] (4) at (6,0){4};
\node[cnode] (5) at (8,0){3};
\node[cnode] (6) at (7,2){1};
\node[cnode] (7) at (7,-2){1};
\draw[-] (1) -- (2);
\draw[-] (2)-- (3);
\draw[-] (3)-- (4);
\draw[-] (5)-- (4);
\draw[-] (6)-- (4);
\draw[-] (5)-- (6);
\draw[-] (7)-- (6);
\draw[-] (4)-- (7);
\draw[-] (5)-- (7);
\node[] (9) at (0,-0.75){};
\node[text width=.2cm](11) at (4,-2){$(X')$};
\end{tikzpicture}}
& \qquad  
& \scalebox{.6}{\begin{tikzpicture}[node distance=2cm, nnode/.style={circle,draw,thick, fill, inner sep=1 pt},cnode/.style={circle,draw,thick,minimum size=1.0 cm},snode/.style={rectangle,draw,thick,minimum size=1.0 cm}, pnode/.style={circle,double,draw,thick, minimum size=1.0cm}]
\node[cnode] (1) at (0,0){1};
\node[cnode] (2) at (2,0){2};
\node[cnode] (3) at (4,0){3};
\node[cnode] (4) at (6,0){4};
\node[cnode] (5) at (8,0){3};
\node[cnode] (6) at (7,2){1};
\node[snode] (7) at (6,-2){1};
\node[snode] (8) at (8,-2){1};
\node[snode] (9) at (9,2){1};
\draw[-] (1) -- (2);
\draw[-] (2)-- (3);
\draw[-] (3)-- (4);
\draw[-] (5)-- (4);
\draw[-] (6)-- (4);
\draw[-] (5)-- (6);
\draw[-] (9)-- (6);
\draw[-] (4)-- (7);
\draw[-] (5)-- (8);
\node[] (9) at (0,-0.75){};
\node[text width=.2cm](11) at (4,-2){$(X')$};
\end{tikzpicture}}
\end{tabular}
\end{center}
Note that we have decoupled a $U(1)$ vector multiplet to write the quiver gauge theory on the RHS. Also, note that we 
have $N-px -m=0 = p(1+x) - N -m=0$, which implies that there are no free hypermultiplets. One can now deploy the 
$S$-type operation technique, the details of which are described below, to construct the 3d mirror of $X'$. The result is 
shown in \figref{fig: pltN-quivers}. Note that the quiver $Y'$ has as gauge group $U(4) \times U(2) \times U(1)$ 
with fundamental and bifundamental hypers. In addition, it involves a single hypermultiplet with charges $(4,1)$ under the 
$U(1)$ subgroup of the $U(4)$ gauge node and the $U(1)$ gauge node respectively. This is consistent with our analysis of the 
general features of the theory $Y'$, as discussed in \Secref{pltNGen}.
The moduli space dimensions and global symmetries of the dual theories are 
summarized in Table \ref{Tab:pltNEx}.

\begin{table}[htbp]
\begin{center}
\begin{tabular}{|c|c|c|}
\hline
Moduli space data & Theory $X'$ & Theory $Y'$ \\
\hline \hline 
dim\,$\CM_H$ & 7 & 14\\
\hline
dim\,$\CM_C$ & 14 & 7\\
\hline
$G_H$ & $U(1)^3$ & $SU(6) \times U(1)$\\
\hline
$G_C$ & $SU(6) \times U(1)$ & $U(1)^3$ \\
\hline
\end{tabular}
\end{center}
\caption{\footnotesize{Summary table for the moduli space dimensions and global symmetries for the mirror pair in \figref{fig: pltN-quivers}.}}
\label{Tab:pltNEx}
\end{table}


\subsubsection{Derivation via $S$-type operations}

\begin{figure}[htbp]
\begin{center}
\begin{tabular}{ccc}
\scalebox{.7}{\begin{tikzpicture}[node distance=2cm, nnode/.style={circle,draw,thick, fill, inner sep=1 pt},cnode/.style={circle,draw,thick,minimum size=1.0 cm},snode/.style={rectangle,draw,thick,minimum size=1.0 cm}, pnode/.style={circle,double,draw,thick, minimum size=1.0cm}]
\node[cnode] (1) at (0,0){1};
\node[cnode] (2) at (2,0){2};
\node[cnode] (3) at (4,0){3};
\node[cnode] (4) at (6,0){4};
\node[cnode] (5) at (8,0){3};
\node[snode] (6) at (6,-2){2};
\node[snode] (7) at (8,-2){2};
\draw[-] (1) -- (2);
\draw[-] (2)-- (3);
\draw[-] (3)-- (4);
\draw[-] (5)-- (4);
\draw[-] (6)-- (4);
\draw[-] (5)-- (7);
\node[] (9) at (0,-0.75){};
\node[text width=.2cm](11) at (4,-2){$(X)$};
\end{tikzpicture}}
& \qquad \qquad
& \scalebox{.7}{\begin{tikzpicture}[node distance=2cm, nnode/.style={circle,draw,thick, fill, inner sep=1 pt},cnode/.style={circle,draw,thick,minimum size=1.0 cm},snode/.style={rectangle,draw,thick,minimum size=1.0 cm}]
\node[snode] (1) at (0,0){6};
\node[cnode] (2) at (2,0){4};
\node[cnode] (3) at (4,0){2};
\node[cnode] (4) at (6,0){1};
\node[](5) at (2,-2){};
\node[](6) at (4,-2){};
\draw[-] (1) -- (2);
\draw[-] (2)-- (3);
\draw[-] (3)-- (4);
\node[] (9) at (0,-0.75){};
\node[text width=.2cm](11) at (4,-2){$(Y)$};
\end{tikzpicture}}
\end{tabular}
\end{center}
\caption{\footnotesize{A pair of mirror dual linear quivers with unitary gauge groups.}}
\label{LQ-pltNEx}
\end{figure}
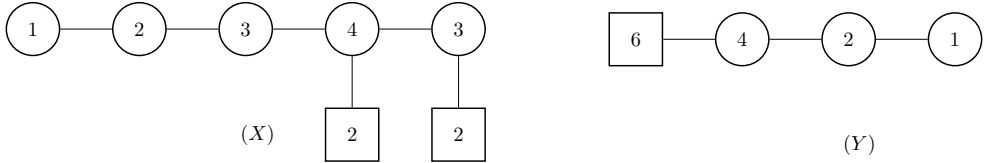

The starting point for obtaining the mirror pair $(X',Y')$ by an $S$-type operation is the linear mirror pair $(X,Y)$, as 
shown in \figref{LQ-pltNEx}. The $S^3$ partition functions of the theories $(X,Y)$ can be written down from the rules 
given in \Secref{SOp-rev} as follows:
\begin{align}
& Z^{(X)}(\vec{m}; \vec{t})=\int \prod^5_{\gamma=1} \Big[ d \vec s^\gamma\Big]\, Z^{(X)}_{\rm int} (\{\vec s^\gamma\},\vec m, \vec t) = \int \prod^5_{\gamma=1} \Big[ d \vec s^\gamma\Big]\, Z^{(X)}_{\rm FI}(\{\vec s^\gamma \}, \vec t)\,Z^{(X)}_{\rm{1-loop}}(\{\vec s^\gamma \}, \vec m ),\\
& Z^{(Y)}(\vec{t}; \vec{m})=\int \prod^{3}_{\gamma'=1} \Big[d\vec \s^{\gamma'}\Big]\, Z^{(Y)}_{\rm int}(\{\vec \s^{\gamma'} \},\vec{t}, \vec{m}) 
=\int \prod^{3}_{\gamma'=1}\,\Big[d\vec \s^{\gamma'}\Big] \,Z^{(Y)}_{\rm FI}(\{\vec \s^{\gamma'} \}, \vec m)\, Z^{(Y)}_{\rm{1-loop}}(\{\vec \s^{\gamma'} \}, 
\vec t \} ),
\end{align}
where $\gamma, \gamma'$ label the gauge nodes of the linear quivers $X,Y$ respectively, with the count increasing from left to right in both cases. 
The functions $Z^{(X)}_{\rm int}$ and $Z^{(Y)}_{\rm int}$ are the respective matrix model integrands for $X$ and $Y$, each consisting of an FI term and 
a one-loop term with contributions from vector multiplets and hypermultiplets. We will, in particular, need the FI term of $Y$, which is given as
\be
Z^{(Y)}_{\rm FI}(\{\vec \s^{\gamma'} \}, \vec t)= e^{2\pi i (m_1- m_2)\,\tr \vec \s^1}\,e^{2\pi i (m_2- m_3)\,\tr \vec \s^2} \, e^{2\pi i (m_3- m_4)\,\s^3}.
\ee

Mirror symmetry of $X$ and $Y$ implies that
\begin{align}
Z^{(X)}(\vec m ; \vec t) =C_{XY}(\vec m, \vec t) \, Z^{(Y)}(\vec t; - \vec m),
\end{align}
where $C_{XY}$ is a contact term, which reminds us that the duality holds for unrestricted $\vec m$ and $\vec t$, 
only when we turn on a certain background Chern-Simons sector. We will not need the precise form of the contact 
term to read off the gauge group and matter content of theories we will construct using $S$-type operations.

\begin{figure}[htbp]
\begin{center}
\begin{tabular}{ccc}
\scalebox{.7}{\begin{tikzpicture}[node distance=2cm, nnode/.style={circle,draw,thick, fill, inner sep=1 pt},cnode/.style={circle,draw,thick,minimum size=1.0 cm},snode/.style={rectangle,draw,thick,minimum size=1.0 cm}, pnode/.style={circle,double,draw,thick, minimum size=1.0cm},srnode/.style={red,rectangle,draw,thick,minimum size=1.0 cm}]
\node[cnode] (1) at (0,0){1};
\node[cnode] (2) at (2,0){2};
\node[cnode] (3) at (4,0){3};
\node[cnode] (4) at (6,0){4};
\node[cnode] (5) at (8,0){3};
\node[snode] (6) at (6,-2){1};
\node[snode] (7) at (8,-2){1};
\node[srnode] (8) at (6,2){1};
\node[srnode] (9) at (8,2){1};
\draw[-] (1) -- (2);
\draw[-] (2)-- (3);
\draw[-] (3)-- (4);
\draw[-] (5)-- (4);
\draw[-] (6)-- (4);
\draw[-] (5)-- (7);
\draw[-] (4)-- (8);
\draw[-] (5)-- (9);
\node[] (10) at (0,-0.75){};
\node[text width=.2cm](11) at (4,-2){$(X)$};
\end{tikzpicture}}
& \qquad  
& \scalebox{.7}{\begin{tikzpicture}[node distance=2cm, nnode/.style={circle,draw,thick, fill, inner sep=1 pt},cnode/.style={circle,draw,thick,minimum size=1.0 cm},snode/.style={rectangle,draw,thick,minimum size=1.0 cm}]
\node[snode] (1) at (0,0){6};
\node[cnode] (2) at (2,0){4};
\node[cnode] (3) at (4,0){2};
\node[cnode] (4) at (6,0){1};
\node[](5) at (2,-2){};
\node[](6) at (4,-2){};
\draw[-] (1) -- (2);
\draw[-] (2)-- (3);
\draw[-] (3)-- (4);
\node[] (9) at (0,-0.75){};
\node[text width=.2cm](11) at (4,-2){$(Y)$};
\end{tikzpicture}}\\
 \scalebox{.7}{\begin{tikzpicture}
\draw[->] (15,-3) -- (15,-5);
\node[text width=0.1cm](20) at (14.5, -4) {$\CO^\alpha_{\CP}$};
\end{tikzpicture}}
&\qquad \qquad 
& \scalebox{.7}{\begin{tikzpicture}
\draw[->] (15,-3) -- (15,-5);
\node[text width=0.1cm](29) at (15.5, -4) {$\wt{\CO}^\alpha_{\CP}$};
\end{tikzpicture}}\\
\scalebox{.7}{\begin{tikzpicture}[node distance=2cm, nnode/.style={circle,draw,thick, fill, inner sep=1 pt},cnode/.style={circle,draw,thick,minimum size=1.0 cm},snode/.style={rectangle,draw,thick,minimum size=1.0 cm}, pnode/.style={circle,double,draw,thick, minimum size=1.0cm}]
\node[cnode] (1) at (0,0){1};
\node[cnode] (2) at (2,0){2};
\node[cnode] (3) at (4,0){3};
\node[cnode] (4) at (6,0){4};
\node[cnode] (5) at (8,0){3};
\node[cnode] (6) at (7,2){1};
\node[snode] (7) at (6,-2){1};
\node[snode] (8) at (8,-2){1};
\node[snode] (9) at (9,2){1};
\draw[-] (1) -- (2);
\draw[-] (2)-- (3);
\draw[-] (3)-- (4);
\draw[-] (5)-- (4);
\draw[-] (6)-- (4);
\draw[-] (5)-- (6);
\draw[-] (9)-- (6);
\draw[-] (4)-- (7);
\draw[-] (5)-- (8);
\node[] (9) at (0,-0.75){};
\node[text width=.2cm](11) at (4,-2){$(X')$};
\end{tikzpicture}}
&\qquad \qquad 
& \scalebox{.7}{\begin{tikzpicture}[node distance=2cm, nnode/.style={circle,draw,thick, fill, inner sep=1 pt},cnode/.style={circle,draw,thick,minimum size=1.0 cm},snode/.style={rectangle,draw,thick,minimum size=1.0 cm}]
\node[snode] (1) at (0,0){6};
\node[cnode] (2) at (2,0){4};
\node[cnode] (3) at (4,0){2};
\node[cnode] (4) at (6,0){1};
\node[](5) at (2,-2){};
\node[](6) at (4,-2){};
\draw[-] (1) -- (2);
\draw[-] (2)-- (3);
\draw[-] (3)-- (4);
\draw[-, thick, blue] (2)-- (2,1);
\draw[-, thick, blue] (2,1)-- (6,1);
\draw[-, thick, blue] (6,1)-- (4);
\node[] (9) at (0,-0.75){};
\node[text width=.2cm](10) at (2.2, 0.75){4};
\node[text width=.2cm](11) at (6.2, 0.75){1};
\node[text width=.2cm](11) at (4,-2){$(Y')$};
\end{tikzpicture}}
\end{tabular}
\caption{\footnotesize{The construction of the quiver gauge theory $(X')$ from the theory $(X)$ by the action of the $S$-type operation 
${\CO}_{\vec \CP}$. The flavor nodes to be identified in the quiver $(X)$ are shown in red. The quiver gauge theory $(Y')$ is constructed 
from the theory $(Y)$ by the action of the dual operation $\wt{\CO}_{\vec \CP}$.}}
\label{SOp-pltNEx}
\end{center}
\end{figure}
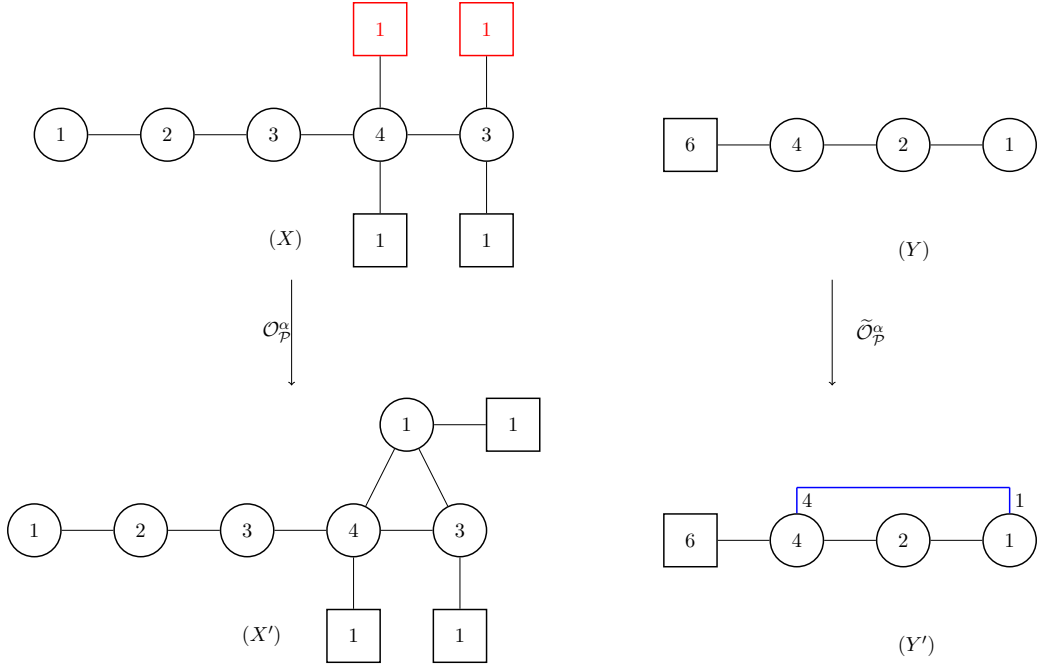

The quiver $X'$ can be obtained from the linear quiver $X$ by implementing an $S$-type operation $\CO_{\vec \CP}$, 
which constitutes a single elementary Abelian operation of the identification-flavoring-gauging type, as shown in 
\figref{SOp-pltNEx}. The mass parameters $\{u_i\}$ associated the $S$-type operation are given as
\be
u_1=m_2, \, u_2=m_3,\, v_1=m_1\,v_2=m_4. \label{SOp-uv1}
\ee
The partition function of the theory $X'= \CO_{\vec \CP}(X)$ is then given as
\begin{align}
Z^{\CO_{\vec \CP}(X)}= \int\, d{u}\, \CZ_{\CO_{\vec \CP}(X)}(u, u_1,u_2, \eta, m^f, \vec \mu) \cdot Z^{(X, \vec\CP)}(u_1, u_2, v_1,v_2; \vec t), 
\end{align}
where the operator $\CZ_{\CO_{\vec \CP}(X)}$ can be constructed from the general prescription in \eref{CZ-OP} and has the following form:
\begin{align}
 \CZ_{\CO_{\vec \CP}(X)} =  \int\, \prod^2_{i=1}\, du_i\, \frac{e^{2\pi i \eta u}\,\prod^2_{i=1}\delta(u_i -u -\mu_i)}{\ch{(u- m^f)}} .
\end{align}
Using the above expression, the partition function of the theory $X'$ can be explicitly written as
\begin{align} \label{Pf-pltNEx}
Z^{(X')} = &\int\,du\, \prod^5_{\gamma=1} \Big[ d \vec s^\gamma\Big]\,Z^{(X')}_{\rm FI}(\{\vec s^\gamma, u \}, \vec t, \eta)\, \prod^5_{\gamma=1} Z^{\rm vector}_{\rm{1-loop}}(\vec s^\gamma)\,\prod^{4}_{\gamma=1} Z^{\rm bif}_{\rm{1-loop}}(\vec s^\gamma, \vec s^{\gamma +1},0)\nn \\
& \times Z^{\rm fund}_{\rm{1-loop}}(\vec s^4, v_1)\, Z^{\rm fund}_{\rm{1-loop}}(\vec s^5, v_2)\,Z^{\rm bif}_{\rm{1-loop}}(\vec s^4, u, \mu_1)\,
Z^{\rm bif}_{\rm{1-loop}}(\vec s^5, u, \mu_2)\,Z^{\rm fund}_{\rm{1-loop}}(u,m^f),
\end{align}
where $Z^{(X')}_{\rm FI}= e^{2\pi i \eta u}\, Z^{(X)}_{\rm FI}$. The matrix integral on the RHS manifestly corresponds to the quiver 
$X'$ in \figref{fig: pltN-quivers}. By shifting the integration variables, one can check that the matrix integral can be written in terms 
of three independent real masses, which live in the Cartan subalgebra of the Higgs branch global symmetry group of $X'$, i.e. $G^{X'}_H=U(1)^3$.
The Coulomb branch global symmetry can be read off from the quiver itself. It has a linear subquiver of 5 balanced nodes giving an $SU(6)$ factor
and a single unbalanced gauge node which gives an additional $U(1)$ factor, giving $G^{X'}_C=SU(6) \times U(1)$.

The dual theory can be read off from the general equations in \eref{PF-wtOPgen}-\eref{CZ-wtOP}, with the function $ \CZ_{\wt{\CO}_{\vec \CP}(Y)}$ being 
given as
\begin{align}
& \CZ_{\wt{\CO}_{\vec \CP}(Y)} = \int \, du \, \CZ_{\CO_{\vec \CP}(X)} \cdot \prod^{2}_{i=1}\, e^{2\pi i (g^i(\{\vec \s^k\}, \vec \CP) + b^{il}t_l)\,u_i},\\
& \prod^{2}_{i=1}\, e^{2\pi i (g^i(\{\vec \s^k\}, \vec \CP) + b^{il}t_l)\,u_i} = e^{2\pi i u_1 (\tr \vec \s^1- \tr \vec \s^2 + b^{1l} t_l)}\,e^{2\pi i u_2 (\tr \vec \s^2-\s^3 + b^{2l} t_l)}.
\end{align}
Integrating over $u$, the function $\CZ_{\wt{\CO}_{\vec \CP}(Y)} $ can be written in the following form:
\be
\CZ_{\wt{\CO}_{\vec \CP}(Y)} = e^{2\pi i \,\mu_i b^{il}t_l}\,\frac{e^{2\pi i m^f\,(\eta + \tr \vec \s^1-\s^3 )}}{\ch{(\eta + \tr \vec \s^1-\s^3)}}
\,e^{2\pi i \mu_1 \tr \vec \s^1}\, e^{-2\pi i (\mu_1 - \mu_2)\tr \vec \s^2}\, e^{-2\pi i \mu_2 \s^3}.
\ee

After some minor rearrangement of terms, the dual partition function can be written as (up to contact terms)
\begin{align}
& Z^{\wt{\CO}_{\vec \CP}(Y)}(\vec{m}'; \vec \eta')=  \int \prod^{3}_{\gamma'=1} \Big[d\vec \s^{\gamma'}\Big]\, Z^{\rm hyper}_{\rm{1-loop}}(\vec \s^1, \s^3, \eta)\,Z^{(Y)}_{\rm int}(\{\vec \s^{\gamma'} \},\vec{t}, \{-v_1, -(\mu_1+ m^f), -(\mu_2+ m^f), -v_2\}), \label{Pf-pltNEx-dual} \\
& Z^{\rm hyper}_{\rm{1-loop}}(\vec \s^1, \s^3, \eta) = \frac{1}{\ch{(\tr \vec \s^1-\s^3 + \eta)}}. \label{Pf-hyper1}
\end{align}
The resultant dual quiver gauge theory can be read off from the matrix model integrand in \eref{Pf-pltNEx-dual} -- it is given by 
the quiver $Y$ with an additional hypermultiplet which is charged under the $U(1)$ subgroup of the gauge group $U(4)$ as well 
under the gauge group $U(1)$. The respective charges (4 and 1) can be read off from the RHS of \eref{Pf-hyper1}. Therefore, 
one can identify the 3d mirror of $X'$ as the quiver gauge theory $Y'$ in \figref{SOp-pltNEx}. In addition, by shifting the 
integration variables, one can check that the dual partition function is a function of 6 independent real masses, which live in 
the Cartan subalgebra of the Higgs branch global symmetry of $Y'$. i.e. $G^{Y'}_{H}=SU(6) \times U(1)$. The Coulomb branch 
symmetry can be read off from the quiver $Y'$. Since none of the gauge nodes are balanced, it is given as $G^{Y'}_{C}=U(1)^3$.

\subsection{An example with $p> N$ : $D_9(SU(3))$}\label{pgtNEx}

In this section, we consider a specific example of the class of quiver gauge theories discussed in \Secref{pgtNGen} and explicitly 
determine its 3d mirror following the general prescription given in \Secref{SOp-rev}. 

\subsubsection{The unitary Lagrangian}

Consider the quiver gauge theory $X'$ in \figref{3dmirr-pgtN} for $p=9, N=3$. The various parameters that encode the quiver 
data are given as:
\begin{align}
m=3 , \quad n=1, \quad q=3.
\end{align}

This corresponds to the quiver gauge theory:

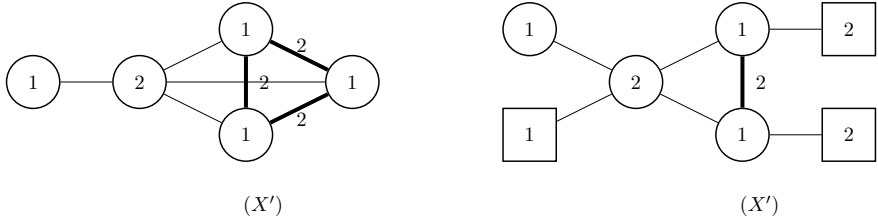
\begin{center}
\begin{tabular}{ccc}
\scalebox{.7}{\begin{tikzpicture}[
cnode/.style={circle,draw,thick, minimum size=1.0cm},snode/.style={rectangle,draw,thick,minimum size=1cm}]
\node[cnode] (9) at (0,0){1};
\node[cnode] (10) at (6,0){1};
\node[cnode] (11) at (2, 0){2};
\node[cnode] (12) at (4, 1){1};
\node[cnode] (13) at (4, -1){1};
\draw[-] (9) -- (11);
\draw[-] (10) -- (11);
\draw[-] (12) -- (11);
\draw[-] (13) -- (11);
\draw[line width=0.75mm, black] (12) -- (10);
\draw[line width=0.75mm, black] (13) -- (10);
\draw[line width=0.75mm, black] (12) to (13);
\node[text width=0.1cm](20) at (4.3,0){$2$};
\node[text width=0.1cm](21) at (5,0.7){$2$};
\node[text width=0.1cm](21) at (5,-0.7){$2$};
\node[text width=0.1cm](31)[below=0.5 cm of 13]{$(X')$};
\end{tikzpicture}}
& \qquad \qquad & 
\scalebox{.7}{\begin{tikzpicture}[
cnode/.style={circle,draw,thick, minimum size=1.0cm},snode/.style={rectangle,draw,thick,minimum size=1cm}]
\node[cnode] (9) at (0,1){1};
\node[snode] (10) at (0,-1){1};
\node[cnode] (11) at (2, 0){2};
\node[cnode] (12) at (4, 1){1};
\node[cnode] (13) at (4, -1){1};
\node[snode] (14) at (6, 1){$2$};
\node[snode] (15) at (6, -1){$2$};
\draw[-] (9) -- (11);
\draw[-] (10) -- (11);
\draw[-] (12) -- (11);
\draw[-] (13) -- (11);
\draw[-] (12) -- (14);
\draw[-] (13) -- (15);
\draw[line width=0.75mm, black] (12) to (13);
\node[text width=0.1cm](20) at (4.3,0){$2$};
\node[text width=0.1cm](31)[below=0.5 cm of 13]{$(X')$};
\end{tikzpicture}}
\end{tabular}
\end{center}

\begin{figure}[htbp]
\begin{center}
\begin{tabular}{ccc}
\scalebox{.7}{\begin{tikzpicture}[
cnode/.style={circle,draw,thick, minimum size=1.0cm},snode/.style={rectangle,draw,thick,minimum size=1cm}]
\node[cnode] (9) at (0,1){1};
\node[snode] (10) at (0,-1){1};
\node[cnode] (11) at (2, 0){2};
\node[cnode] (12) at (4, 1){1};
\node[cnode] (13) at (4, -1){1};
\node[snode] (14) at (6, 1){$2$};
\node[snode] (15) at (6, -1){$2$};
\draw[-] (9) -- (11);
\draw[-] (10) -- (11);
\draw[-] (12) -- (11);
\draw[-] (13) -- (11);
\draw[-] (12) -- (14);
\draw[-] (13) -- (15);
\draw[line width=0.75mm, black] (12) to (13);
\node[text width=0.1cm](20) at (4.5,0){$2$};
\node[text width=0.1cm](31)[below=0.5 cm of 13]{$(X')$};
\end{tikzpicture}}
&\qquad  \qquad
& \scalebox{.7}{\begin{tikzpicture}[node distance=2cm,cnode/.style={circle,draw,thick,minimum size=8mm},snode/.style={rectangle,draw,thick,minimum size=8mm},pnode/.style={rectangle,red,draw,thick,minimum size=8mm}]
\node[cnode] (1) at (-3,0) {$2$};
\node[snode] (2) at (-5,0) {$3$};
\node[cnode] (3) at (-1,2) {$1$};
\node[cnode] (4) at (-2,-2) {$1$};
\node[cnode] (5) at (0,-2) {$1$};
\node[cnode] (6) at (1,0) {$1$};
\node[cnode] (7) at (3,0) {$1$};
\node[snode] (8) at (5,0) {$1$};
\draw[thick] (1) -- (2);
\draw[thick, blue] (1) -- (-3.5,0.5);
\draw[thick, blue] (3) -- (-1.5, 2.5);
\draw[thick, blue] (-3.5,0.5) -- (-1.5, 2.5);
\draw[thick] (1) -- (4);
\draw[thick] (4) -- (5);
\draw[thick] (5) -- (6);
\draw[thick] (6) -- (7);
\draw[thick] (7) -- (8);
\draw[thick] (3) -- (6);
\node[text width=0.1cm](31) at (-3.1,0.6){2};
\node[text width=0.1cm](32) at (-1.6,2){1};
\node[text width=0.1cm](30) at (-1,-3){$(Y')$};
\end{tikzpicture}}
\end{tabular}
\caption{\footnotesize{The quiver $X'$ is the 3d mirror of $D_9(SU(3))$ reduced on a circle, constructed from a 
class $\CS$ prescription. The quiver $Y'$ is the 3d mirror of $X'$, i.e. a Lagrangian realization of the SCFT obtained 
by the reduction of $D_9(SU(3))$ on a circle.}}
\label{fig: pgtN-quivers}
\end{center}
\end{figure}
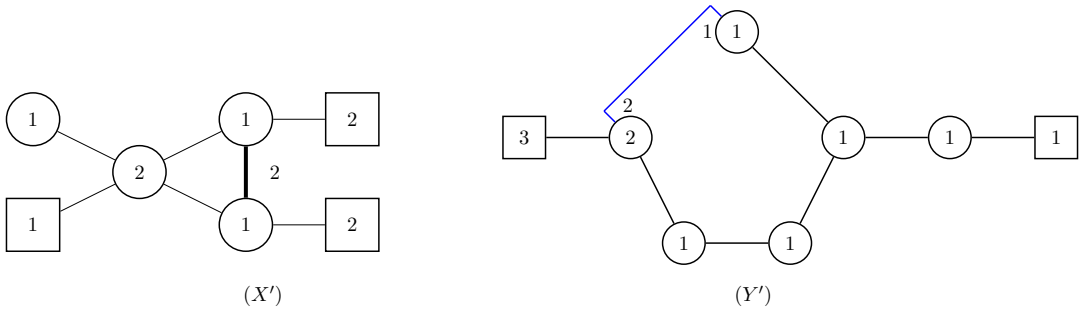

Note that we have decoupled a $U(1)$ vector multiplet to write the quiver gauge theory on the RHS. Also, note that we 
have $N-m=0$, which implies that there are no free hypermultiplets. We can again use the 
$S$-type operation technique, the details of which are described below, to construct the 3d mirror of $X'$. The result is 
shown in \figref{fig: pgtN-quivers}. Note that the quiver $Y'$ has as gauge group $U(2) \times U(1)^5$ 
with fundamental and bifundamental hypers. In addition, it involves a single hypermultiplet with charges $(2,1)$ under the 
subgroup $U(1) \subset U(2)$ and one of the $U(1)$ gauge groups respectively. This is consistent with our analysis of the 
general features of the theory $Y'$, as discussed in \Secref{pgtNGen}.
The moduli space dimensions and global symmetries of the dual theories are 
summarized in Table \ref{Tab:pgtNEx}.

\begin{table}[htbp]
\begin{center}
{%
\begin{tabular}{|c|c|c|}
\hline
Moduli space data & Theory $X'$ & Theory $Y'$ \\
\hline \hline 
dim\,$\CM_H$ & 7 & 5\\
\hline
dim\,$\CM_C$ & 5 & 7\\
\hline
$G_H$ & $SU(2)^3 \times U(1)^3$ & $SU(3) \times U(1)^2$\\
\hline
$G_C$ & $SU(3) \times U(1)^2$ & $SU(2)^3 \times U(1)^3$ \\
\hline
\end{tabular}}
\end{center}
\caption{\footnotesize{Summary table for the moduli space dimensions and global symmetries for the mirror pair in \figref{fig: pgtN-quivers}.}}
\label{Tab:pgtNEx}
\end{table}

\subsubsection{Derivation via $S$-type operations}

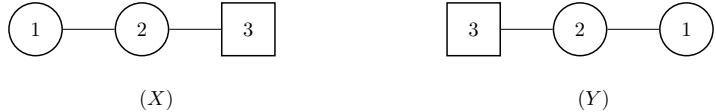
\begin{figure}[htbp]
\begin{center}
\scalebox{.7}{\begin{tikzpicture}[cnode/.style={circle,draw,thick, minimum size=1.0cm},snode/.style={rectangle,draw,thick,minimum size=1cm}]
\node[cnode] (1) at (0,0){1};
\node[cnode] (2) at (2,0){2};
\node[snode] (3) at (4,0){3};
\draw[-] (1) -- (2);
\draw[-] (2) -- (3);
\node[text width=0.1cm](20)[below=0.5 cm of 2]{$(X)$};
\end{tikzpicture}
\qquad \qquad \qquad \qquad 
\begin{tikzpicture}[
cnode/.style={circle,draw,thick, minimum size=1.0cm},snode/.style={rectangle,draw,thick,minimum size=1cm}]
\node[snode] (1) at (0,0){3};
\node[cnode] (2) at (2,0){2};
\node[cnode] (3) at (4,0){1};
\node[text width=0.1cm](20)[below=0.5 cm of 2]{$(Y)$};
\draw[-] (1) -- (2);
\draw[-] (2)-- (3);
\end{tikzpicture}}
\end{center}
\caption{\footnotesize{The starting point for the $S$-type operation construction is the self-dual theory $T(U(3))$.}}
\label{LQ-pgtNEx}
\end{figure}

The starting point for obtaining the mirror pair $(X',Y')$ by an $S$-type operation will again be a linear mirror pair $(X,Y)$. 
As discussed in \Secref{pgtNGen}, we choose $X$ to be the self-dual theory $T(U(3))$ as shown in \figref{LQ-pgtNEx}.
The partition functions of the theories $(X,Y)$ are given as
\begin{align}
& Z^{(X)}(\vec{m}; \vec{t})=\int \prod^2_{\gamma=1} \Big[ d \vec s^\gamma\Big]\, Z^{(X)}_{\rm int} (\{\vec s^\gamma\},\vec m, \vec t) = \int \prod^2_{\gamma=1} \Big[ d \vec s^\gamma\Big]\, Z^{(X)}_{\rm FI}(\{\vec s^\gamma \}, \vec t)\,Z^{(X)}_{\rm{1-loop}}(\{\vec s^\gamma \}, \vec m ),\\
& Z^{(Y)}(\vec{t}; \vec{m})=\int \prod^{2}_{\gamma'=1} \Big[d\vec \s^{\gamma'}\Big]\, Z^{(Y)}_{\rm int}(\{\vec \s^{\gamma'} \},\vec{t}, \vec{m}) 
=\int \prod^{2}_{\gamma'=1}\,\Big[d\vec \s^{\gamma'}\Big] \,Z^{(Y)}_{\rm FI}(\{\vec \s^{\gamma'} \}, \vec m)\, Z^{(Y)}_{\rm{1-loop}}(\{\vec \s^{\gamma'} \}, 
\vec t \} ),
\end{align}
where $\gamma, \gamma'$ label the gauge nodes of the linear quivers $X,Y$ respectively, with the count increasing from left to right in both cases. 
The FI contribution for $Y$ in the matrix integral is given as
\be
Z^{(Y)}_{\rm FI}(\{\vec \s^{\gamma'} \}, \vec t)= e^{2\pi i (m_1- m_2)\,\tr \vec \s^1}\,e^{2\pi i (m_2- m_3)\,\s^2}.
\ee

The fact that $T(U(3))$ is self-mirror implies that
\begin{align}
Z^{(X)}(\vec m ; \vec t) =C_{XY}(\vec m, \vec t) \, Z^{(Y)}(\vec t; - \vec m),
\end{align}
where $C_{XY}$ is a contact term, and as before, we will not need the precise form of the contact 
term to read off the gauge group and matter content of theories that we will construct using $S$-type operations.

\begin{figure}[htbp]
\begin{center}
\begin{tabular}{ccc}
\scalebox{.7}{\begin{tikzpicture}[
cnode/.style={circle,draw,thick, minimum size=1.0cm},snode/.style={rectangle,draw,thick,minimum size=1cm}]
\node[cnode] (9) at (0,1){1};
\node[snode] (10) at (0,-1){1};
\node[cnode] (11) at (2, 0){2};
\node[snode, red] (12) at (4, 1){1};
\node[snode] (13) at (4, -1){1};
\draw[-] (9) -- (11);
\draw[-] (10) -- (11);
\draw[-] (12) -- (11);
\draw[-] (13) -- (11);
\node[text width=0.1cm](21)[above=0.2 cm of 9]{3};
\node[text width=0.1cm](23)[above=0.2 cm of 12]{1};
\node[text width=0.1cm](24)[below=0.05 cm of 13]{2};
\node[text width=0.1cm](31)[below=0.5 cm of 11]{$(X)$};
\end{tikzpicture}}
& \qquad  \qquad 
& \scalebox{.7}{\begin{tikzpicture}[
cnode/.style={circle,draw,thick, minimum size=1.0cm},snode/.style={rectangle,draw,thick,minimum size=1cm}]
\node[snode] (1) at (0,0){3};
\node[cnode] (2) at (2,0){2};
\node[cnode] (3) at (4,0){1};
\node[text width=0.1cm](20)[below=0.5 cm of 2]{$(Y)$};
\draw[-] (1) -- (2);
\draw[-] (2)-- (3);
\end{tikzpicture}}\\
\scalebox{.7}{\begin{tikzpicture}
\draw[->] (15,-3) -- (15,-5);
\node[text width=0.1cm](20) at (14.2, -4) {$\CO^1_{\CP_1}$};
\end{tikzpicture}}
&\qquad \qquad 
& \scalebox{.7}{\begin{tikzpicture}
\draw[->] (15,-3) -- (15,-5);
\node[text width=0.1cm](29) at (15.5, -4) {$\wt{\CO}^1_{\CP_1}$};
\end{tikzpicture}}\\
\scalebox{.7}{\begin{tikzpicture}[
cnode/.style={circle,draw,thick, minimum size=1.0cm},snode/.style={rectangle,draw,thick,minimum size=1cm}]
\node[snode] (9) at (0,1){1};
\node[snode] (10) at (0,-1){1};
\node[cnode] (11) at (2, 0){2};
\node[cnode] (12) at (4, 1){1};
\node[snode, red] (13) at (4, -1){1};
\node[snode] (14) at (6, 2.5){2};
\node[snode, red] (15) at (6, 1){1};
\node[snode, red] (16) at (6, -0.5){1};
\draw[-] (9) -- (11);
\draw[-] (10) -- (11);
\draw[-] (12) -- (11);
\draw[-] (13) -- (11);
\draw[-] (12) -- (14);
\draw[-] (12) -- (15);
\draw[-] (12) -- (16);
\node[text width=0.1cm](21)[above=0.2 cm of 9]{3};
\node[text width=0.1cm](23)[above=0.2 cm of 12]{1};
\node[text width=0.1cm](24)[below=0.05 cm of 13]{2};
\node[text width=1cm](31)[below=0.5 cm of 13]{$\CO^1_{\CP_1}(X)$};
\end{tikzpicture}}
&\qquad \qquad 
& \scalebox{.7}{\begin{tikzpicture}[node distance=2cm,cnode/.style={circle,draw,thick,minimum size=8mm},snode/.style={rectangle,draw,thick,minimum size=8mm},pnode/.style={rectangle,red,draw,thick,minimum size=8mm}]
\node[cnode] (1) at (-3,0) {$2$};
\node[snode] (2) at (-5,0) {$3$};
\node[cnode] (4) at (-2,-2) {$1$};
\node[cnode] (5) at (0,-2) {$1$};
\node[cnode] (6) at (1,0) {$1$};
\node[cnode] (7) at (3,0) {$1$};
\node[snode] (8) at (5,0) {$1$};
\draw[thick] (1) -- (2);
\draw[thick] (1) -- (4);
\draw[thick] (4) -- (5);
\draw[thick] (5) -- (6);
\draw[thick] (6) -- (7);
\draw[thick] (7) -- (8);
\node[text width=0.1cm](30) at (-1,-3){$\wt{\CO}^1_{\CP_1}(Y)$};
\end{tikzpicture}}\\
\scalebox{.7}{\begin{tikzpicture}
\draw[->] (15,-3) -- (15,-5);
\node[text width=0.1cm](20) at (14.2, -4) {$\CO^2_{\CP_2}$};
\end{tikzpicture}}
&\qquad \qquad 
& \scalebox{.7}{\begin{tikzpicture}
\draw[->] (15,-3) -- (15,-5);
\node[text width=0.1cm](29) at (15.5, -4) {$\wt{\CO}^2_{\CP_2}$};
\end{tikzpicture}}\\
\scalebox{.7}{\begin{tikzpicture}[
cnode/.style={circle,draw,thick, minimum size=1.0cm},snode/.style={rectangle,draw,thick,minimum size=1cm}]
\node[cnode] (9) at (0,1){1};
\node[snode] (10) at (0,-1){1};
\node[cnode] (11) at (2, 0){2};
\node[cnode] (12) at (4, 1){1};
\node[cnode] (13) at (4, -1){1};
\node[snode] (14) at (6, 1){$2$};
\node[snode] (15) at (6, -1){$2$};
\draw[-] (9) -- (11);
\draw[-] (10) -- (11);
\draw[-] (12) -- (11);
\draw[-] (13) -- (11);
\draw[-] (12) -- (14);
\draw[-] (13) -- (15);
\draw[line width=0.75mm, black] (12) to (13);
\node[text width=0.1cm](20) at (4.5,0){$2$};
\node[text width=0.1cm](21)[above=0.2 cm of 9]{3};
\node[text width=0.1cm](23)[above=0.2 cm of 12]{1};
\node[text width=0.1cm](24)[below=0.05 cm of 13]{2};
\node[text width=0.1cm](31)[below=0.5 cm of 13]{$(X')$};
\end{tikzpicture}}
&\qquad \qquad 
& \scalebox{.7}{\begin{tikzpicture}[node distance=2cm,cnode/.style={circle,draw,thick,minimum size=8mm},snode/.style={rectangle,draw,thick,minimum size=8mm},pnode/.style={rectangle,red,draw,thick,minimum size=8mm}]
\node[cnode] (1) at (-3,0) {$2$};
\node[snode] (2) at (-5,0) {$3$};
\node[cnode] (3) at (-1,2) {$1$};
\node[cnode] (4) at (-2,-2) {$1$};
\node[cnode] (5) at (0,-2) {$1$};
\node[cnode] (6) at (1,0) {$1$};
\node[cnode] (7) at (3,0) {$1$};
\node[snode] (8) at (5,0) {$1$};
\draw[thick] (1) -- (2);
\draw[thick, blue] (1) -- (-3.5,0.5);
\draw[thick, blue] (3) -- (-1.5, 2.5);
\draw[thick, blue] (-3.5,0.5) -- (-1.5, 2.5);
\draw[thick] (1) -- (4);
\draw[thick] (4) -- (5);
\draw[thick] (5) -- (6);
\draw[thick] (6) -- (7);
\draw[thick] (7) -- (8);
\draw[thick] (3) -- (6);
\node[text width=0.1cm](31) at (-3.1,0.6){2};
\node[text width=0.1cm](32) at (-1.6,2){1};
\node[text width=0.1cm](30) at (-1,-3){$(Y')$};
\end{tikzpicture}}
\end{tabular}
\caption{\footnotesize{The construction of the quiver gauge theory $X'$ of \figref{fig: pgtN-quivers} by a sequence of 
two elementary Abelian $S$-type operation. The labels $i=1,2$ on the flavor nodes correspond to the mass parameters 
$u_i$. At each step, the flavor node on which the elementary operation acts is shown in blue.}}
\label{SOp-pgtNEx}
\end{center}
\end{figure}
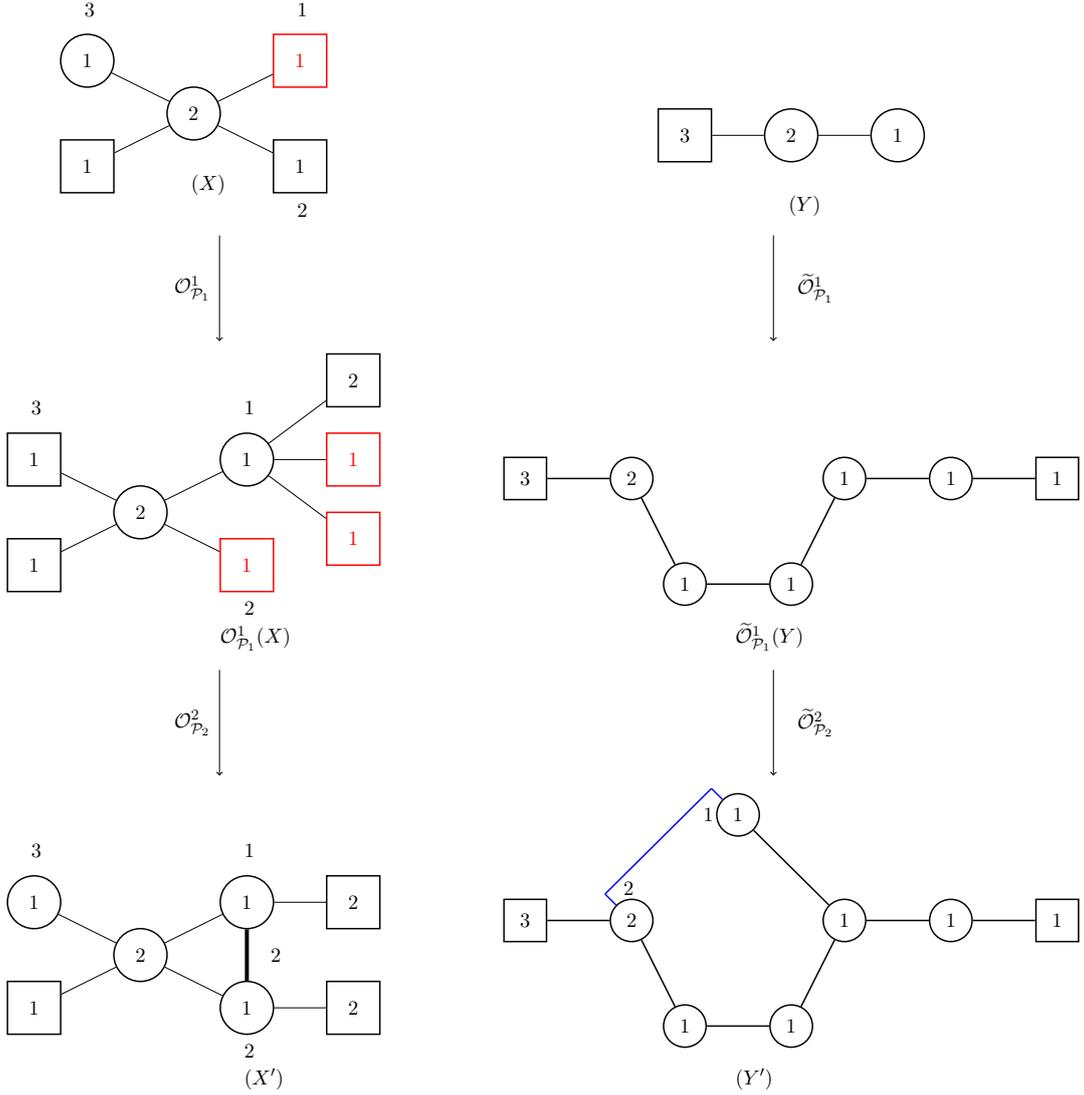

The quiver $X'$ can be obtained from the linear quiver $X$ by implementing an $S$-type operation $\CO_{\vec \CP}$, 
which includes two elementary Abelian $S$-type operations, i.e.
\begin{align}
\CO_{\vec \CP} (X)=\CO^2_{\CP_2} \circ \CO^1_{\CP_1}(X), \label{SOp-2a}
\end{align}
where $ \CO^i_{\CP_i}$ ($i=1,2$) are elementary Abelian $S$-type operations shown in \figref{SOp-pgtNEx}.
Note that $\CO^1_{\CP_1}$ is a flavoring-gauging operation, while $\CO^2_{\CP_2}$ is an identification-flavoring-gauging 
operation. The mass parameters $\{u_i\}$ associated the $S$-type operation $\CO^i_{\CP_i}$, are given as
\be
u_1=m_3, \, u_2=m_2, v=m_1. \label{SOp-2b}
\ee
The partition function of the theory $X'= \CO_{\vec \CP}(X)$ is then given as
\begin{align}
Z^{\CO_{\vec \CP}(X)}= \int\, \prod^2_{i=1}\, d{u}_i\, \CZ_{\CO_{\vec \CP}(X)}(\vec u, \vec \eta, \vec m^f, \vec x^f, \vec \mu) \cdot Z^{(X, \vec\CP)}(\vec u, v; \vec t), 
\end{align}
where the operator $\CZ_{\CO_{\vec \CP}(X)}$ can be constructed from the general prescription in \eref{CZ-OP} and has the following form:
\begin{align}
 \CZ_{\CO_{\vec \CP}(X)}= &  \CZ_{\CO^2_{\CP_2} (\CO^1_{\CP_1}(X))} \cdot \CZ_{\CO^1_{\CP_1}(X)} \nn \\
 = & \Big( \int\, \prod^2_{b=1}\, dm^f_b\, \frac{e^{2\pi i \eta_2 u_2}\,\prod^2_{b=1}\delta(m^f_b -u_2- \mu_b)}{\prod^2_{c=1}\,\ch{(u_2-x^f_c)}}\Big)\cdot\frac{e^{2\pi i \eta_1 u_1}}{\prod^4_{a=1}\, \ch{(u_1-m^f_a)}},
\end{align}
where $m^f_a$ ($a=1,\ldots,4$) and $x^f_c$ ($c=1,2$) are the masses of hypers added in the two flavoring operations. 
Note that the second equality involves specific choices of the permutation matrices $\CP_i$. 
Using the above expression, the partition function of the theory $X'$ can be explicitly written as
\begin{align} \label{Pf-pltNEx}
Z^{(X')} = &\int\,\prod^2_{i=1}\,du_i\, \prod^2_{\gamma=1} \Big[ d \vec s^\gamma\Big]\,Z^{(X')}_{\rm FI}(\{\vec s^\gamma, \vec u \}, \vec t, \vec \eta)\, Z^{\rm vector}_{\rm{1-loop}}(\vec s^2)\, Z^{\rm bif}_{\rm{1-loop}}(\vec s^1, \vec s^{2},0)\, Z^{\rm fund}_{\rm{1-loop}}(\vec s^2, v)\nn \\
& \times \prod^2_{i=1}\,Z^{\rm bif}_{\rm{1-loop}}(\vec s^2, u_i, 0)\,\prod^2_{a=1}Z^{\rm fund}_{\rm{1-loop}}(u_1, m^f_{2+a})\, 
\prod^2_{c=1}Z^{\rm fund}_{\rm{1-loop}}(u_2, x^f_{c})\, \prod^2_{b=1} Z^{\rm bif}_{\rm{1-loop}}(u_1, u_2, \mu_b),
\end{align}
where $Z^{(X')}_{\rm FI}= e^{2\pi i \eta_2 u_2}\,e^{2\pi i \eta_1 u_1}\, Z^{(X)}_{\rm FI}$. The matrix integral on the RHS manifestly corresponds to the quiver 
$X'$ in \figref{SOp-pgtNEx}. By shifting the integration variables, one can check that the matrix integral can be written in terms 
of 6 independent real masses, which live in the Cartan subalgebra of the Higgs branch global symmetry group of $X'$, i.e. 
$G^{X'}_H=SU(2)^3 \times U(1)^3$. The Coulomb branch global symmetry can be read off from the quiver itself. 
It has a linear subquiver of 2 balanced nodes giving an $SU(3)$ factor and two unbalanced gauge node which 
gives an additional $U(1)^2$ factor, giving $G^{X'}_C=SU(3) \times U(1)^2$.

The dual theory can be read off from the general equations in \eref{PF-wtOPgen}-\eref{CZ-wtOP}, with the function $ \CZ_{\wt{\CO}_{\vec \CP}(Y)}$ being 
given as
\begin{align}
& \CZ_{\wt{\CO}_{\vec \CP}(Y)} = \CZ_{\wt\CO^2_{\CP_2} (\wt\CO^1_{\CP_1}(Y))} \cdot \CZ_{\wt\CO^1_{\CP_1}(Y)} 
= \int \, \prod^2_{i=1}\, d{u}_i \, \CZ_{\CO_{\vec \CP}(X)} \cdot \prod^{2}_{i=1}\, e^{2\pi i (g^i(\{\vec \s^k\}, \vec \CP) + b^{il}t_l)\,u_i} \nn\\
&=\int \, du_2 \Big( \int\, \prod^2_{b=1}\, dm^f_b\, \frac{e^{2\pi i u_2 ( \tr \s^1 - \s^2 + \eta_2+ b^{2l}t_l)}\,\prod^2_{b=1}\delta(m^f_b -u_2- \mu_b)}{\prod^2_{c=1}\,\ch{(u_2-x^f_c)}}\Big)\cdot\int \, du_1\frac{e^{2\pi i u_1(\s^2 + \eta_1 +b^{1l}t_l)}}{\prod^4_{a=1}\, \ch{(u_1-m^f_a)}},
\end{align}
where the functions $g^i(\{\vec \s^k\}, \vec \CP)$ are given as:
\be
\prod^{2}_{i=1}\, e^{2\pi i (g^i(\{\vec \s^k\}, \vec \CP) + b^{il}t_l)\,u_i} = e^{2\pi i u_2 ( \tr \s^1 - \s^2 + b^{2l}t_l)}\,e^{2\pi i u_1 (\s^2 +b^{1l}t_l)}.
\ee

In the first step, integrating over $u_1$, the function $\CZ_{\wt{\CO}^1_{\CP_1}(Y)} $ can be written in the following form:
\begin{align}
\CZ_{\wt{\CO}^1_{\CP_1}(Y)} =&  \int \prod^3_{j=1}\,d\tau_j\, \frac{e^{2\pi i m^f_1 (\s^2 + \eta_1 + b^{1l} t_l)}}{\ch{(\s^2 -\tau_1 +\eta_1 + b^{1l} t_l)}}
\cdot \frac{\prod^3_{j=1}e^{-2\pi i \tau_j (m^f_j - m^f_{j+1}) }}{\prod^2_{j=1}\ch{(\tau_i -\tau_{i+1})}\,\ch{\tau_3}} \\
=&  \int \prod^3_{j=1}\,d\tau_j\, e^{2\pi i m^f_1 (\s^2 + b^{1l} t_l)}\, Z^{\rm bif}_{\rm{1-loop}}(\tau_1, \s^2, \eta_1 + b^{1l} t_l)\,
Z^{\CT[3]}_{\rm int} (\vec \tau, -\vec m^f, 0), 
\end{align}
where $\CT[N]$ is a linear chain of $N$ $U(1)$ gauge nodes with bifundamental hypers and a single fundamental hyper at one end.
After some minor rearrangement of terms, the dual partition function can be written as (up to contact terms)
\begin{align}
Z^{\wt{\CO}^1_{\CP_1}(Y)}(\vec{m}'; \vec \eta')= & \int \prod^{2}_{\gamma'=1} \Big[d\vec \s^{\gamma'}\Big]\, \prod^3_{j=1}\,d\tau_j\,Z^{(Y)}_{\rm int}(\{\vec \s^{\gamma'} \},\vec{t}, \{-v_1, -u_2, - m^f_1\}) \nn \\
&\times Z^{\rm bif}_{\rm{1-loop}}(\tau_1, \s^2, \eta_1 + b^{1l} t_l)\,Z^{\CT[3]}_{\rm int} (\vec \tau, -\vec m^f, 0) , \label{Pf-pgtNEx-dual1} 
\end{align}
The resultant dual quiver gauge theory can be read off from the matrix model integrand in \eref{Pf-pgtNEx-dual1}. It is given by 
the quiver $Y$ and the quiver $\CT[3]$ connected by a single bifundamental hyper, where the latter is charged under the 
$U(1)$ gauge factor of $Y$ and one of the $U(1)$ gauge nodes of $\CT[3]$, as shown in \figref{SOp-pgtNEx}.\\

In the final step, integrating over $u_2$, the function $\CZ_{\wt{\CO}_{\vec \CP}(Y)}$ can be written in the following form, 
after a change of variables $ \wt{\tau}\to -\wt{\tau} + \tau_2$:
\begin{align}
\CZ_{\wt{\CO}_{\vec \CP}(Y)} =& \int d\wt{\tau}\,\prod^3_{j=1}\,d\tau_j\, e^{2\pi i x^f_1 (\tr \vec \s^1-s^2 + (b^{2l} t_l +\eta_2))}\, e^{-2\pi i \wt{\tau}(x^f_1 -x^f_2)}\,
e^{2\pi i( \mu_1 + x^f_1) (\s^2 + \eta_1 + b^{1l}t_l)}\nn \\
\times & Z^{\rm bif}_{\rm{1-loop}}(\wt{\tau}, \tau_2, 0)\, Z^{\rm hyper}_{\rm{1-loop}}(\vec \s^1, \wt{\tau}, \sum_i(\eta_i + b^{il} t_l))\,
Z^{\rm bif}_{\rm{1-loop}}({\tau}_1, \s^2, \eta_1 + b^{1l} t_l)\nn \\
\times & Z^{\CT[3]}_{\rm int} (\vec \tau, -\{\mu_1 + x^f_1, \mu_2 + x^f_2, m^f_3, m^f_4 \}, 0),
\end{align}
where the function $Z^{\rm hyper}_{\rm{1-loop}}$ is given as:
\be
Z^{\rm hyper}_{\rm{1-loop}} = \frac{1}{\ch{(\tr \vec \s^1 - \wt{\tau} - \sum_i(\eta_i + b^{il} t_l))}}.
\ee
After some minor rearrangement of terms, the dual partition function can be written as (up to contact terms)
\begin{align}
Z^{\wt{\CO}_{\vec \CP}(Y)}(\vec{m}'; \vec \eta') &=\int \prod^{2}_{\gamma'=1} \Big[d\vec \s^{\gamma'}\Big]\, \prod^3_{j=1}\,d\tau_j\, d\wt{\tau}\,
e^{-2\pi i \wt{\tau}(x^f_1 -x^f_2)}\,Z^{(Y)}_{\rm int}(\{\vec \s^{\gamma'} \},\vec{t}, -\{v_1, x^f_1, \mu_1 + x^f_1\}) \nn \\
&\times Z^{\rm hyper}_{\rm{1-loop}}(\vec \s^1, \wt{\tau}, \sum_i(\eta_i + b^{il} t_l))\, Z^{\rm bif}_{\rm{1-loop}}(\wt{\tau}, \tau_2, 0)\,Z^{\rm bif}_{\rm{1-loop}}(\tau_1, \s^2, \eta_1 + b^{1l} t_l)\nn \\
&\times Z^{\CT[3]}_{\rm int} (\vec \tau, -\{ \mu_1+ x^f_1, \mu_2+ x^f_2 , m^f_3, m^f_4 \}, 0). \label{Pf-pgtNEx-dual2} 
\end{align}
The resultant dual quiver gauge theory can be read off from the matrix model integrand in \eref{Pf-pgtNEx-dual1}, 
and manifestly reproduces the quiver $Y'$ in  \figref{SOp-pgtNEx}. 
By shifting the integration variables, one can check that the dual partition function is a function of 4 independent real masses, which live in 
the Cartan subalgebra of the Higgs branch global symmetry of $Y'$. i.e. $G^{Y'}_{H}=SU(3) \times U(1)^2$. The Coulomb branch 
symmetry can be read off from the quiver $Y'$. It includes three isolated balanced gauge nodes giving a factor $SU(2)^3$, and 
three unbalanced nodes giving a factor $U(1)^3$, so that $G^{Y'}_{C}=SU(2)^3 \times U(1)^3$.

\section{An infinite class of IR dualities}\label{U2-USU}

From \Secref{pltNGen} and \Secref{pgtNGen}, we showed that the 3d SCFT obtained by compactifying $D_p(SU(N)$ on a 
circle has a Lagrangian description in terms of a quiver gauge theory. The gauge group of the theory in question has 
only unitary factors, and involves a set of Abelian hypermultiplets in addition to fundamental and bifundamental 
hypers. Comparing our results with \cite{Closset:2020afy} suggests that we have an IR duality for every $D_p(SU(N)$, where the 
dual theories belong to the following classes of quiver gauge theory respectively:
\begin{itemize}

\item Theory $\CT$: A linear chain of unitary and special unitary gauge groups with fundamentals and bifundamentals.

\item Theory $\CT^\vee$: A generically non-linear quiver involving only unitary gauge groups, decorated with Abelian matter 
in addition to fundamental/bifundamental hypers.

\end{itemize}

In this section, we present a couple of simple examples of this conjectured duality, and check that the three-sphere partition 
function of the dual theories do agree.

\subsection{An IR duality for $SU(N)$ with $2N-1$ flavors}\label{U2-USU-basic}

\begin{figure}[htbp]
\begin{center}
\begin{tabular}{ccc}
\scalebox{.8}{\begin{tikzpicture}[node distance=2cm, nnode/.style={circle,draw,thick, fill, inner sep=1 pt},cnode/.style={circle,draw,thick,minimum size=1.0 cm},snode/.style={rectangle,draw,thick,minimum size=1.0 cm}, pnode/.style={circle,double,draw,thick, minimum size=1.0cm}]
\node[pnode] (1) at (0,0){$N$};
\node[snode] (2) at (0,-2){$2N-1$};
\draw[-] (1) -- (2);
\node[text width=0.1cm](20)[below=0.5 cm of 2]{$(\CT)$};
\end{tikzpicture}}
& \qquad   \qquad
&\scalebox{.8}{\begin{tikzpicture}[node distance=2cm, nnode/.style={circle,draw,thick, fill, inner sep=1 pt},cnode/.style={circle,draw,thick,minimum size=1.0 cm},snode/.style={rectangle,draw,thick,minimum size=1.0 cm}]
\node[cnode] (1) at (0,0){$N-1$};
\node[snode] (2) at (0,-2){$2N-1$};
\node[snode,blue] (3) at (3,0){$1$};
\draw[-] (1) -- (2);
\draw[-, thick, blue] (1)-- (3);
\node[text width=1cm](10) at (1.2, 0.2){$N-1$};
\node[text width=1cm](11) at (0, 1.1){$\eta=0$};
\node[text width=0.1cm](20)[below=0.5 cm of 2]{$(\CT^\vee)$};
\end{tikzpicture}}\\
 \qquad   \qquad
 & \scalebox{.7}{\begin{tikzpicture}
\draw[->] (15,-3) -- (15,-5);
\node[text width=3 cm](29) at (16, -4) {3d mirror};
\end{tikzpicture}}
 & \qquad   \qquad
 \end{tabular}
  \end{center}
 \begin{center}
  \scalebox{0.7}{\begin{tikzpicture}[
cnode/.style={circle,draw,thick,minimum size=4mm},snode/.style={rectangle,draw,thick,minimum size=8mm},pnode/.style={rectangle,red,draw,thick,minimum size=8mm}]
\node[cnode] (1) {$1$};
\node[cnode] (2) [right=.5cm  of 1]{$2$};
\node[cnode] (3) [right=.5cm of 2]{$3$};
\node[cnode] (4) [right=1cm of 3]{$N -1$};
\node[cnode] (6) [right=1 cm of 4]{$N-1$};
\node[cnode] (7) [right=1cm of 6]{{$3$}};
\node[cnode] (8) [right=0.5cm of 7]{$2$};
\node[cnode] (9) [right=0.5cm of 8]{$1$};
\node[cnode] (13) [above=0.5cm of 4]{$1$};
\node[snode] (14) [left=0.5cm of 13]{1};
\node[snode] (15) [above=0.5cm of 6]{$1$};
\draw[-] (1) -- (2);
\draw[-] (2)-- (3);
\draw[dashed] (3) -- (4);
\draw[-] (4) --(6);
\draw[dashed] (6) -- (7);
\draw[-] (7) -- (8);
\draw[-] (8) --(9);
\draw[-] (4) -- (13);
\draw[-] (13) -- (14);
\draw[-] (6) -- (15);
\end{tikzpicture}}
\end{center}
\caption{The simplest example of an IR duality discussed above. The partition functions of the two theories $\CT$ and $\CT^\vee$ agree only when $\eta=0$. 
The 3d mirror for the two theories $\CT$ and $\CT^\vee$ is also shown.}
\label{IRdual-Ex1}
\end{figure}
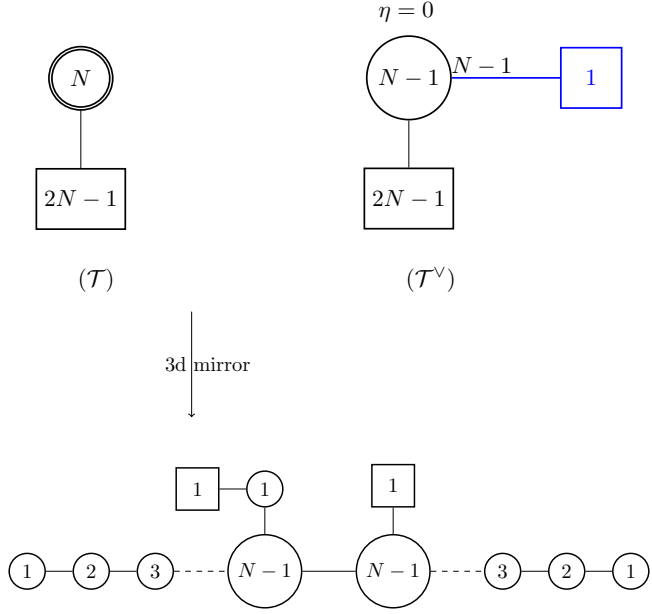

Consider the simplest possible duality of the type mentioned above -- an $SU(N)$ gauge theory with $2N-1$ fundamental hypers, 
and a $U(N-1)$ gauge theory with $2N-1$ fundamental hypers and a single hypermultiplet charged under $U(1) \subset U(N-1)$ 
with charge $N-1$, as shown in \figref{IRdual-Ex1}. This specific duality is not associated with any $D_p(SU(N))$ theory, but 
will play a central role in checking the dualities of the class described above. Firstly, note that the quaternionic dimensions of the  
Higgs and the Coulomb branches do agree:
\begin{align}
& {\rm dim}\CM^{(\CT)}_H = {\rm dim}\CM^{(\CT^\vee)}_H =N^2 -N+1,\\
& {\rm dim}\CM^{(\CT)}_C = {\rm dim}\CM^{(\CT^\vee)}_C = N-1,
\end{align}
and that the Higgs branch global symmetries $G^{(\CT)}_H = G^{(\CT^\vee)}_H =U(2N-1)$ manifestly agree. The theory 
$\CT^\vee$ has a $U(1)$ Coulomb branch global symmetry manifest in the UV, while the theory $\CT$ has trivial 
Coulomb branch global symmetry in the UV. However, it can be shown \cite{Giacomelli:2020ryy} that $\CT$ has an emergent $U(1)$ 
Coulomb branch global symmetry in the IR.

Note that the first non-trivial duality in this family, for $N=2$, is the familiar duality between an $SU(2)$ 
theory with three fundamental hypers and a $U(1)$ theory with four fundamental hypers \cite{Kapustin:1998fa}.\\

The equality of the partition functions of the two theories can be demonstrated as follows.
Recall that a $U(N)$ gauge theory with $N_f=2N-1$  fundamental hypers is an ugly theory in the Gaiotto-Witten sense.
This implies that the IR theory factorizes into an SCFT, which in turn has a good UV description, 
and a single twisted hypermultiplet. The good SCFT in question is the IR SCFT of a $U(N-1)$ 
gauge theory with $N_f=2N-1$ fundamental hypers. In terms of the partition function, this translates to 
the following identity (derived in \cite{Kapustin:2010mh} with a sign mistake that we have corrected below): 
\be \label{ugly-id}
Z^{U(N)}_{2N-1}(\vec m, \eta) = \Big(\frac{e^{2\pi i \eta \sum^{2N-1}_{i=1} m_i}}{\ch{\eta}}\Big) \cdot Z^{U(N-1)}_{2N-1}(\vec m, -\eta).
\ee
The partition function of an $SU(N)$ theory can be obtained from the partition function of the 
unitary theory on the LHS by integrating over $\eta$, i.e.
\be 
Z^{SU(N)}_{2N-1}(\vec m) = \int \, d\eta \, Z^{U(N)}_{2N-1}(\vec m, \eta) =  \int \, [d \vec s] \, \delta(\tr \vec s)\, \frac{\prod_{j<k}\, \sinh^2{\pi(s_j- s_k)}}{\prod^{N}_{j=1} \prod^{2N-1}_{i=1}\ch{(s_j- m_i)}}.
\ee
Integrating both sides of the identity over $\eta$, we then have 
\begin{align}\label{Id-1}
Z^{SU(N)}_{2N-1}(\vec m) = & \int \, d\eta \, \Big(\frac{e^{2\pi i \eta \sum^{2N-1}_{i=1} m_i}}{\ch{\eta}}\Big) \, \int \, [d \vec \s] \, e^{-2\pi i \eta \, \tr \s} \,\frac{\prod_{j<k}\, \sinh^2{\pi(\s_j- \s_k)}}{\prod^{N-1}_{j=1} \prod^{2N-1}_{i=1}\ch{(\s_j- m_i)}} \nn \\
= & \int \, [d \vec \s] \,d\eta \, \Big(\frac{e^{2\pi i \eta (\sum^{2N-1}_{i=1} m_i -\tr \s)}}{\ch{\eta}}\Big) \, \frac{\prod_{j<k}\, \sinh^2{\pi(\s_j- \s_k)}}{\prod^{N-1}_{j=1} \prod^{2N-1}_{i=1}\ch{(\s_j- m_i)}} \nn \\
= &  \int \, [d \vec \s] \ \frac{1}{\ch{(\tr \s - \sum^{2N-1}_{i=1} m_i)}}\,\frac{\prod_{j<k}\, \sinh^2{\pi(\s_j- \s_k)}}{\prod^{N-1}_{j=1} \prod^{2N-1}_{i=1}\ch{(\s_j- m_i)}}.
\end{align}
The final form of the matrix integral can be identified as the partition function of a $U(N-1)$ theory with $2N-1$ fundamental 
hypers and a single hyper charged under $U(1) \subset U(N-1)$ with charge $N-1$. We therefore have the following relation 
between the two partition functions :
\be \label{Id-0}
\boxed{Z^{SU(N)}_{2N-1}(\vec m) = Z^{U(N-1)}_{2N-1,\, 1^{N-1}} (\vec m, \tm= \sum_i m_i, \eta=0),}
\ee
where $\tm$ is the real mass for the Abelian hypermultiplet. Note that the equality of the partition 
function holds only when the FI parameter of the $U(N-1)$ vector multiplet is tuned to zero. Also, 
note that all the masses can be shifted by a real parameter without changing the partition function 
relation (this does not even induce a contact term since the FI parameter is set to zero). The 
$2N-1$ masses for the $SU(N)$ gauge theory live in the Cartan subalgebra of the Higgs 
branch global symmetry $U(2N-1)$. The $2N-1$ independent masses for the $U(N-1)$ gauge theory 
also live in the Cartan subalgebra of the Higgs branch global symmetry $U(2N-1)$.\\

\begin{figure}[htbp]
\begin{center}
\begin{tabular}{ccc}
 \scalebox{0.7}{\begin{tikzpicture}[
cnode/.style={circle,draw,thick,minimum size=4mm},snode/.style={rectangle,draw,thick,minimum size=8mm},pnode/.style={rectangle,red,draw,thick,minimum size=8mm}]
\node[cnode] (1) {$1$};
\node[cnode] (2) [right=.5cm  of 1]{$2$};
\node[cnode] (3) [right=.5cm of 2]{$3$};
\node[cnode] (4) [right=1cm of 3]{$N -1$};
\node[cnode] (6) [right=1 cm of 4]{$N-1$};
\node[cnode] (7) [right=1cm of 6]{{$3$}};
\node[cnode] (8) [right=0.5cm of 7]{$2$};
\node[cnode] (9) [right=0.5cm of 8]{$1$};
\node[snode, red] (13) [above=0.5cm of 4]{$1$};
\node[snode] (15) [above=0.5cm of 6]{$1$};
\draw[-] (1) -- (2);
\draw[-] (2)-- (3);
\draw[dashed] (3) -- (4);
\draw[-] (4) --(6);
\draw[dashed] (6) -- (7);
\draw[-] (7) -- (8);
\draw[-] (8) --(9);
\draw[-] (4) -- (13);
\draw[-] (6) -- (15);
\end{tikzpicture}}
& \qquad  
&\scalebox{.8}{\begin{tikzpicture}[node distance=2cm, nnode/.style={circle,draw,thick, fill, inner sep=1 pt},cnode/.style={circle,draw,thick,minimum size=1.0 cm},snode/.style={rectangle,draw,thick,minimum size=1.0 cm}]
\node[cnode] (1) at (0,0){$N-1$};
\node[snode] (2) at (0,-2){$2N-1$};
\draw[-] (1) -- (2);
\end{tikzpicture}}\\
 \scalebox{.7}{\begin{tikzpicture}
\draw[->] (15,-3) -- (15,-5);
\node[text width=0.1cm](20) at (14.5, -4) {$\CO^\alpha_{\CP}$};
\end{tikzpicture}}
&\qquad \qquad 
& \scalebox{.7}{\begin{tikzpicture}
\draw[->] (15,-3) -- (15,-5);
\node[text width=0.1cm](29) at (15.5, -4) {$\wt{\CO}^\alpha_{\CP}$};
\end{tikzpicture}}\\
 \scalebox{0.7}{\begin{tikzpicture}[
cnode/.style={circle,draw,thick,minimum size=4mm},snode/.style={rectangle,draw,thick,minimum size=8mm},pnode/.style={rectangle,red,draw,thick,minimum size=8mm}]
\node[cnode] (1) {$1$};
\node[cnode] (2) [right=.5cm  of 1]{$2$};
\node[cnode] (3) [right=.5cm of 2]{$3$};
\node[cnode] (4) [right=1cm of 3]{$N -1$};
\node[cnode] (6) [right=1 cm of 4]{$N-1$};
\node[cnode] (7) [right=1cm of 6]{{$3$}};
\node[cnode] (8) [right=0.5cm of 7]{$2$};
\node[cnode] (9) [right=0.5cm of 8]{$1$};
\node[cnode] (13) [above=0.5cm of 4]{$1$};
\node[snode] (14) [left=0.5cm of 13]{1};
\node[snode] (15) [above=0.5cm of 6]{$1$};
\draw[-] (1) -- (2);
\draw[-] (2)-- (3);
\draw[dashed] (3) -- (4);
\draw[-] (4) --(6);
\draw[dashed] (6) -- (7);
\draw[-] (7) -- (8);
\draw[-] (8) --(9);
\draw[-] (4) -- (13);
\draw[-] (13) -- (14);
\draw[-] (6) -- (15);
\end{tikzpicture}}
&\qquad \qquad 
& \scalebox{.8}{\begin{tikzpicture}[node distance=2cm, nnode/.style={circle,draw,thick, fill, inner sep=1 pt},cnode/.style={circle,draw,thick,minimum size=1.0 cm},snode/.style={rectangle,draw,thick,minimum size=1.0 cm}]
\node[cnode] (1) at (0,0){$N-1$};
\node[snode] (2) at (0,-2){$2N-1$};
\node[snode,blue] (3) at (3,0){$1$};
\draw[-] (1) -- (2);
\draw[-, thick, blue] (1)-- (3);
\node[text width=1cm](10) at (1.2, 0.2){$N-1$};
\end{tikzpicture}}
\end{tabular}
\caption{\footnotesize{The construction of the 3d mirror for $\CT^\vee$ in \figref{IRdual-Ex1}. The $S$-type operation 
${\CO}_{\vec \CP}$ is a flavoring-gauging operation. The dual operation $\wt{\CO}_{\vec \CP}$ amounts to adding 
a single hypermultiplet charged under the $U(1) \subset U(N-1)$ gauge group with charge $N-1$.}}
\label{SOp-IRdual-Ex1}
\end{center}
\end{figure}
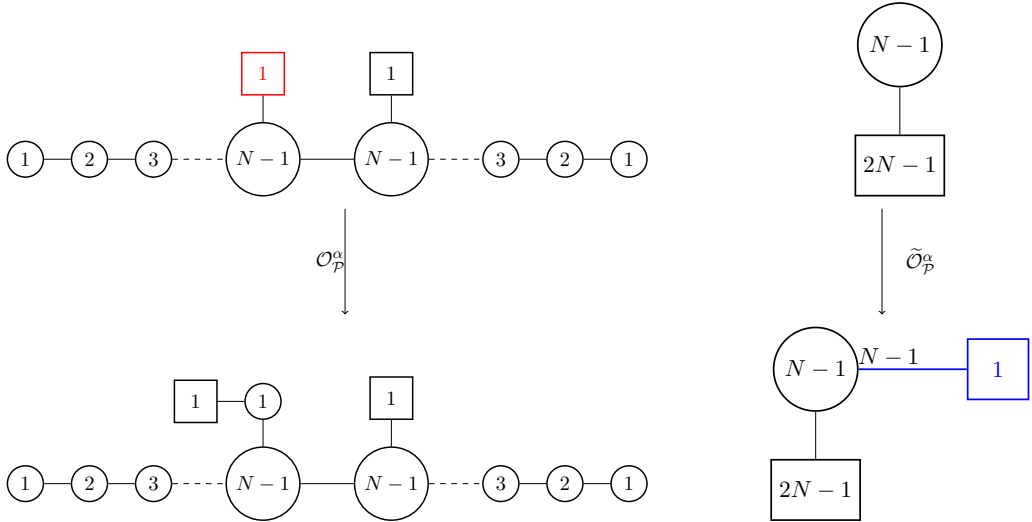

The 3d mirror for the theories $\CT$ and $\CT^\vee$ is shown in \figref{IRdual-Ex1}. This fact can be understood 
from an $S$-type operation construction starting from the mirror linear quiver pairs in the first row of \figref{SOp-IRdual-Ex1}. 
An Abelian flavoring-gauging operation on the $U(1)$ flavor node (marked in red) and its dual leads to the quiver 
gauge theory $\CT^\vee$ and its 3d mirror. The latter is then also the 3d mirror of $\CT$, since we have already 
shown (at least at the level of the three-sphere partition function) that $\CT$ and $\CT^\vee$ are IR dual.

\subsection{IR dual pairs associated with $D_p(SU(N))$}\label{IRdual-Dp}

\subsubsection{General strategy for obtaining the duality}\label{IRdual-Dp-gen}

We can now write down the general prescription for relating the quiver $\CT$ and the quiver $\CT^\vee$, associated 
with the 3d SCFT that arises from the circle reduction of $D_p(SU(N))$, for a given $(p,N)$. Firstly, let us recall the 
general structure of 3d quiver $\CT$ that arise from the analysis of \cite{Closset:2020afy, Giacomelli:2020ryy}. For $p \geq N$, we have
\begin{align} \label{DpN-CT-pgtN}
\Big[D_p(SU(N))\Big]_{3d} = & \Big[N\Big] - (N-1) - \ldots - (N-n+1) - (N-n)^{q-n} - {SU(N-n)} \nn \\
& - (N-n-1) - \ldots - (N-2n)^{q-n} - {SU(N-2n)} - \ldots  \ldots - {SU(2n)} \nn \\
&  - (2n-1) - \ldots - (n+1) - (n)^{q-n} - {SU(n)} - (n-1) - \ldots -2 -1,
\end{align}
where the integers $m$, $n$ and $q$ are given in \eref{pgtN-pars}.
For $p \leq N$, the 3d quiver $\CT$ has the form
\begin{align} \label{DpN-CT-pltN}
\Big[D_p(SU(N))\Big]_{3d} = &  \Big[N\Big] - (N-(x+1)) - \ldots - (N-m_B(x+1)) - C[N-n] - {SU(N-n)} \nn \\
& - (N-n -(x+1)) - \ldots - (N-n -m_B(x+1)) - C[N-2n] -  { SU(N-2n)}  \nn \\
& - \ldots - C[n] - {SU(n)} - (n-(x+1)) - \ldots - (n-m_B(x+1)) - C[0],
\end{align}
where the various parameters are given in \eref{pltN-pars}, and $C[Y]$ (for a positive integer $Y$) is a linear 
chain of unitary gauge groups:
\be
C[Y] = (Y + (m_A-1)x) - (Y + (m_A-2)x) - \ldots - (Y+2x) - (Y+x).
\ee

The quiver $\CT^\vee$ consisting of unitary gauge groups can then be obtained from the quiver $\CT$ using 
the following steps:

\begin{enumerate}

\item Note that all the $SU(N_c)$ factors in the 3d quiver $\CT$ have $2N_c-1$ flavors. One can therefore use 
the identity \eref{Id-0} to replace each $SU(N_c)$ factor with a $U(N_c-1)$ factor with an additional Abelian 
hypermultiplet. The resultant quiver has only unitary gauge groups with a set of Abelian hypermultiplets.

\item To the left of the replaced $SU(N_c)$ node, there will generically be a unitary node $U(N'_c)$ with 
$2N'_c -1$ fundamental hypers, in addition to an Abelian hyper. Note that the presence of the Abelian hypermultiplet 
ensures that the unitary node is not ugly. At the level of the matrix integral, one can treat the $U(N'_c)$ node as a 
$U(1) \times SU(N'_c)$ node, where the $SU(N'_c)$ factor has $2N'_c -1$ fundamental hypers. Therefore, one can apply 
the identity \eref{Id-0} again to replace the $SU(N'_c)$ factor with a $U(N'_c-1)$ factor\footnote{This is in fact the partition 
function realization of another duality that we will discuss in \cite{Dey:2022pqr}.}. 

\item This procedure is repeated from right to left along the quiver tail until we encounter a unitary gauge 
node $U(N''_c)$ with $N''_f \geq 2N'_c$ fundamental hypers. This is the quiver $\CT^\vee$ that we derived 
from the class $\CS$ mirror of $D_p(SU(N))$ by the $S$-type operations.

\end{enumerate}

\subsubsection{Example I : Dual pairs from $D_4(SU(6))$}\label{IRdual-Dp-Ex1}

Now, let us consider the IR duality between the two Lagrangian theories associated with the 3d SCFT that arises from the 
circle reduction of $D_4(SU(6))$. The quiver gauge theories are shown in \figref{IRdual-Ex2} -- the theory $\CT$ arises from the 
dimensional reduction in \cite{Closset:2020afy, Giacomelli:2020ryy} while the theory $\CT^\vee$ was derived from the $S$-type operation construction in 
\Secref{pltNEx}. The 3d mirror of these theories is given in \figref{pltNEx-3dmirr}. The quaternionic dimensions of the  
Higgs and the Coulomb branches of the theories are given as
\begin{align}
& {\rm dim}\CM^{(\CT)}_H = {\rm dim}\CM^{(\CT^\vee)}_H =14,\\
& {\rm dim}\CM^{(\CT)}_C = {\rm dim}\CM^{(\CT^\vee)}_C = 7,
\end{align}
and the Higgs branch global symmetries $G^{(\CT)}_H = G^{(\CT^\vee)}_H =SU(6) \times U(1)$ of the theories manifestly agree. 
The Coulomb branch symmetry for $\CT^\vee$ is manifestly $G^{(\CT^\vee)}_C= U(1)^3$, which agrees with the Higgs branch 
global symmetry of the 3d mirror in \figref{pltNEx-3dmirr}. However, for the theory $\CT$, one only has a  $U(1)^2$ 
Coulomb branch symmetry manifest in the UV Lagrangian. Similar to the example of $SU(N)$ gauge theory above, the theory 
has an additional $U(1)$ global symmetry which arises as an emergent symmetry in the IR. One can confirm this for example 
from a Hilbert Series computation in \cite{Giacomelli:2020ryy}.\\

\begin{figure}[htbp]
\begin{center}
\begin{tabular}{ccc}
\scalebox{.8}{\begin{tikzpicture}[node distance=2cm, nnode/.style={circle,draw,thick, fill, inner sep=1 pt},cnode/.style={circle,draw,thick,minimum size=1.0 cm},snode/.style={rectangle,draw,thick,minimum size=1.0 cm}, pnode/.style={circle,double,draw,thick, minimum size=1.0cm}]
\node[snode] (1) at (0,0){6};
\node[cnode] (2) at (2,0){4};
\node[pnode] (3) at (4,0){3};
\node[cnode] (4) at (6,0){1};
\draw[-] (1) -- (2);
\draw[-] (2)-- (3);
\draw[-] (3)-- (4);
\node[text width=.2cm](12) at (4,-2){$(\CT)$};
\node[text width=0.5cm](11) at (2, -1.1){$\eta_1$};
\node[text width=0.5cm](12) at (6, -1.1){$\eta_3$};
\end{tikzpicture}}
& \qquad   \qquad
& \scalebox{.8}{\begin{tikzpicture}[node distance=2cm, nnode/.style={circle,draw,thick, fill, inner sep=1 pt},cnode/.style={circle,draw,thick,minimum size=1.0 cm},snode/.style={rectangle,draw,thick,minimum size=1.0 cm}]
\node[snode] (1) at (0,0){6};
\node[cnode] (2) at (2,0){4};
\node[cnode] (3) at (4,0){2};
\node[cnode] (4) at (6,0){1};
\draw[-] (1) -- (2);
\draw[-] (2)-- (3);
\draw[-] (3)-- (4);
\draw[-, thick, blue] (2)-- (2,1);
\draw[-, thick, blue] (2,1)-- (6,1);
\draw[-, thick, blue] (6,1)-- (4);
\node[text width=.2cm](10) at (2.2, 0.75){4};
\node[text width=.2cm](11) at (6.2, 0.75){1};
\node[text width=.2cm](12) at (4,-2){$(\CT^\vee)$};
\node[text width=0.5cm](12) at (2, -1.1){$\eta_1$};
\node[text width=0.5cm](13) at (4, -1.1){$\eta_3$};
\node[text width=0.5cm](14) at (6, -1.1){-$\eta_3$};
\end{tikzpicture}}
\end{tabular}
\caption{The two Lagrangians for the 3d SCFT that arise from the circle reduction of the $D_4(SU(6))$.}
\label{IRdual-Ex2}
\end{center}
\end{figure}
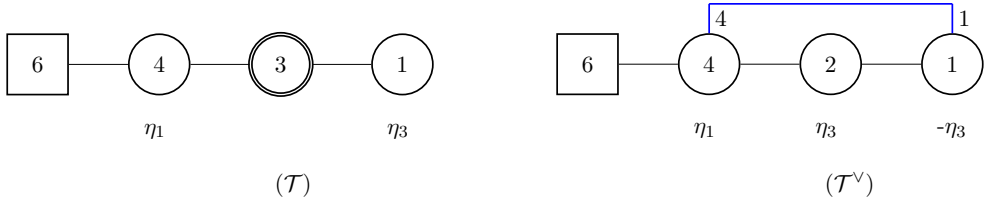

\begin{figure}[htbp]
\begin{center}
\scalebox{.7}{\begin{tikzpicture}[node distance=2cm, nnode/.style={circle,draw,thick, fill, inner sep=1 pt},cnode/.style={circle,draw,thick,minimum size=1.0 cm},snode/.style={rectangle,draw,thick,minimum size=1.0 cm}, pnode/.style={circle,double,draw,thick, minimum size=1.0cm}]
\node[cnode] (1) at (0,0){1};
\node[cnode] (2) at (2,0){2};
\node[cnode] (3) at (4,0){3};
\node[cnode] (4) at (6,0){4};
\node[cnode] (5) at (8,0){3};
\node[cnode] (6) at (7,2){1};
\node[snode] (7) at (6,-2){1};
\node[snode] (8) at (8,-2){1};
\node[snode] (9) at (9,2){1};
\draw[-] (1) -- (2);
\draw[-] (2)-- (3);
\draw[-] (3)-- (4);
\draw[-] (5)-- (4);
\draw[-] (6)-- (4);
\draw[-] (5)-- (6);
\draw[-] (9)-- (6);
\draw[-] (4)-- (7);
\draw[-] (5)-- (8);
\node[] (9) at (0,-0.75){};
\end{tikzpicture}}
\caption{The 3d mirror of the theories $\CT$ and $\CT^\vee$ in \figref{IRdual-Ex2}.}
\label{pltNEx-3dmirr}
\end{center}
\end{figure}

Let us now demonstrate how the partition functions of the theory $\CT$ and $\CT^\vee$ are related. The partition 
function for $\CT$ is given as 
\begin{align}\label{PF-CT-1}
Z^{(\CT)}(\vec m, \vec \eta) = & \int\,\Big[d\vec \s^1\Big]\,\Big[d\vec s^2\Big]\, d\vec \s^3\, Z_{\rm FI}(\vec\s^1, \eta_1)\, Z_{\rm FI}(\s^3, \eta_3)\,\delta(\tr s^2)\,
\prod_{\gamma=1,3}\, Z^{\rm vec}_{\rm 1-loop}(\vec \s^\gamma) \nn \\
& \times Z^{\rm vec}_{\rm 1-loop}(\vec s^2) \, Z^{\rm fund}_{\rm 1-loop}(\vec \s^1, \vec m)\, Z^{\rm bif}_{\rm 1-loop}(\vec \s^1, \vec s^2, 0)\,
Z^{\rm bif}_{\rm 1-loop}(\vec s^2, \s^3,0),
\end{align}
where $\vec \s^1, \s^3$ denote the integration variables associated with the $U(4)$ and $U(1)$ gauge groups respectively, while $\vec s^2$ denotes 
the integration variable associated with the $SU(3)$ gauge group. Isolating the $\vec s^2$-dependent part of the matrix integral, and using the 
relation \eref{Id-1} (for the special case of $N=3)$, we have the following identity:
\begin{align}
& \int \, [d \vec s^2] \, \delta(\tr \vec s^2)\, Z^{\rm vec}_{\rm 1-loop}(\vec s^2)\, Z^{\rm bif}_{\rm 1-loop}(\vec \s^1, \vec s^2, 0)\, Z^{\rm bif}_{\rm 1-loop}(\vec s^2, \s^3,0) \nn \\
&=   \int \, [d \vec \s^2] \, \frac{1}{\ch{(\tr \s^2 - \tr \s^1 - \s^3)}}\,Z^{\rm vec}_{\rm 1-loop}(\vec \s^2)\, Z^{\rm bif}_{\rm 1-loop}(\vec \s^2, \vec \s^1, 0)\,Z^{\rm bif}_{\rm 1-loop}(\vec \s^2, \s^3, 0).
\end{align}
Substituting the above identity in \eref{PF-CT-1}, and performing the change of variable $\s^3 \to -\s^3 + \tr \s^2$, the partition function of the theory $\CT$ is 
given as:
\begin{align} \label{PF-CT-2}
Z^{(\CT)}(\vec m, \vec \eta) = & \int \prod^{3}_{\gamma'=1}\,\Big[d\vec \s^{\gamma'}\Big] \, Z_{\rm FI}(\vec\s^1, \eta_1)\,Z_{\rm FI}(\s^3, \eta_3) \,Z_{\rm FI}(\s^3, -\eta_3)\,\prod^3_{\gamma=1}\, Z^{\rm vec}_{\rm 1-loop}(\vec \s^\gamma) \nn \\
& \times \, Z^{\rm bif}_{\rm 1-loop}(\vec \s^2, \vec \s^1, 0)\,Z^{\rm bif}_{\rm 1-loop}(\vec \s^2, \s^3, 0)\,
Z^{\rm fund}_{\rm 1-loop}(\vec \s^1, \vec m)\,  Z^{\rm hyper}_{\rm 1-loop} (\vec \s^1, \s^3,0),
\end{align} 
where $Z^{\rm hyper}_{\rm 1-loop}$ is the 1-loop contribution of a single hyper to the matrix integral integrand given as
\be
Z^{\rm hyper}_{\rm 1-loop} (\vec \s^1, \s^3,0) =\frac{1}{\ch{(\tr \s^1 -  \s^3)}}.
\ee
The RHS of \eref{PF-CT-2} can be identified with the partition function of theory $\CT^\vee$, which leads to the following relation 
between the partition functions of the theories $\CT$ and $\CT^\vee$:
\be
\boxed{Z^{(\CT)}(\vec m; \eta_1, \eta_3) = Z^{(\CT^\vee)}(\vec m; \eta_1, \eta_3, -\eta_3),}
\ee
where $\vec m$ are the masses of the fundamental hypermultiplets on both sides, $\eta_1, \eta_3$ are the FI parameters of the 
$U(4)$ and the $U(1)$ gauge group factors of the theory $\CT$ respectively, while $\eta_1, \eta_3, -\eta_3$ are the FI parameters of the 
$U(4), U(2)$ and the $U(1)$ gauge group factors of the theory $\CT^\vee$ respectively. Note that the partition functions do not agree 
for completely generic FI parameters of $\CT^\vee$ -- they only do so if the FI parameters of the $U(2)$ and the $U(1)$ gauge group factors of $\CT^\vee$ 
are related as above.

\subsubsection{ Example II : Dual pairs from $D_9(SU(3))$}\label{IRdual-Dp-Ex2}

Now, let us consider the IR duality between the two Lagrangian theories associated to the 3d SCFT that arises from the 
circle reduction of $D_9(SU(3))$. The quiver gauge theories are shown in \figref{IRdual-Ex3} -- the theory $\CT$ arises from the 
dimensional reduction in \cite{Closset:2020afy, Giacomelli:2020ryy} while the theory $\CT^\vee$ was derived from the $S$-type operation construction in 
\Secref{pgtNEx}. The 3d mirror of these theories is given in \figref{pgtNEx-3dmirr}. The quaternionic dimensions of the  
Higgs and the Coulomb branches of the theories are given as
\begin{align}
& {\rm dim}\CM^{(\CT)}_H = {\rm dim}\CM^{(\CT^\vee)}_H =5,\\
& {\rm dim}\CM^{(\CT)}_C = {\rm dim}\CM^{(\CT^\vee)}_C = 7,
\end{align}
and the Higgs branch global symmetries $G^{(\CT)}_H = G^{(\CT^\vee)}_H =SU(3) \times U(1)^2$ of the theories manifestly agree. 
The Coulomb branch symmetry for $\CT^\vee$ is manifestly $G^{(\CT^\vee)}_C= SU(2)^3 \times U(1)^3$, which agrees with the Higgs branch 
global symmetry of the 3d mirror in \figref{pgtNEx-3dmirr}. As before, for the theory $\CT$, the full Coulomb branch symmetry 
arises in the IR as an emergent symmetry.  This can be confirmed from a Hilbert Series computation, for example, as discussed 
in \cite{Giacomelli:2020ryy}.\\

\begin{figure}[htbp]
\begin{center}
\begin{tabular}{ccc}
\scalebox{.7}{\begin{tikzpicture}[
cnode/.style={circle,draw,thick, minimum size=1.0cm},snode/.style={rectangle,draw,thick,minimum size=1cm},pnode/.style={circle,double,draw,thick, minimum size=1.0cm}]
\node[cnode] (1) at (0,0){2};
\node[snode] (2) at (0,-2){3};
\node[cnode] (3) at (2,0){2};
\node[pnode] (4) at (4,0){2};
\node[cnode] (5) at (6,0){1};
\node[cnode] (6) at (8,0){1};
\node[snode] (7) at (8,-2){1};
\draw[-] (1) -- (2);
\draw[-] (1)-- (3);
\draw[-] (3)-- (4);
\draw[-] (4)-- (5);
\draw[-] (5)-- (6);
\draw[-] (6)-- (7);
\node[text width=0.1cm](30) at (4,-3){$(\CT)$};
\end{tikzpicture}}
&\qquad  \qquad
&\scalebox{.7}{\begin{tikzpicture}[node distance=2cm,cnode/.style={circle,draw,thick,minimum size=8mm},snode/.style={rectangle,draw,thick,minimum size=8mm},pnode/.style={rectangle,red,draw,thick,minimum size=8mm}]
\node[cnode] (1) at (-3,0) {$2$};
\node[snode] (2) at (-5,0) {$3$};
\node[cnode] (3) at (-1,2) {$1$};
\node[cnode] (4) at (-2,-2) {$1$};
\node[cnode] (5) at (0,-2) {$1$};
\node[cnode] (6) at (1,0) {$1$};
\node[cnode] (7) at (3,0) {$1$};
\node[snode] (8) at (5,0) {$1$};
\draw[thick] (1) -- (2);
\draw[thick, blue] (1) -- (-3.5,0.5);
\draw[thick, blue] (3) -- (-1.5, 2.5);
\draw[thick, blue] (-3.5,0.5) -- (-1.5, 2.5);
\draw[thick] (1) -- (4);
\draw[thick] (4) -- (5);
\draw[thick] (5) -- (6);
\draw[thick] (6) -- (7);
\draw[thick] (7) -- (8);
\draw[thick] (3) -- (6);
\node[text width=0.1cm](31) at (-3.1,0.6){2};
\node[text width=0.1cm](32) at (-1.6,2){1};
\node[text width=0.1cm](30) at (-1,-3){$(\CT^\vee)$};
\end{tikzpicture}}
\end{tabular}
\caption{The two Lagrangians for the 3d SCFT that arise from the circle reduction of the $D_9(SU(3))$.}
\label{IRdual-Ex3}
\end{center}
\end{figure}

\begin{figure}[htbp]
\begin{center}
\scalebox{.7}{\begin{tikzpicture}[
cnode/.style={circle,draw,thick, minimum size=1.0cm},snode/.style={rectangle,draw,thick,minimum size=1cm}]
\node[cnode] (9) at (0,1){1};
\node[snode] (10) at (0,-1){1};
\node[cnode] (11) at (2, 0){2};
\node[cnode] (12) at (4, 1){1};
\node[cnode] (13) at (4, -1){1};
\node[snode] (14) at (6, 1){$2$};
\node[snode] (15) at (6, -1){$2$};
\draw[-] (9) -- (11);
\draw[-] (10) -- (11);
\draw[-] (12) -- (11);
\draw[-] (13) -- (11);
\draw[-] (12) -- (14);
\draw[-] (13) -- (15);
\draw[line width=0.75mm, black] (12) to (13);
\node[text width=0.1cm](20) at (4.5,0){$2$};
\end{tikzpicture}}
\caption{The 3d mirror of the theories $\CT$ and $\CT^\vee$ in \figref{IRdual-Ex3}.}
\label{pgtNEx-3dmirr}
\end{center}
\end{figure}

Let us now demonstrate how the partition functions of the theory $\CT$ and $\CT^\vee$ are related. The partition 
function for $\CT$ is given as 
\begin{align}\label{PF-CT-3}
Z^{(\CT)}(\vec m, \vec \eta) = & \int\, \prod^{5}_{\gamma=1}\,\Big[d\vec \s^{\gamma}\Big]_{\gamma \neq 3}\,\Big[d\vec s^3\Big]\, \prod^{5}_{\gamma=1}\,Z_{\rm FI}(\vec\s^\gamma, \eta_\gamma)|_{\gamma \neq 3}\,\delta(\tr s^3)\, \prod^2_{\gamma=1}\, Z^{\rm vec}_{\rm 1-loop}(\vec \s^\gamma) \nn \\
& \times Z^{\rm vec}_{\rm 1-loop}(\vec s^3) \, Z^{\rm fund}_{\rm 1-loop}(\vec \s^1, \vec m^1)\, Z^{\rm bif}_{\rm 1-loop}(\vec \s^1, \vec \s^2, 0)\,
Z^{\rm bif}_{\rm 1-loop}(\vec \s^2, \vec s^3,0)\, \nn \\
&  \times Z^{\rm bif}_{\rm 1-loop}(\vec s^3, \s^4, 0)\, Z^{\rm bif}_{\rm 1-loop}(\s^4, \s^5, 0)\, Z^{\rm fund}_{\rm 1-loop}(\s^5, m^5), 
\end{align}
where $\{\vec \s^\gamma\}_{\gamma \neq 3}$ denote the integration variables associated with the unitary gauge groups, while $\vec s^3$ denotes 
the integration variable associated with the $SU(2)$ gauge group. Isolating the $\vec s^3$-dependent part of the matrix integral, and using the 
relation \eref{Id-1} (for the special case of $N=2)$, we have the following identity:
\begin{align}
& \int \, [d \vec s^3] \, \delta(\tr \vec s^3)\, Z^{\rm vec}_{\rm 1-loop}(\vec s^3)\, Z^{\rm bif}_{\rm 1-loop}(\vec s^3, \vec \s^2, 0)\, Z^{\rm bif}_{\rm 1-loop}(\vec s^3, \s^4,0) \nn \\
&=   \int \, d  \s^3 \, \frac{1}{\ch{(\s^3 - \tr \s^2 - \s^4)}}\,Z^{\rm vec}_{\rm 1-loop}(\vec \s^3)\, Z^{\rm bif}_{\rm 1-loop}(\s^3, \vec \s^2, 0)\,
Z^{\rm bif}_{\rm 1-loop}(\s^3, \s^4, 0).
\end{align}
Substituting the above identity in \eref{PF-CT-3}, the partition function for $\CT$ can be written as
\begin{align}\label{PF-CT-4}
Z^{(\CT)}(\vec m, \vec \eta) = & \int\, \prod^{5}_{\gamma=1}\,\Big[d\vec \s^{\gamma}\Big]\, \prod^{5}_{\gamma=1}\,Z_{\rm FI}(\vec\s^\gamma, \eta_\gamma)|_{\gamma \neq 3} \, \prod^2_{\gamma=1}\, Z^{\rm vec}_{\rm 1-loop}(\vec \s^\gamma)\,Z^{\rm fund}_{\rm 1-loop}(\vec \s^1, \vec m^1) \nn \\
 \times & \, \prod^{4}_{\gamma=1} Z^{\rm bif}_{\rm 1-loop}(\vec \s^\gamma, \vec \s^{\gamma +1}, 0)\,
Z^{\rm fund}_{\rm 1-loop}(\s^5, m^5)\, Z^{\rm hyper}_{\rm 1-loop}(\s^3, \vec \s^2, \s^4), 
\end{align}
where $Z^{\rm hyper}_{\rm 1-loop}$ is the contribution of a single hypermultiplet charged under $U(1)_2 \subset U(2)_2$, $U(1)_3$ and $U(1)_4$ 
to the matrix model integrand:
\be
Z^{\rm hyper}_{\rm 1-loop}(\s^3, \vec \s^2, \s^4) = \frac{1}{\ch{(\tr \s^2 - \s^3 + \s^4)}}. 
\ee
After a simple change of variables : $\s^4 \to -\s^4 + \s^3$ and $\s^5 \to -\s^5 + \s^3$, one can check that the matrix integral on the RHS of 
\eref{PF-CT-4} is the partition function of the following quiver:

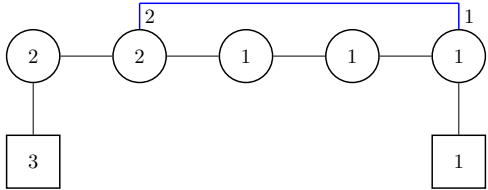
\begin{center}
\scalebox{.7}{\begin{tikzpicture}[
cnode/.style={circle,draw,thick, minimum size=1.0cm},snode/.style={rectangle,draw,thick,minimum size=1cm},pnode/.style={circle,double,draw,thick, minimum size=1.0cm}]
\node[cnode] (1) at (0,0){2};
\node[snode] (2) at (0,-2){3};
\node[cnode] (3) at (2,0){2};
\node[cnode] (4) at (4,0){1};
\node[cnode] (5) at (6,0){1};
\node[cnode] (6) at (8,0){1};
\node[snode] (7) at (8,-2){1};
\draw[-] (1) -- (2);
\draw[-] (1)-- (3);
\draw[-] (3)-- (4);
\draw[-] (4)-- (5);
\draw[-] (5)-- (6);
\draw[-] (6)-- (7);
\draw[-, thick, blue] (3)-- (2,1);
\draw[-, thick, blue] (2,1)-- (8,1);
\draw[-, thick, blue] (8,1)-- (6);
\node[text width=.2cm](10) at (2.2, 0.75){2};
\node[text width=.2cm](11) at (8.2, 0.75){1};
\end{tikzpicture}}
\end{center}

Note that after the change of variables, $\s^4$ labels the rightmost $U(1)$ gauge node, while $\s^5$ labels the $U(1)$ gauge node immediately to its left.
The contribution of the Abelian hypermultiplet to the matrix integral also transforms as:
\be
Z^{\rm hyper}_{\rm 1-loop}(\s^3, \vec \s^2, \s^4) \to Z^{\rm hyper}_{\rm 1-loop}(\vec \s^2, \s^4) = \frac{1}{\ch{(\tr \s^2 - \s^4)}}.
\ee
The second $U(2)$ gauge node from the left has three fundamental hypers and a single hyper charged only under the $U(1)$ subgroup with charge 2. 
One can again use the identity \eref{Id-1} to rewrite the matrix integral in a way that the desired duality becomes manifest.

To do this, let us focus on the $\vec \s^2$-dependent part of the matrix integral in \eref{PF-CT-4}. We can rewrite the $U(2)$ matrix integral 
explicitly as a $U(1) \times SU(2)$ integral by the following the change of variables:
\be
u =\frac{1}{2}\, \tr \vec \s^2, \qquad s = \frac{1}{2}\,(\s^2_1 - \s^2_2).
\ee
In terms of the new integration variables, the $\vec \s^2$-dependent part of the matrix integral can be written as
\begin{align} \label{Id-2}
& \int\,\Big[d\vec \s^{2}\Big]\, Z_{\rm FI}(\vec\s^2, \eta_2)\, Z^{\rm vec}_{\rm 1-loop}(\vec \s^2) \, Z^{\rm bif}_{\rm 1-loop}(\vec \s^1, \vec \s^{2}, 0)\,
 Z^{\rm bif}_{\rm 1-loop}(\vec \s^2, \s^{3}, 0)\, Z^{\rm hyper}_{\rm 1-loop}(\vec \s^2, \s^4) \nn \\
=&  2\,\int\,\frac{d\,s}{2!}\,\frac{\sinh^2{\pi(2s)}}{\prod_j \ch{(s \pm (\s^1_j-u))}\,\ch{(s \pm (\s^3- u))}}\cdot  \int\, du\, \frac{e^{4\pi i \eta_2 u}}{\ch{(2u - \s^4)}} \nn \\
=&2\,  \int\, dv\, \frac{1}{\ch{(v- \tr \s^1 + \s^3 +u)}\,\prod_j \ch{(v+u -\s^1_j)}\,\ch{(v-u +\s^3)}}\nn \\
& \times \int\, du\, \frac{e^{4\pi i \eta_2 u}}{\ch{(2u - \s^4)}},
\end{align}
where for the final equality we have used \eref{Id-1} (for $N=2$) with the masses of the three hypers in the fundamental representation of 
$SU(2)$ as $m_1=: \s^1_1 - u$, $m_2=: \s^1_2 - u$, and $m_3=: -\s^3 +u$. Substituting the identity \eref{Id-2} in \eref{PF-CT-4}, and 
implementing the following change of integration variables:
\begin{align}
& \bs^0 = u + v + \s^3, \\
& \bs^1_i =\s^1_i,\\
& \bs^2= u +v, \\
& \bs^3= -\s^5 +\s^3 + \bs^2 = -\s^5 + \s^3+ u +v, \\
& \bs^4= - \s^4 +\s^3 + \bs^2 = -\s^4 +\s^3 + u +v,\\
& \bs^5= v-u + \s^3,
\end{align}
the partition function of $\CT$ can be written in the following form\footnote{Note that the Jacobian factor of 1/2 precisely cancels the factor of 2 sitting in 
front of the integral in \eref{Id-2}.}:
\begin{align}\label{PF-CT-5}
Z^{(\CT)}(\vec m, \vec \eta) = & \int\, \prod^{5}_{\gamma=0}\,\Big[d\vec \bs^{\gamma}\Big]\,\prod^{5}_{\gamma=0}\,Z_{\rm FI}(\vec\bs^\gamma, \overline{\eta}_\gamma)\,
Z^{\rm vec}_{\rm 1-loop}(\vec \bs^1)\,Z^{\rm fund}_{\rm 1-loop}(\vec \bs^1, \vec m^1)\, Z^{\rm hyper}_{\rm 1-loop}(\bs^0, \vec \bs^1, m^5) \nn \\
& \times \,\prod^{4}_{\gamma=1} Z^{\rm bif}_{\rm 1-loop}(\vec \bs^\gamma, \vec \bs^{\gamma +1}, 0)\, Z^{\rm bif}_{\rm 1-loop}(\vec \bs^0, \vec \bs^{4}, 0)\, 
Z^{\rm fund}_{\rm 1-loop}(\bs^5, m^5),
\end{align}
where the contribution of the Abelian hypermultiplet to the integrand is given as
\be
Z^{\rm hyper}_{\rm 1-loop}(\bs^0, \vec \bs^1, m^5) = \frac{1}{\ch{(\bs^0 - \tr \bs^1 + m^5)}}.
\ee
The RHS of \eref{PF-CT-5} can now be identified as the partition function of the theory $\CT^\vee$, with the gauge nodes labelled as:

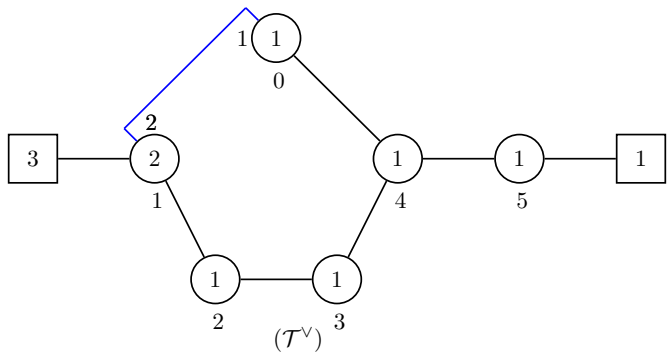
\begin{center}
\scalebox{.8}{\begin{tikzpicture}[node distance=2cm,cnode/.style={circle,draw,thick,minimum size=8mm},snode/.style={rectangle,draw,thick,minimum size=8mm},pnode/.style={rectangle,red,draw,thick,minimum size=8mm}]
\node[cnode] (1) at (-3,0) {$2$};
\node[snode] (2) at (-5,0) {$3$};
\node[cnode] (3) at (-1,2) {$1$};
\node[cnode] (4) at (-2,-2) {$1$};
\node[cnode] (5) at (0,-2) {$1$};
\node[cnode] (6) at (1,0) {$1$};
\node[cnode] (7) at (3,0) {$1$};
\node[snode] (8) at (5,0) {$1$};
\draw[thick] (1) -- (2);
\draw[thick, blue] (1) -- (-3.5,0.5);
\draw[thick, blue] (3) -- (-1.5, 2.5);
\draw[thick, blue] (-3.5,0.5) -- (-1.5, 2.5);
\draw[thick] (1) -- (4);
\draw[thick] (4) -- (5);
\draw[thick] (5) -- (6);
\draw[thick] (6) -- (7);
\draw[thick] (7) -- (8);
\draw[thick] (3) -- (6);
\node[text width=0.1cm](31) at (-3.1,0.6){2};
\node[text width=0.1cm](32) at (-1.6,2){1};
\node[text width=0.1cm](30) at (-1,-3){$(\CT^\vee)$};
\node[text width=0.1cm](31) at (-3.1,0.6){2};
\node[text width=0.1cm](40) at (-3,-0.7){1};
\node[text width=0.1cm](41) at (-2,-2.7){2};
\node[text width=0.1cm](42) at (0,-2.7){3};
\node[text width=0.1cm](43) at (1,-0.7){4};
\node[text width=0.1cm](44) at (3,-0.7){5};
\node[text width=0.1cm](45) at (-1, 1.3){0};
\end{tikzpicture}}
\end{center}

The FI parameters of $\CT^\vee$ are given in terms of the FI parameters of $\CT$ as follows:

\begin{align}
& \overline{\eta}_0= \eta_2,\\
& \overline{\eta}_1= \eta_1, \\
& \overline{\eta}_2= - (\eta_4 + \eta_5), \\
& \overline{\eta}_3=  \eta_5, \\
& \overline{\eta}_4= \eta_4,\\
&\overline{\eta}_5= -\eta_2.
\end{align}
The fundamental hypermultiplet masses $\vec m^1$ of the $U(2)$ gauge group in $\CT$ are mapped to the masses $\vec{\overline{m}}^1$ of 
fundamental hypers for the $U(2)$ gauge group, i.e. $\vec{\overline{m}}^1 = \vec m^1$.
The mass of the fundamental hyper for $U(1)_5$ (i.e. the gauge group labelled by 
$\bs^5$) and the mass of the Abelian hyper are given as:
\be
\overline{m}^5 = -m^5, \qquad \overline{m}_{\rm hyper} = m^5.
\ee

Therefore, the partition functions of the theories $\CT$ and $\CT^\vee$ are related as follows:
\be
\boxed{Z^{(\CT)}(\vec m^1, m^5; \eta_1, \eta_2, \eta_4, \eta_5 ) = Z^{(\CT^\vee)}(\vec{\overline{m}}^1, \overline{m}^5, \overline{m}_{\rm hyper} ; \vec{\overline{\eta}}),}
\ee
where the relation between themasses and the FI parameters of $\CT$ and $\CT^\vee$ are given above.
As we observed in the previous example, the partition functions do not agree for completely generic FI parameters 
of $\CT^\vee$ -- they only do so if the FI parameters of $\CT^\vee$ obey two non-trivial relations so that the number 
of parameters match on both sides.

\section* {Acknowledgements}
The author would like to thank the organizers of the Simons Summer Workshop 2021, where part of the work was completed.
The author would also like to thank  Stefano Cremonesi, Amihay Hanany, Noppadol Mekareeya, Jaewon Song, Cumrun Vafa 
and Dan Xie for comments and discussion. 
This work is partially supported at the Johns Hopkins University by NSF grant PHY-1820784, and the Simons Collaboration on 
Global Categorical Symmetries.

\bibliography{cpn1-1}
\bibliographystyle{JHEP}

\end{document}